# Self-regulated emergence of heavy-tailed weight distributions in evolving complex network architectures

Jia Li[1,2*], Cees van Leeuwen[1,2], Roman Bauer[3], Ilias Rentzeperis[4,5]

[1] Brain and Cognition, KU Leuven, Leuven, Belgium

[2] Center for Cognitive Science, University of Kaiserslautern-Landau, Kaiserslautern, Germany

[3] NICE Research Group, Computer Science Research Centre, University of Surrey, Guildford, United Kingdom

[4] Institute of Optics, Spanish National Research Council (CSIC), Madrid, Spain

[5] Université Paris-Saclay, CNRS, CentraleSupélec, Laboratoire des Signaux et Systèmes, Paris, France

[*] **Correspondence:** jia.li@kuleuven.be

**Abstract**

The nervous system continuously adjusts connection strengths and reorganizes its structure to form and maintain complex connectivity patterns with heavy-tailed weight distributions. We propose a parsimonious model in which structural and synaptic plasticity are driven by common diffusion dynamics. Synaptic plasticity alone generates heavy-tailed weight distributions, but only when activity spreading remains predominantly local. However, when combined with structural plasticity through adaptive rewiring, the model also generates these distributions with more extensive activity flow. Furthermore, adaptive rewiring produces complex network structures with convergent-divergent circuits. These circuits contain motifs that are pervasive in nervous systems and are responsible for context-sensitive signal propagation and enhanced signal to noise ratios. Our model robustly reproduces these results across diverse dynamical regimes while capturing key connectivity features of both *C. elegans* and mouse brain networks. These findings suggest that the underlying principles are shared across species of varying complexity.

Word count: 144

# Introduction

The nervous system can be considered as a network of interconnected neurons whose connections are weighted according to their synaptic strength. Invariably, the network features a few strong connections among many weaker ones, resulting in a skewed, heavy-tailed weight distribution (1–6). Generative models have proposed various mechanisms for producing such a distribution through iterative weight adjustment (7–13). However, these models tend to overlook the fact that the weight dynamics in biological systems occur within continuously evolving network structures. Structural plasticity, the rewiring of connectivity structures through the addition or pruning of synapses, constantly alters brain networks (14–17). We propose that connection weights and connectivity structures co-evolve adaptively. We present a model in which plasticity mechanisms, regulated by diffusion dynamics, drive the adaptive update of network connectivity weights and structure. This yields both heavy-tailed weight distributions and complex network structures consistent with biological connectomes of various complexities.

The proposed model is based on adaptive rewiring, a parsimonious mechanism for structural plasticity (18, 19). Adaptive rewiring dynamically creates shortcuts between highly interactive neural units while pruning underused connections. When driven by spontaneous activity, adaptive rewiring generates and maintains modular, small-world networks with a rich-club core (20–22), thereby sharing crucial organizational features with biological neural networks (18, 19, 23–28). While these global topological properties provide useful insights for characterizing networks, including undirected binary ones, they may also present certain limitations. First, some of these measures provide a generic characterization of network organization. For example, two networks with markedly different structural properties may exhibit similar small-world values (25). Second, connections in biological brains are directional and vary in strength. These properties strongly influence function and may be overlooked when using simplified binary or undirected network models.

To address these issues, recent work has extended adaptive rewiring to weighted and directed networks (29–31). In these networks, signal propagation is modeled by advection (32) and consensus (33) dynamics, both of which are generalizations of diffusion on a graph. In the context of communication in brain networks, diffusion models abstract away biophysical details to focus on network information flow. This reveals relationships between regions or neurons that are not necessarily directly connected. These models propose that signals propagate along multiple paths, in a manner analogous to random walks (34). Diffusion dynamics have been shown to predict functional connectivity, i.e., statistical dependencies between patterns of activity in different regions, based on the underlying anatomical connectivity (35). Thus, adaptive rewiring and adaptive weight adjustment can both be viewed as functional network optimization.

Driven by advection/consensus dynamics, adaptive rewiring gives rise to connectivity patterns that resemble key motifs of synaptic circuits ubiquitous in the nervous system. Advection produces convergent motifs, in which a unit receives multiple inputs (Fig. 1A), while consensus produces divergent motifs that broadcast their output to many nodes (Fig. 1B). Both types of circuits support complex computations and information transmission (36). For instance, converging inputs that are similar and fluctuate independently produce outputs with an improved signal-to-noise ratio. Conversely, divergent motifs efficiently broadcast signals by distributing outputs to many parallel pathways. Combining consensus and advection dynamics in a dynamically evolving network model generates convergent-divergent units, which enable context-sensitive computing in feed-forward sensory processing (36). These units integrate inputs via convergent hubs, process information in an internal core, a module consisting of intermediate nodes on the paths from convergent to divergent hubs, and distribute outputs via divergent hubs (Fig. 1C). Although predominantly viewed as instruments of feedforward processing, these units may also contain recurrent connections (Fig. 1D; 31, 37).

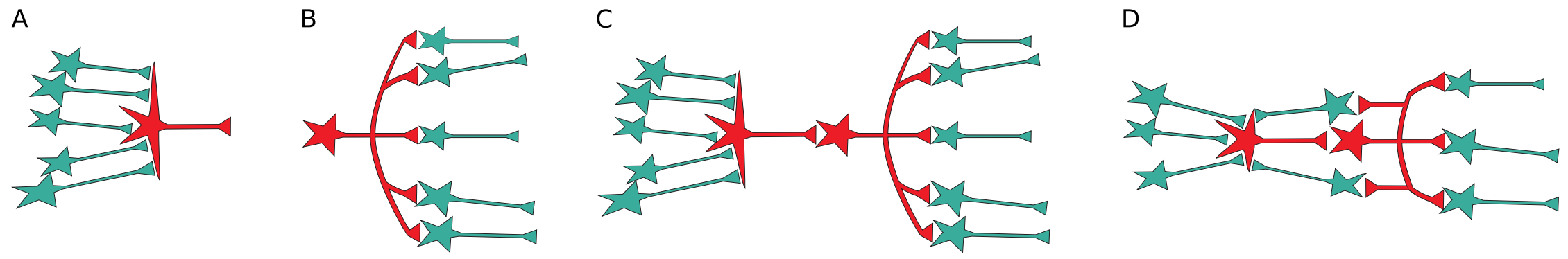

**Fig. 1. Convergent and divergent circuit motifs.** (A) A convergent circuit: a neuron (in red) receives multiple inputs. The star-like terminations represent the dendrites, and the endings on the opposite side represent axon terminals. (B) A divergent circuit: a neuron (in red) sends its signal to many neurons. (C) A feedforward convergent-divergent circuit. (D) A recurrent convergent-divergent circuit.

So far, adaptive rewiring models have not taken synaptic plasticity into account; fixed weights have been used since rewiring can be performed without adjusting them. However, incorporating weight adjustment into these models should provide a robust explanation for why networks exhibit heavy-tailed weight distributions. To that end, we provide a dual-adaptive algorithm combining adaptive rewiring and weight adjustment. The model parsimoniously represents both synaptic and structural plasticity as processes driven by the same diffusion dynamics (Fig. 2). These processes are Hebbian-like in a broad sense: interactions associated with high diffusion between nodes are reinforced through the strengthening or addition of connections, whereas interactions associated with low diffusion are weakened through synaptic attenuation or pruning. However, the model is not Hebbian in the classical sense of activity-dependent synaptic strengthening based on correlated pre- and postsynaptic firing, nor does it implement spike-timing-dependent plasticity (STDP), long-term potentiation (LTP), or long-term depression (LTD) mechanisms explicitly.

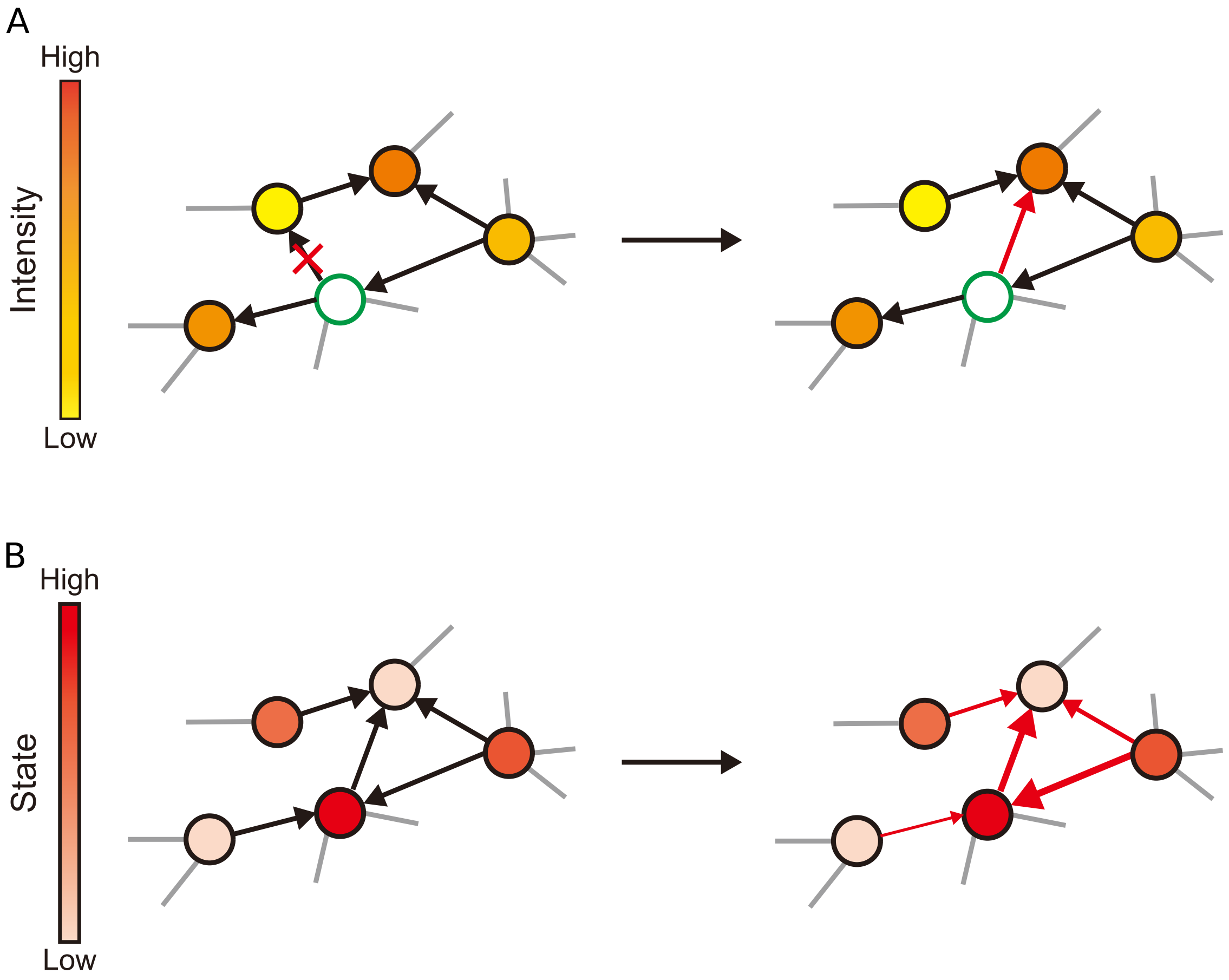

**Fig 2. Schema of adaptive rewiring and adaptive weight adjustment.** (A) In each adaptive rewiring step, a random node (green) is selected. Its connection to the neighbor with which it interacts least (marked with a red cross) is pruned, and a new connection (red arrows) is added to the node with which it interacts most but is indirectly linked to. (B) In each adaptive weight adjustment step, all connections update their weights according to a Hebbian-like rule (Equation 10), with increments proportional to the product of the sending and receiving nodes' states, followed by synaptic normalization to prevent unbounded weight growth.

Diffusion in the model drives the equilibration of node states (concentrations), either toward a common value in consensus dynamics or toward connectivity-dependent values in advection dynamics that preserve the sum of concentrations. In this respect, the framework may be viewed as a minimal and abstract approximation of homeostatic processes regulating network activity. Experimental evidence further suggests that some forms of plasticity do not operate strictly according to Hebbian or anti-Hebbian rules but instead contribute to the stabilization and homeostasis of neuronal dynamics (16).

We demonstrate that adaptive weight adjustment alone, without rewiring, can already produce heavy-tailed weight distributions and rudimentary modular and clustering structures. However, this only occurs under the extremely unrealistic restriction where network flow remains predominantly local and has only weak influence beyond immediate neighbors. When combined with adaptive rewiring, where the network weights and its complex structure change concurrently, the network develops heavy-tailed weight distributions across a significantly wider range of diffusion. Moreover, adaptive weight adjustment does not abolish the convergent-divergent structures resulting from adaptive rewiring.

Having established these model features, we proceed by fitting the dual-adaptive model to the weight distributions and convergent-divergent units extracted from two neuronal connectomes, *C. elegans* (38) and mouse retina (39). The model successfully reproduces their heavy-tailed weight distributions and qualitatively captures the denser and heavier internal cores in their convergent-divergent units. These results suggest Hebbian-like principles driven by diffusion dynamics as a candidate mechanism governing the evolution of both the strength and structure of biological network connections.

# Methods

We focus on neuron-level connectivity, where synaptic rewiring and weight adjustment shape the network architecture. A network is represented by its adjacency matrix $A = [A_{ij}]$, where $A_{ij} = w_{ij}$ is the weight of the directed connection from $j$ (tail) to $i$ (head), denoted as $j \rightarrow i$ (Fig. S1). If $j \rightarrow i$ exists, $w_{ij} > 0$; otherwise, $w_{ij} = 0$. Unless otherwise stated, the network has 100 nodes and 912 randomly assigned connections. Initial connection weights are sampled from a normal distribution and normalized so that their sum equals the number of connections. A full account of the network settings, mathematical formulations and analytical derivations can be found in Supplementary Methods.

## Diffusion dynamics on the directed network

The local state, $x_i$, of each node $i$ is referred to as its concentration. Activity flow on the directed network is modeled using either advection (32) or consensus (33) dynamics, with the dynamical regime selected probabilistically at each iteration of the algorithm (see below, Adaptive plasticity rules). Both are based on the heat (diffusion) equation, which describes how a quantity spreads across a medium over time. On an undirected graph, i.e. a symmetric adjacency matrix, $A$, with each connection having equal weight in both directions, the diffusion process has an explicit solution:

$$\boldsymbol{x}(\tau) = e^{-L\tau}\boldsymbol{x}(0) \tag{1}$$

where $\boldsymbol{x}(0)$ and $\boldsymbol{x}(\tau)$ denote the nodes' concentration before and after diffusion at time τ, and $L = D - A$ is the graph Laplacian with $D$ the diagonal matrix of node strengths (or node degrees for binary connections). The graph Laplacian is the discrete analogue of the Laplace Beltrami operator ($\nabla^2 f$), quantifying the rate of change of flow across network edges. Mathematically, advection and consensus (and diffusion) differ in their use of distinct Laplacian matrices (see Supplementary Methods). In the undirected case, both dynamics reduce to standard heat diffusion (Equation 1).

In consensus dynamics, the concentration of node $i$ evolves based on the weighted sum of the differences between the concentrations of itself and its in-neighbors (Fig. 3A):

$$\dot{x}_i(t) = \sum_{\{\forall j | j \to i\}} w_{ij}\left(x_j(t) - x_i(t)\right) \tag{2}$$

From Equation (2), the solution for all nodes can be written compactly as

$$\boldsymbol{x}(t_{end}) = c(\tau)\boldsymbol{x}(t_{begin}) \tag{3}$$

where $\tau = t_{end} - t_{begin}$, $c(\tau) = \exp(-L_{in}\tau)$, $L_{in}$ is the in-degree Laplacian, and $\boldsymbol{x}(t)$ the node concentrations at time $t$.

In advection dynamics, the change of concentration of node $i$ depends on the inflow it receives from its in-neighbors and the outflow to its out-neighbors (Fig. 3B):

$$\dot{x}_i(t) = \sum_{\{\forall j | j \to i\}} w_{ij} x_j(t) - \sum_{\{\forall k | i \to k\}} w_{ki} x_i(t) \quad (4)$$

From Equation (4), the solution for all nodes can be written compactly as

$$\boldsymbol{x}(t_{end}) = a(\tau)\boldsymbol{x}\big(t_{begin}\big) \quad (5)$$

where $\tau = t_{end} - t_{begin}$, $a(\tau) = \exp(-L_{out}\tau)$, $L_{out}$ is the out-degree Laplacian, and $\boldsymbol{x}(t)$ the node concentrations at time $t$.

We refer to $c(\tau)$ as the consensus kernel and $a(\tau)$ as the advection kernel, both reflecting the intensity of interaction between nodes during the time interval $\big(t_{begin}, t_{end}\big)$. In the consensus case, the nodes eventually converge to a common value for sufficiently large $\tau$. As the name implies, consensus models have traditionally been used to describe how groups of individuals reach a common opinion (40) or how multi-agent systems converge to a common quantity value (41). In the case of advection, the nodes do not necessarily converge to a common value, but their dynamics are state sum conservative: $\sum_{i=1}^{n} x_i(\tau) = \sum_{i=1}^{n} x_i(0)$ (see (32) for a more detailed mathematical treatment). For both advection and consensus dynamics, the connectivity structure influences how nodes' states evolve, with networks with short path lengths typically exhibiting faster diffusion (Fig. 3C; Fig. S2). Due to their boundedness (state values do not diverge to infinity), convergence properties in consensus, and conservation of sum total of concentrations in advection, these dynamics may be viewed as abstractions of homeostatic mechanisms, although the present model is intentionally minimal and does not aim to capture all aspects of neural activity, similarly to other recent models (9).

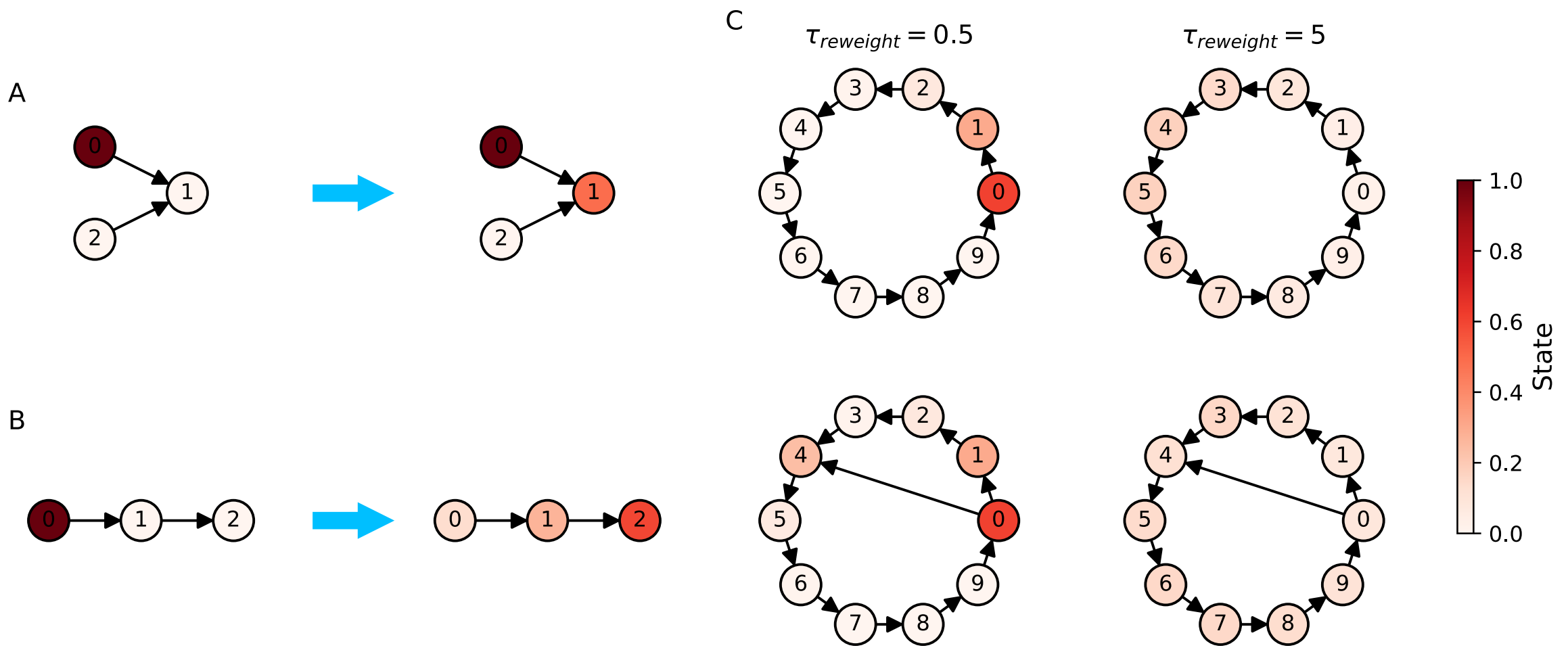

**Fig 3. The dynamics of concentration spread according to consensus and advection dynamics.** (A) When consensus is applied, the concentration of node 1 is adjusted according to its in-degree neighborhood concentrations, weighted by the connections. (B) When advection is applied, the concentration of node 1 increases by the weighted sum of its in-degree neighborhood concentrations (inflow) and decreases by the weighted sum of its out-degree neighborhood (outflow). (C) Concentration diffuses on the network as time passes based on consensus dynamics. The source node was node 0 and was initialized with state 1. Other nodes were initialized with state 0. Fig. S2 provides the same evolution based on advection dynamics.

## Adaptive plasticity rules

The model encompasses two types of plasticity, adaptive rewiring and adaptive weight adjustment. We first describe separately the two plasticity types and then the self-organizing algorithm that includes both.

### Adaptive rewiring

Adaptive rewiring describes the process of pruning underused connections and connecting nodes with high indirect interaction, as quantified by activity flow. At each rewiring step, a node $v$ is randomly selected from the set of nodes whose in- and out-degrees are neither zero nor $n-1$. With a given probability $p_{in}$, an in-connection of node $v$ is rewired; otherwise, an out-connection is rewired. At each step of rewiring, the initial concentration vector is set to $\boldsymbol{e}_v$, where all node concentrations are initialized to 0 except for node $v$, which is set to 1. The dynamics evolves over a time interval of length $\tau_{rewire}$ before rewiring occurs. When rewiring in-connections, we assume that interaction intensity is represented by the consensus kernel, whereas for out-connections, it is represented by the advection kernel (31). $\tau_{rewire}$ is set to 1 in all simulations.

Both in case of rewiring in- and out-connections, the connection with the lowest kernel value among the existing neighbors is removed, and a new connection is added to the non-neighbor with the highest kernel value. For in-connection rewiring of the selected node $v$, the connection $k \to v$ is cut where

$$k = argmin_{u \in N_{in}(v)} \{c(\tau_{rewire})_{vu}\}$$

and a new connection $l \to v$ is added, where

$$l = argmax_{u \in N_{in}^{c}(v)} \{c(\tau_{rewire})_{vu}\}$$

Here, $N_{in}(v)$ denotes the in-neighbors of node $v$, and $N_{in}^{c}(v)$ the set of nodes not connected to $v$ via in-connections. Similarly, when rewiring an out-connection of node $v$, we use the advection kernel to choose node $k$ (Equation 6) and $l$ (Equation 7).

$$k = argmin_{u \in N_{out}(v)} \{a(\tau_{rewire})_{uv}\} \tag{6}$$

$$l = argmax_{u \in N_{out}^{c}(v)} \{a(\tau_{rewire})_{uv}\} \tag{7}$$

where $N_{out}(v)$ denotes the out-neighbors of $v$, and $N_{out}^{c}(v)$ the set of nodes not connected via out-connections.

## Adaptive weight adjustment

Compared to adaptive rewiring, adaptive weight adjustment is a more gradual process. At each weight adjustment step, a node $u$, referred to as the candidate node, is randomly selected from the network. The initial concentration vector, $\boldsymbol{x}(t_{begin})$, is set to $\boldsymbol{e}_u$, where all node concentrations are initialized to 0 except for node $u$, which is set to 1. After the dynamics evolves for a time interval of length $\tau_{reweight}$, the weights of all connections are updated according to the following Hebbian rule:

$$\Delta w_{ij} = \eta \cdot x_i(t_{end}) \cdot x_j(t_{end}) \tag{8}$$

where $\eta$ is the learning rate, $t_{end} = t_{begin} + \tau_{reweight}$, and $x_i(t_{end})$ and $x_j(t_{end})$ are the concentration of node $i$ and $j$, respectively, at the time when weight adjustment occurs. We assume that $r$ steps of weight adjustment happen between every two consecutive steps of adaptive rewiring; that is, $\tau_{rewire} = r \cdot \tau_{reweight}$. $r$ is assumed to be no less than 1, and since $\tau_{rewire}$ is fixed to be 1, $\tau_{reweight}$ is between 0 and 1. The concentrations $x(t_{end})$ are calculated by either consensus (Equation 3) or advection dynamics (Equation 5). This choice is made before each simulation run and remains fixed throughout the run. For computational efficiency, $\eta$ is set to 10 for all simulations. We confirmed that varying $\eta$ over two orders of magnitude did not alter the main qualitative conclusion (Supplementary Note 1).

To prevent unbounded weight growth, we apply synaptic normalization to keep the in- (13, 42, 43) or out-strength (44) of each node constant with each weight adjustment step. In-strength normalization could be understood as competition for shared resources between activated and inactivated synapses onto a postsynaptic neuron (45–47). A similar competitive mechanism is governing out-strength normalization.

Synaptic normalization is implemented as follows. For in-strengths, the new weight $w'_{ij}$ of connection $j \to i$ is given by

$$w'_{ij} = \frac{s_{in}(i)}{\tilde{s}_{in}(i)} \cdot (w_{ij} + \Delta w_{ij})$$

where $s_{in}(i) = \sum_{k \in N_{in}(i)} w_{ik}$ is the in-strength before the Hebbian operation, and $\tilde{s}_{in}(i) = \sum_{k \in N_{in}(i)}(w_{ik} + \Delta w_{ik})$ represents the unnormalized in-strength immediately after the Hebbian rule is applied. Analogously, with the out-strengths normalization, the new weight $w'_{ij}$ is

$$w'_{ij} = \frac{s_{out}(j)}{\tilde{s}_{out}(j)} \cdot (w_{ij} + \Delta w_{ij})$$

where $s_{out}(j)$ and $\tilde{s}_{out}(j)$ are again the pre-Hebbian and unnormalized post-Hebbian out-strengths, respectively.

Normalization in adaptive weight adjustment is selectively paired, as in rewiring, with consensus or advection dynamics. Adaptive weight adjustment is either of type “C-out”, which refers to consensus dynamics in combination with out-connection normalization, or “A-in”, which employs advection dynamics with in-connection normalization.

## The Dual-Adaptive Algorithm

We combine adaptive rewiring with adaptive weight adjustment into a dual-adaptive algorithm. The condition of adaptive weight adjustment (“C-out” or “A-in”) stays fixed throughout the course of a run. The combined dynamics algorithm consists of the following steps:

**Step 1:** Select a random node $u \in V$. Set its initial concentration to 1 and other nodes’ initial concentrations to 0, i.e., $\boldsymbol{x}(t_{begin}) = \boldsymbol{e}_u$. Perform weight adjustment based on the selected dynamics.

**Step 2:** Repeat step 1 until $r$ steps of adaptive weight adjustment have been performed.

**Step 3:** Sample $p$ from a uniform distribution $U[0,1]$. If $p < p_{in}$, select a random node $v \in V$ such that its in-degree is neither zero nor $n - 1$ and rewire its in-connection. Otherwise, select a random node $v \in V$ such that its out-degree is neither zero nor $n - 1$ and rewire its out-connection. In rewiring steps, connection weights are fixed.

**Step 4:** Return to step 1 until $N$ steps of adaptive weight adjustment have been performed.

The tuning parameter of the algorithm is $\tau_{reweight}$, which specifies the duration over which the dynamics evolve prior to updating the weights at each adjustment step. $p_{in}$, the probability of rewiring the in-connection, is set to 0.5 unless stated otherwise.

## Statistics and Reproducibility

All statistical analyses were performed using custom-made Python scripts and Jupyter notebooks. We used the same set of 30 randomly initialized networks for all simulations designed to demonstrate model behavior. For each condition and parameter combination, one replicate was defined as the simulated network obtained by applying the corresponding algorithm to one of these initial networks. For simulations fitted to empirical connectomes, 20 simulated networks matching the number of nodes and connections in the corresponding empirical connectome were generated for each dataset.

Weight distributions of connections in simulated networks were characterized by fitting candidate distributions (lognormal, Weibull, gamma, exponential, inverse Gaussian, and inverse gamma) using maximum likelihood estimation and comparing goodness-of-fit using the Kolmogorov-Smirnov statistic. Differences in fit between candidate distributions were assessed using Wilcoxon signed-rank tests with a significance level of 0.05 (48). The upper tails of the weight distributions were further characterized by fitting power-law functions to logarithmically binned weights exceeding the median, using linear regression in log-log scale.

The fit of simulated weight distributions to those of empirical connectomes was assessed using a Monte Carlo goodness-of-fit test based on the Kolmogorov-Smirnov statistic (49). We further assessed whether the internal cores of convergent-divergent units in the simulated networks were denser and had higher average connection weights than expected from the corresponding network-wide averages. For each simulated network, the density and average connection weight of the internal core of each convergent-divergent unit were first normalized by the corresponding network-wide average, and the normalized values were then averaged across units. The resulting network-level values from 20 simulation instances were tested against the null value of 1 using one-sided Wilcoxon signed-rank tests. The significance level was set to 0.05.

Detailed descriptions of the distribution fitting, power-law fitting, and Monte Carlo goodness-of-fit test are provided in Supplementary Methods.

## Results

### Control conditions: Adaptive weight adjustment alone produces heavy-tailed distributions of connection weights and rudimentary brain-like connectivity

We first consider the effects of adaptive weight adjustment without rewiring (control algorithm 1, see Supplementary Methods). We systematically varied the parameter $\tau_{reweight}$, the time interval that the dynamics takes place before weight adjustment. In general, its value determines how far advection with in-connection normalization (A-in condition), and consensus with out-connection normalization (C-out condition) is allowed to percolate from each node to the rest of the network before the next weight adjustment. Smaller values restrict activity flow mainly around nearby nodes, while higher values allow more flow spreading to topologically more distant nodes. Adaptive weight adjustment alone generates heavy-tailed weight distributions only for small ($\tau_{reweight} \approx 0.05$; Fig. S3; Best fits are obtained for Weibull distributions with shape parameters < 1, Fig. S5) and intermediate values ($\tau_{reweight} \approx 0.1$; Fig. 4A, Fig. S4A; Best fit: lognormal distributions, Fig. S5). For intervals that allow the dynamics to spread wider ($\tau_{reweight} \approx 0.5$; Best fit: gamma distributions, Fig. S5), the connection weights become more homogeneous and lose the heavy tail (Fig. 4B, Fig.S4B). In log-log scale, the upper tails of the distributions are approximately linear (Fig. 4C, Fig. S4C), suggesting that these tails follow a power-law. The power-law exponent increases, i.e. connectivity strength drops off faster, as $\tau_{reweight}$ increases, stabilizing for A-in after $\tau_{reweight} > 0.4$ and continuing to grow for C-out until $\tau_{reweight} > 0.6$ (Fig. 4D).

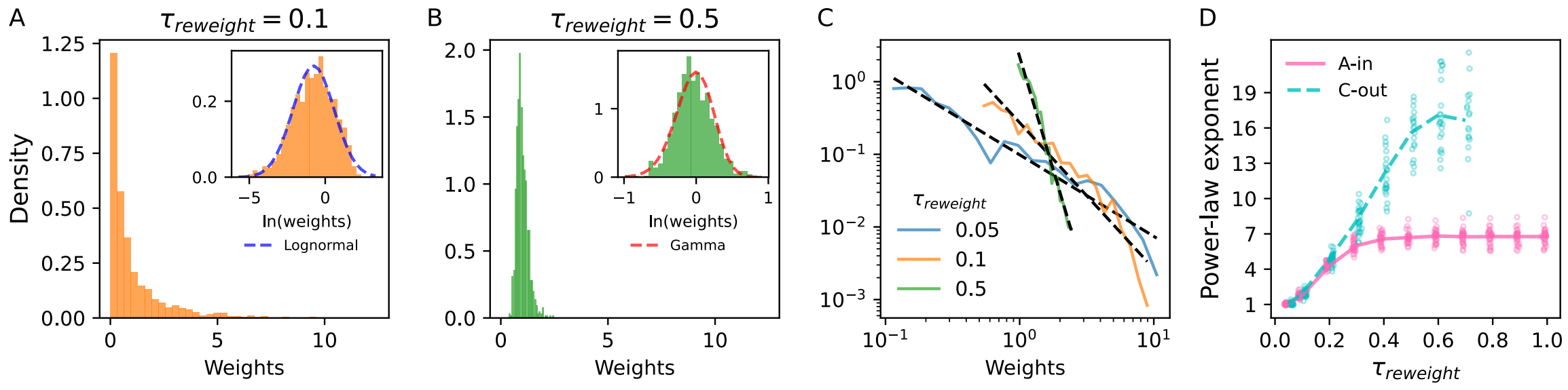

**Fig 4. Adaptive weight adjustment alone produces heavy-tailed weights only for small $\tau_{reweight}$ values.** (A, B) Distributions of connection weights of single representative simulated networks generated at $\tau_{reweight}$ = 0.1 (A) and $\tau_{reweight}$ = 0.5 (B) in the A-in condition.

(C) Power-law fits to the upper tails of the distributions of connection weights for $\tau_{reweight} = 0.05, 0.1, 0.5$ in the A-in condition. Colored lines represent upper tails of distributions of connection weights from single representative simulated networks, and black dashed lines are the best power-law fits. (D) Best-fit power-law exponents as a function of $\tau_{reweight}$. Points represent values from individual simulated networks with $R^2 > 0.85$, and lines connect the means of these values at different $\tau_{reweight}$ values. A $\tau_{reweight}$ value was included only when more than half of the 30 simulated networks met $R^2 > 0.85$. Simulations were run for $N = 2 \times 10^5$ steps of adaptive weight adjustment. The corresponding stationarity analysis showed stabilization before the final simulation step used for analysis (Fig. S6).

To probe how adaptive weight adjustment without rewiring affects the networks' initially random connectivity, we compiled several summary connectivity metrics. Each measure was normalized by the average value obtained from 100 null networks, in which weights were shuffled while preserving the connectivity structure. If weights are randomly distributed on network connections, the normalized measures will be 1. We found deviations from the randomly distributed case only for localized interactions (small $\tau_{reweight}$ values), characterized by increased clustering coefficient, small-world, and modularity indices (50, 51) and reduced efficiency (52) (Fig. 5). When the process of adaptive weight adjustment is interrupted by random rewiring (control algorithm 2, see Supplementary Methods), adaptive weight adjustment's influence does not change (Fig. S7-9).

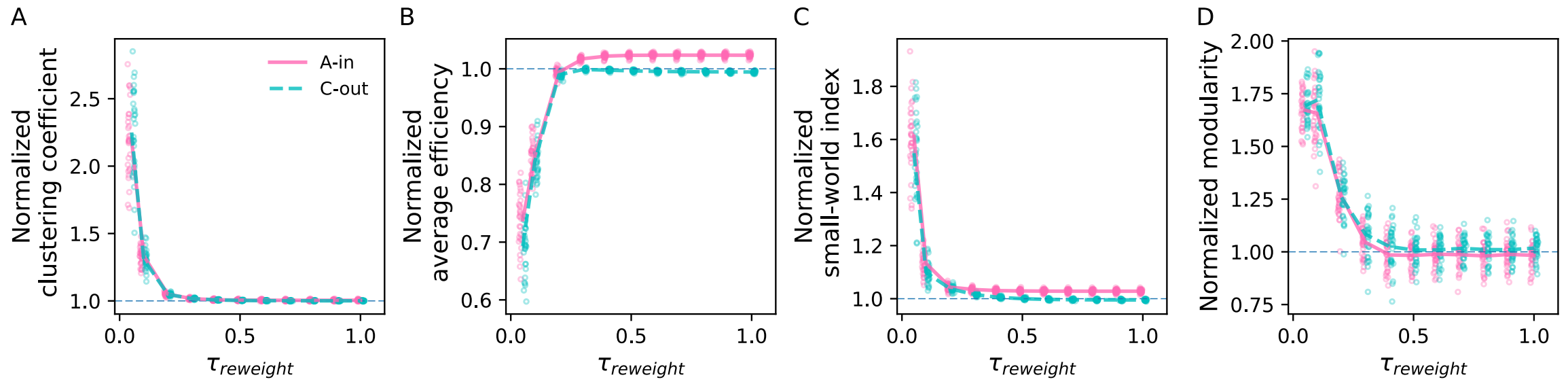

**Fig 5. Adaptive weight adjustment induces brain-like topological features for small $\tau_{reweight}$ values.** (A) Normalized clustering coefficient, (B) normalized average efficiency, (C) normalized small-world index and (D) normalized modularity as functions of $\tau_{reweight}$. Points represent values from 30 individual simulated networks and lines connect the means across networks at different $\tau_{reweight}$ values.

Our results so far show that adaptive weight adjustment alone produces heavy-tailed weights only for small $\tau_{reweight}$ values, i.e., when interactions are largely restricted to neighboring nodes (Fig. S10A, D). This result could be explained by how the diffusion processes, advection and consensus, spread on the random connectivity structure of the network. For small enough $\tau_{reweight}$ values under which diffusion mainly concentrates around the neighbors of the candidate nodes, node concentrations and subsequently weight increments (Equation 8) become approximately proportional to connection weights (Fig. S10B, C, E, F). In this regime, adaptive weight adjustment effectively mimics multiplicative dynamics where stronger connections receive larger weight increments, leading to heavy-tailed weight distributions (4).

Due to the short path lengths characteristic of random networks, modest increases in $\tau_{reweight}$ allow nodes to interact across broad regions of the network (Fig. S10G, J). This reduces heterogeneity of node concentrations (Fig. S11), but the mechanism differs between consensus and advection. Consensus dynamics reduces concentration differences by updating each node toward the weighted average of its upstream neighbors' concentrations. Abundant recurrent paths in random networks makes these local averaging interactions effectively mutual, driving node concentrations toward a common value. Advection dynamics is different. It conserves the total state and redistributes it along directed connections until a balance between incoming and outgoing flow is reached. Because the connectivity of random directed networks is close to balanced, the advection equilibrium is close to but not perfectly uniform.

Thus, with increasing $\tau_{reweight}$, both consensus and advection promote concentrations to equilibrate among many nodes (Fig. S10H, K), which in turn weakens the correlation between weight increments and existing weights (Fig. S10I, L) and essentially disrupts multiplicative dynamics. Consequently, weight heterogeneity is precipitously reduced when $\tau_{reweight}$ increases. The equilibrium of advection is nevertheless not perfectly uniform, consistent with the saturation of power-law exponent for A-in after $\tau_{reweight} > 0.4$ (Fig. 4D) and the persistent elevation of normalized average efficiency (Fig. 5B).

## The combination of adaptive rewiring and adaptive weight adjustment generates heavy-tailed weights for a wide range of $\tau_{reweight}$ values

We next consider the emergence of heavy-tailed weight distributions for the dual-adaptive algorithm where we combine adaptive weight adjustment with adaptive rewiring. Adaptive rewiring driven by advection and consensus dynamics promotes hub formation, where most nodes become polarized, exhibiting either high in-degree and low out-degree or low in-degree and high out-degree (31). This polarization leads to a hierarchical structure, where diffusion flows primarily from upstream to downstream nodes, while feedback from downstream to upstream is comparatively reduced, in contrast to the abundant feedback loops in random networks. For consensus-based weight adjustment, reduced feedback weakens the mutual averaging that otherwise drives node concentrations rapidly toward a common value in random networks. For advection-based weight adjustment, the more heterogeneous structure, in which some nodes are dominated by incoming flow and others by outgoing flow, moves the flow-balance equilibrium further away from a uniform profile. Based on the analysis in the previous section, we hypothesized that adaptive rewiring could enable adaptive weight adjustment to generate heavy-tailed weights for a wider range of $\tau_{reweight}$ values.

Our results validate this hypothesis: the dual-adaptive algorithm facilitates the emergence of heavy-tailed weights for a wider range of $\tau_{reweight}$ values compared to solely adaptive weight adjustment (Fig. 6A, B; Fig. S12A, B; for best fits see Fig. S13). Increasing $\tau_{reweight}$ does not substantially change the decay rate of the upper tails of weight distributions, in contrast to when only adaptive weight adjustment was used (Compare Fig. 6C, D and Fig. S12C with Fig. 4C, D and Fig. S4, S7). This robustness indicates that the topology induced by adaptive rewiring counteracts the homogenizing tendency of consensus and advection dynamics. The hierarchical structure, in which downstream nodes exert minimal influence on upstream nodes, reduces recurrent feedback for consensus and moves the equilibrium away from a uniform profile for advection. Thus, adaptive rewiring preserves these concentration differences as $\tau_{reweight}$ increases (Fig. S15A, B). As a result, the correlation between connection weights and weight increments persists across a broad range of $\tau_{reweight}$ values (Fig. S15C, D), maintaining the conditions necessary for the emergence of heavy-tailed weights.

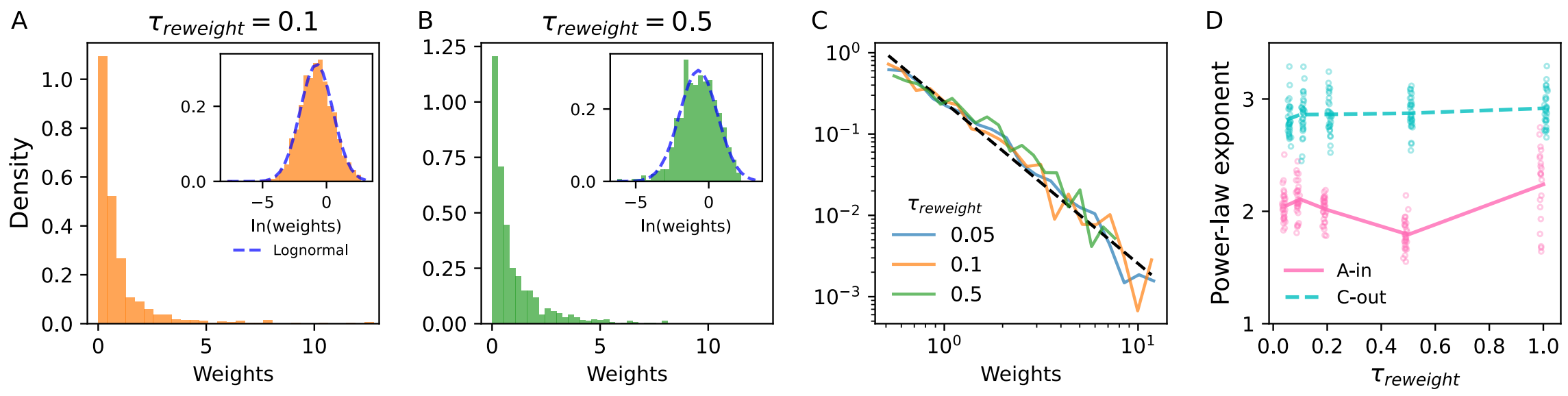

**Fig 6. Combined with adaptive rewiring, adaptive weight adjustment produces heavy-tailed weight distributions for a wider range of $\tau_{reweight}$.** (A, B) Distribution of connection weights of single representative simulated networks generated at $\tau_{reweight} = 0.1$ (A) and $\tau_{reweight} = 0.5$ (B) in the A-in condition. (C) Power-law fit of upper tails of connection weights from networks generated at $\tau_{reweight} = 0.05, 0.1, 0.5$ in the A-in condition. Colored lines represent upper tails of distributions of connection weights from single representative simulated networks, and the black dashed line indicates the best power-law fit for $\tau_{reweight} = 0.1$. (D) Best-fit power-law exponents as a function of $\tau_{reweight}$. Points represent values from individual simulated networks with $R^2 > 0.85$, and lines connect the means of these values at

different $\tau_{reweight}$ values. A $\tau_{reweight}$ value was included only when more than half of the 30 simulated networks met $R^2 > 0.85$. All networks were generated at $p_{in} = 0.5$. Simulations were run for $N = 10^6$ steps of adaptive weight adjustment. The corresponding stationarity analysis showed stabilization before the final simulation step used for analysis (Fig. S14).

With the dual-adaptive model, when varying the probability of rewiring the in-connection, $p_{in}$, the weight distributions remain heavy-tailed across most parameter settings (Fig. 7A, B; Fig. S16, S17). As $p_{in}$ increases, the upper tail becomes heavier (slower decay) in the 'A-in' condition (Fig. 7C) and lighter (faster decay) in the 'C-out' condition (Fig. 7D).

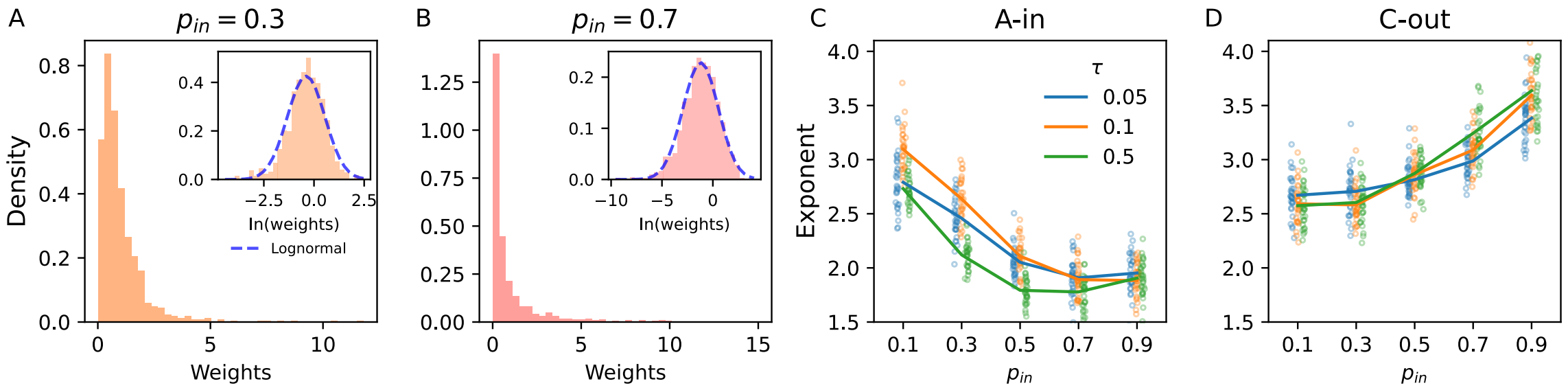

**Fig 7. The probability of rewiring the in-connection, $p_{in}$, modulates the decay rate of the upper tail of the weight distribution in a condition-dependent manner.** (A, B) Distribution of connection weights of single representative simulated networks generated at $\tau_{reweight} = 0.1$ for $p_{in} = 0.3$ (A) and $p_{in} = 0.7$ (B) in the A-in condition. (C, D) Best-fit power-law exponents as a function of $p_{in}$ for the "A-in'' (C) and "C-out'' (D) conditions. Points represent values from individual simulated networks with $R^2 > 0.85$, and lines connect the means of these values at different $\tau_{reweight}$ values. A $\tau_{reweight}$ value was included only when more than half of the 30 simulated networks met $R^2 > 0.85$. Simulations were run for $N = 10^6$ steps of adaptive weight adjustment. The corresponding stationarity analysis showed stabilization before the final simulation step used for analysis (Fig. S18).

## Adaptive weight adjustment compatible with the emergence of convergent-divergent units

Recent studies have shown that in directed networks with fixed weights, adaptive rewiring, combined with a small proportion of random rewiring, produces convergent-divergent units, prominent in feedforward sensory processing (36) and in effectuating context integration in their more general recurrent form (29, 31). Convergent-divergent units receive input from a population of local neurons via convergent hubs, process the information through an internal core of intermediate nodes, and subsequently project the output via divergent hubs to a population of local neurons (Fig. 8A). Convergent hubs are defined as nodes with at least one out-connection and an in-degree exceeding a threshold, and divergent hubs as nodes with at least one in-connection and an out-degree above a threshold.

We investigated whether adaptively adjusting the weights disrupts the organization formed by adaptive rewiring in the fixed-weight networks. We found that, under appropriate variable settings (see the caption in Fig. 8), the dual-adaptive model produces both convergent-divergent units (Fig. S19) and heavy-tailed weight distributions (Fig. S20). Because convergent-divergent units require both the presence of convergent and divergent hubs and the existence of pathways linking these hubs through intermediate nodes, we examined overall connectedness and hub prevalence. These analyses showed that, compared to networks generated by applying only rewiring, the dual-adaptive model preserves overall network connectedness (Fig. S21) and does not eliminate hubs, but instead modulates the prevalence of convergent and divergent hubs in a condition-dependent manner (Fig. S22). The internal cores of convergent-divergent units generated by the dual-adaptive model also had densities broadly comparable to those observed in null networks (Fig. 8B, C). Additional analyses showed that the average weights of these internal cores did not exhibit a consistent, statistically significant increase relative to the null networks (Fig. S23). Together, these results suggest that adaptive weight adjustment does not abolish the hubs and directed pathways required for convergent-divergent organization and remains compatible with the emergence of convergent-divergent units generated by adaptive rewiring.

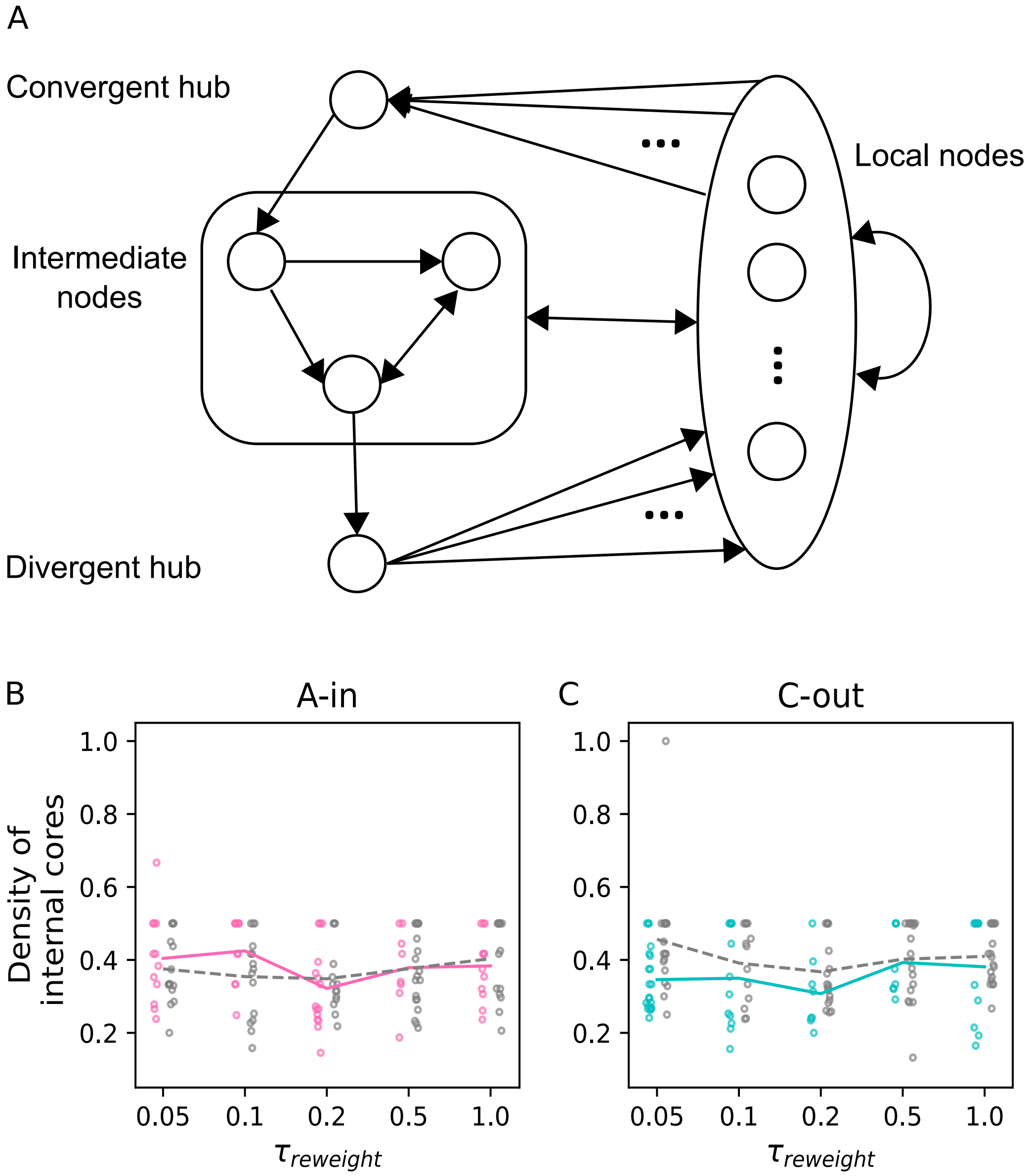

**Fig 8. Adaptive weight adjustment is not affecting the density of the internal core of convergent-divergent units.** (A) A schematic of a convergent-divergent unit. The unit consists of a convergent hub, a divergent hub, and a subnetwork of intermediate nodes connecting them that constitutes the internal core. A convergent hub aggregates inputs from local nodes and relays information to a divergent hub via a subnetwork of intermediate nodes. Adapted from Figure 2 in Li et al. (29) under a CC BY 4.0 license. (B, C) Density of the internal cores of the convergent-divergent units in the A-in (B) and C-out (C) conditions across different $\tau_{reweight}$. Colored points indicate 30 individual simulated networks, and colored lines connect the means across networks at different $\tau_{reweight}$. Grey points and dashed grey lines show the corresponding values for null networks, generated by randomly assigning the final weights of each simulated network to the corresponding initial random network and then applying rewiring alone

with weights held fixed. Points represent values from individual simulated networks and lines connect the means across different $\tau_{reweight}$ values. For this analysis at each rewiring step, adaptive and random rewiring are selected stochastically with a probability of $p_{random} = 0.1$. The threshold for identifying both convergent and divergent hubs is set to 15. Simulations were run for $N = 5 \times 10^5$ steps of adaptive weight adjustment. The corresponding stationarity analysis showed stabilization before the final simulation step used for analysis (Fig. S24).

## Weight distributions and convergent-divergent units of neuronal connectomes fitted with the dual adaptive model

We compared the output of our model to two neuronal connectomes: the nerve ring and ventral ganglion of a young adult *C. elegans* (38, 53) and the inner plexiform layer of the mouse retina (39, 54). The contact area between neurons is used as a proxy for synaptic weight. In both connectomes, the empirical weight distributions are well described by a lognormal form (Fig. 9A, B). Both connectomes exhibit clear convergent and divergent hubs (Fig. 9C, D), and their internal cores contain denser and stronger connections than the network-wide averages (Table 1).

We selected model conditions to match the relative proportions of convergent and divergent hubs in the neuronal connectomes (Fig. S22). When convergent hubs outnumber divergent hubs (as in *C. elegans*; Fig. 9C), we used the "A-in" condition; otherwise (as in the mouse retina; Fig. 9D), we used the "C-out" condition. For each connectome, we ran the dual-adaptive algorithm on 20 independent random directed networks with the same number of nodes and connections as the experimental data and normally distributed weights (Table S1). To improve computational efficiency, we reused the same node states for $M$ iterations of the weight adjustment step (Step 2) before recomputing the kernels. Since each iteration of Step 2 includes $r$ steps of adaptive weight adjustment, this corresponds to reusing the same node states for $M \times r$ steps of adaptive weight adjustment. Likewise, we reused the updated kernels for $M$ iterations of the rewiring step (Step 3). We set $M = 2$ for *C. elegans* and $M = 100$ for the mouse retina.

The simulated weight distributions closely approximate the empirical distributions (Fig. 9E, F). Moreover, the internal core subnetworks generated by the model are denser and heavier than network-wide averages, qualitatively consistent with those observed in the real connectomes (Table 1). However, the proportion of generated convergent and divergent hubs is lower than in the experimental data (compare Fig. 9G, H with Fig. 9C, D), suggesting additional mechanisms might be considered for improving the match.

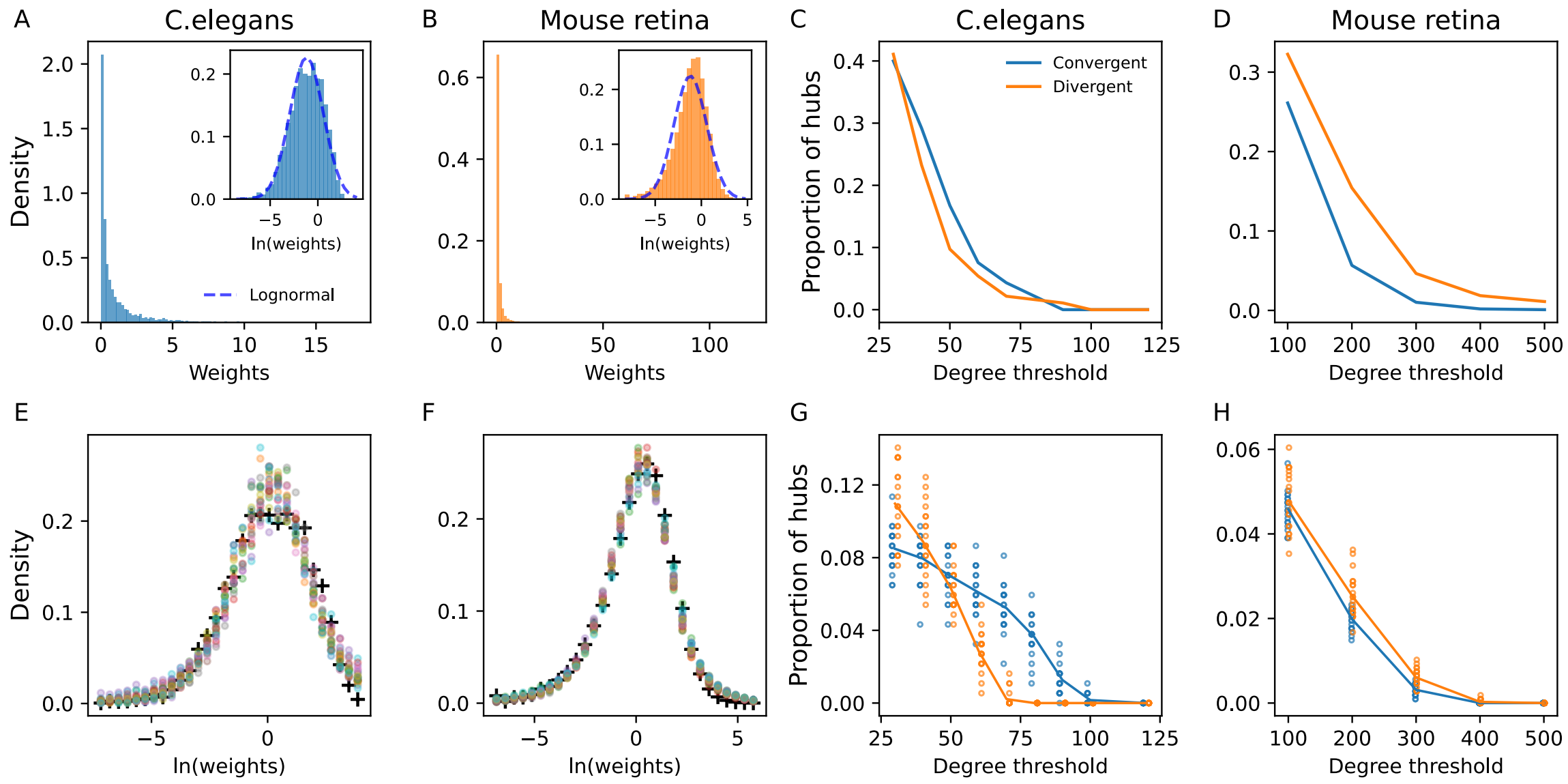

**Fig 9. Simulated networks reproduce the lognormal weight distributions of real neuronal connectomes and qualitatively preserve the relative abundance of convergent and divergent hubs.** (A, B) Empirical weight distributions for *C. elegans* (A) and the mouse retina (B). (C, D) Proportions of convergent and divergent hubs in the two connectomes as a function of degree threshold. (E, F) Log-transformed weight distributions from empirical data (black crosses) and 20 simulated networks (colored dots). Dot colors denote different simulation instances. For comparability, both empirical and simulated weights were shifted so that their peaks align at zero. Monte Carlo goodness-of-fit tests did not reject the null hypothesis, which indicates that the empirical data are typical realizations of the model ($p = 0.29$ for *C. elegans* and $p = 0.19$ for mouse retina). (G, H) Proportions of convergent and divergent hubs in simulated networks as a function of degree threshold. Points represent values from 20 individual simulated networks and lines connect the means across networks at different $\tau_{reweight}$ values. For the *C. elegans* connectome, simulations were run at $p_{\text{random}} = 0.1$, $\tau_{\text{reweight}} = 0.05$, and $p_{\text{in}} = 0.78$; for the mouse retina at $p_{\text{in}} = 0.15$. The degree threshold for identifying convergent and divergent hubs is set to 20% of the total number of neurons. Simulations were run for $N = 10^7$ steps of adaptive weight adjustment for the *C. elegans* dataset and $N = 12 \times 10^7$ steps of adaptive weight adjustment for the mouse retina dataset. The corresponding stationarity analysis showed stabilization before the final simulation step used for analysis (Fig. S25).

| Network | Density (Real) | Density (Model) | Weight (Real) | Weight (Model) |
|---|---|---|---|---|
| *C. elegans* | 2.31 | $1.41 \pm 0.45^{***}$ | 2.04 | $2.78 \pm 1.40^{***}$ |
| Mouse retina | 5.20 | $1.11 \pm 0.19^{*}$ | 1.35 | $2.38 \pm 0.98^{***}$ |

**Table 1. Both real and simulated networks have denser and heavier internal cores**. Average density and average weight of internal cores are normalized by the corresponding network-wide averages. Model values are shown as mean $\pm$ standard deviation across 20 simulation instances. Statistical significance was assessed relative to the network-wide averages. The corresponding $p$ values are *C. elegans* density, $p < 0.001$; *C. elegans* weight, $p < 0.001$; mouse retina density, $p = 0.02$; and mouse retina weight, $p < 0.001$.

# Discussion

We considered how evolving neural networks could develop heavy-tailed weight distributions. In principle, such distributions can emerge from multiplicative stochastic processes such as geometric Brownian Motion (GBM) (55). However, this would be unrealistic for biological systems. The mean and variance of distributions generated by GBM grow exponentially over time (55), whereas biological circuits operate under stabilizing constraints: neuronal activation levels are bounded by homeostatic mechanisms (56), and neural connections develop their strength through synaptic plasticity (57). This understanding has led to generative models for producing these distributions through iterative weight adjustment (7–13). Our proposed model belongs to this category.

The model parsimoniously combines two activity-dependent plasticity mechanisms: adaptive rewiring, which restructures the network's connections, and adaptive weight adjustment, which modifies connection weights. Both forms of adaptation are based on the same diffusion dynamics of consensus and advection. The dual-adaptive algorithm generates a heavy-tailed connection weight distribution for a wider range of dynamics compared to solely adaptive weight adjustment. Furthermore, it gives rise to motif structures that are widespread in neural systems: convergent and divergent circuits and their combination in convergent-divergent units, with internal cores with denser and stronger-than-average interconnections. Our generated data are broadly consistent with real connectomes, yet they also indicate that additional mechanisms may be needed to fully account their complex connectivity.

The Hebbian rule (58) explains the development and maintenance of neural structures by linking neural activity to changes in synaptic strengths. Following its formulation, experimental observations have further solidified this rule with quantitative mechanisms, including LTP (59–61), LTD (62, 63) and STDP (64–67), the latter providing a temporal mechanism for synaptic potentiation and depression. Structural changes in the nervous system include the growth or retraction of neurites as well as of axonal boutons and dendritic spines with the accompanying formation or elimination of synapses (66, see 16 for a review). Adaptive rewiring aims to model these processes with diffusion dynamics offering a way to explore nodes that are not directly connected but have a strong indirect flow of information.

Although the Hebbian rule is typically understood as referring to changes in the strength of existing synapses, experiments also suggest that common activity-based processes induce structural changes (see 69, 70 for reviews). Specifically, the potentiation of synaptic weights (LTP) results in the increase in the number of spines (71) and filopodia (72) and in the enlargement of spine heads (73) which leads to the formation of new synaptic connections; while LTD leads to spine shrinkage (74) which in turn increases the probability of synaptic pruning. This evidence suggests that apparently distinct mechanisms for structural rewiring (formation or elimination of synapses) and synaptic reweighting may arise from common activity-dependent processes. In contrast to previous models that focus exclusively on either synaptic weights or connectivity structure, our framework combines both forms of plasticity under shared dynamics. This allows the model to capture the co-evolution of topology and synaptic strengths, providing a minimalist account for the emergence of heavy-tailed weight distributions and characteristic network motifs within the same framework.

Understanding the mechanisms that generate and maintain neuronal structures in the brain remains a fundamental challenge in neuroscience. Across species, neuronal structures and nervous systems show a characteristic distribution of synaptic strengths with a few strong connections amid a vast majority of weak ones (5, 11, 39–44; for a summary see 9). It has been suggested that this distribution of synaptic strengths facilitates efficient processing, as a minority of strongly connected neurons may support most brain tasks (80), reproduces characteristic features of activity at criticality (81), and enable faster and be more stable responses to perturbations compared to networks with normally distributed synaptic weights (82).

Adaptive weight adjustment alone can produce heavy-tailed distributions and nonrandom topology, but only when the dynamics driving the weight updates is localized, with only little diffusion of activity beyond direct neighbors. Increasing the time of diffusion before weight adjustment favors more distant interactions and, as a result, the network dynamics rapidly homogenizes node concentrations, reducing heterogeneity in network topology and weights. This aligns with the previous observation by Lynn et al. (9), whose algorithm randomly prunes connections and redistributes weights either randomly or following a preferential growth rule weighted by neural interactions. For that model, too, increasing the range of neural interactions reduces the heterogeneity of connection weights. While high clustering emerged at intermediate scales of interactions in their model, it appeared at local interactions in our case. The discrepancy likely results from differences in model design. Unlike the preferential growth rule in Lynn et al., which requires global knowledge of network-wide interactions, adaptive weight adjustment relies solely on local information. Specifically, the update of each connection weight depends on the concentrations of the two connected nodes and the total weights of the connections with their direct neighbors, making our approach more biologically plausible.

By integrating adaptive weight adjustment and rewiring, the dual-adaptive algorithm overcomes the limitations of adaptive weight adjustment. Adaptive rewiring induces structural asymmetries, where nodes increasingly differentiate into in-hubs and out-hubs, which in turn seed weight heterogeneities that are amplified by adaptive weight adjustment. Crucially, this mechanism can operate across a broad timescale of neural interactions, enabling the emergence of heavy-tailed distributions under both restricted and unrestricted activity spreading. This suggests that adaptive rewiring helps counteract the homogenizing tendencies caused by the spreading dynamics, thereby supporting the persistence of functional heterogeneity. As a result, the dual-adaptive algorithm shows tolerance to variation, since it produces the target strength distribution for a wide range of dynamics. The capacity of the model to produce heavy-tailed distributions of connection strengths across different dynamics is consistent with a prominent feature of biological systems, namely their ability to achieve functional outcomes for a constellation of different parameter values (83, 84), a property that also characterizes neural systems (see 85–87, 88 for reviews).

Moreover, the robustness of the model across different diffusion timescales may be relevant biologically, given that classical STDP operates over temporal windows of only tens of milliseconds (66, 67, 89, 90), but behavior and its consequences are often separated by substantially longer intervals. Together with evidence that STDP may require repeated trials to induce robust synaptic changes (17), this suggests that additional plasticity mechanisms operating over longer timescales may contribute to learning and adaptation. Consistent with this view, recent studies in the hippocampus have identified a novel learning mechanisms termed behavioral timescale synaptic plasticity (BTSP), capable of inducing synaptic changes over longer temporal windows and even after a single trial (91–97).

In terms of connectivity, convergent and divergent circuits are pervasive in organisms of different complexities. For instance, for both flies (98) and locusts (99), mushroom body output neurons receive converging input from Kenyon cells to produce a stable stereotypical response. The visual system shows divergence and convergence at different stages of processing: divergence from bipolar cells to retinal ganglion cells (100), convergence of retinal ganglion cells to thalamic neurons (101, 102), and divergence of thalamic inputs to V1 cells (103). Circuits combining convergent and divergent motifs, support diverse computations, notably in visual processing. For example, in mouse V1, somatostatin neurons collect inputs from and project information back to orientation-selective excitatory neurons in layers 2/3, enabling contextual modulation based on center-surround relationships (37). The dual-adaptive algorithm preserves the formation of convergent-divergent units. The algorithm by Lynn et al. (9) produced clustering patterns resembling those observed in biological brains, but it does not generate the more complex topology observed in our model. This limitation likely stems from its exclusive reliance on localized interactions, which is sufficient for the generation of heavy-tailed weight distributions but falls short in supporting higher-order structural organizations. In contrast, our model decouples these processes by implementing two separate Hebbian mechanisms, albeit driven by common dynamics: adaptive rewiring governs the evolution of network topology over longer timescales, while adaptive weight adjustment fine-tunes connection strengths at a faster, more localized scale.

Ongoing advances in image processing and electron microscopy have enabled reconstructions of neuronal structures involving thousands of neurons (38, 39, 5, 104). In order not to get lost in a sea of details, it is useful to examine to what extent a small set of rules or principles and their interactions can explain experimental data. Such an approach may help distinguish variability arising from mechanisms that operate robustly across a broad range of parameter values from variability due to noise (105). Some of these rules are likely conserved across species. To that end, we compared the generated data of our model with neural structures from two species of disparate complexities: the nerve ring and ventral ganglion of *C. elegans* (38) and the inner plexiform layer of the mouse retina (39). *C. elegans* and mice diverged from a common bilaterian ancestor some 600-700 million years ago (106, 107). Common mechanisms between them can thus reflect foundational organization principles.

The model successfully reproduces the heavy-tailed weight distributions for both organisms. It also captures key organizational features reflecting their connectivity structure: convergent and divergent hubs and internal cores with denser and stronger connections than the network average. Although the model underestimates the proportion of hubs, the qualitative agreement suggests that joint adaptive reweighting and rewiring based on diffusion principles are sufficient to recover core aspects of connectome organization. Moreover, an additional mechanism to achieve quantitative precision could readily be envisaged. For instance, an apoptosis mechanism preferentially eliminating non-hub nodes may increase the proportion of hubs in the remaining network to an appropriate level.

Other models have replicated some of the large-scale connectivity features in the brain by implementing distance-dependent rules, whereby the probability of connecting two areas decreases exponentially with interareal distance (108–110) or also by including homophilic attachment rules, in which neurons or areas with shared features have a higher probability of connecting (111–113). Moreover, the seminal work of Barabási and Albert (114) could generate scale-free properties, including rich club structure, both features of nervous systems. Our model (as well as that of Lynn's et al. (9)) differs from these approaches in that the networks do not grow in terms of nodes or connections but instead reorganize existing connections and connection strengths. This distinction reflects two different developmental phases: during early development, growth processes are decisive in network formation; whereas in later development and adulthood, synaptic plasticity refines connectivity structure and synaptic weight distributions (115–117).

In a conceptually related study, Sporns and Kötter (118) probed structural and functional motifs as building blocks of neural assembly and generated networks with several brain-like topological features. As in this study, their model rewired random networks while keeping the numbers of nodes and connections fixed. However, their rewiring was instructed by an evolutionarily motivated global objective that maximized functional motif number, whereas in our case rewiring and reweighting were driven by diffusion-mediated interactions that aim to abstract plasticity mechanisms. Moreover, convergent–divergent structures of the type studied here have, to our knowledge, received little attention in previous modeling studies.

A distinct use of generative models concerns inference of connectivity from neural recordings. These approaches posit a model of how the observed neural activity is generated and use empirical recordings to infer latent connectivity (119). For example, Gosti et al. (120) trained a recurrent Hopfield mass model on MEG time series to estimate effective connectivity. Unlike these studies, our model does not use empirical neural time series to infer connectivity; instead, it examines whether specific rules or mechanisms are sufficient to generate characteristic network topologies and synaptic-weight distributions.

The dual-adaptive algorithm is a simplified model, whose minimalism is similar to that of other recent studies (9). Simplification allowed us to isolate the role of spontaneous activity in the emergence of key organizational features of brain-like networks. For example, the present model does not aim to capture phenomena such as bursts of activation, neuronal avalanches, or other dynamics often associated with systems operating near criticality (121). This simplification is intentional. Throughout neuroscience, simplified models have been instrumental in identifying mechanisms sufficient to produce observed phenomena. For instance, classical point-neuron models neglect dendritic morphology and nonlinear dendritic processing, yet have provided important insights into sensory coding and the emergence of receptive field properties (122). Similarly, studies that seek to investigate additional biological mechanisms often introduce a limited set of features in a controlled manner, while still retaining a schematic representation of biological reality (for example (123) for the case of dendritic nonlinearities on point neuron models). This approach makes it possible to identify the specific mechanisms responsible for the emerging properties of the system. Along these lines, future work can introduce additional degrees of freedom and thereby higher complexity to bring the model closer to biological reality.

First, the model presupposes a network with fixed numbers of nodes and connections, omitting important developmental phenomena such as neurogenesis, programmed cell death, and axonal growth. Recent studies show that dynamic axon expansion based on guidance cues can produce modular small-world networks with lognormal weight and scale-free degree distributions (8), while neurite branching may also contribute to these distributions (76). Future studies could initialize networks using biologically inspired growth rules, then apply the dual-adaptive algorithm to investigate how activity-dependent plasticity operates on, and potentially reshapes, pre-established structures.

Second, additional biological mechanisms contribute to the quantitative organization of convergent–divergent units beyond what is captured by the current model. For example, connection probability in many neural systems decreases with anatomical distance (124, 125), imposing strong spatial constraints that are not considered here. We have shown earlier how these constraints interact constructively with adaptive rewiring (29). Neurons in the retina and other systems also exhibit layer-specific and type-specific connectivity preferences (126, 127). Incorporating these mechanisms may improve the model's ability to quantitatively match the density and weight of internal cores observed in real connectomes.

## Data availability

The data generated by the models and the empirical neuronal connectomes used in this study can be found in the following OSF repository: https://osf.io/c5hny/overview?view_only=b9711204968d430dbbd2f812a4b033e5. The data are publicly accessible and no password is required. The repository wiki provides information on which datasets were used for each figure and analysis.

# Code availability

All code is available at https://github.com/rentzi/adaptive_rewire_reweight.git

# References

1. J. I. Arellano, Ultrastructure of dendritic spines: correlation between synaptic and spine morphologies. *Front. Neurosci.* **1**, 131–143 (2007).

2. M. Chini, M. Hnida, J. K. Kostka, Y.-N. Chen, I. L. Hanganu-Opatz, Preconfigured architecture of the developing mouse brain. *Cell Rep.* **43** (2024).

3. S. Lefort, C. Tomm, J.-C. Floyd Sarria, C. C. H. Petersen, The Excitatory Neuronal Network of the C2 Barrel Column in Mouse Primary Somatosensory Cortex. *Neuron* **61**, 301–316 (2009).

4. Y. Loewenstein, A. Kuras, S. Rumpel, Multiplicative Dynamics Underlie the Emergence of the Log-Normal Distribution of Spine Sizes in the Neocortex In Vivo. *J. Neurosci.* **31**, 9481–9488 (2011).

5. L. K. Scheffer, *et al.*, A connectome and analysis of the adult Drosophila central brain. *eLife* **9**, e57443 (2020).

6. S. Song, P. J. Sjöström, M. Reigl, S. Nelson, D. B. Chklovskii, Highly Nonrandom Features of Synaptic Connectivity in Local Cortical Circuits. *PLoS Biol.* **3**, e68 (2005).

7. A. A. Koulakov, T. Hromadka, A. M. Zador, Correlated Connectivity and the Distribution of Firing Rates in the Neocortex. *J. Neurosci.* **29**, 3685–3694 (2009).

8. Y. Liu, *et al.*, A generative model of the connectome with dynamic axon growth. *Netw. Neurosci.* 1–47 (2024). https://doi.org/10.1162/netn_a_00397.

9. C. W. Lynn, C. M. Holmes, S. E. Palmer, Heavy-tailed neuronal connectivity arises from Hebbian self-organization. *Nat. Phys.* **20**, 484–491 (2024).

10. N. Rößler, *et al.*, Skewed distribution of spines is independent of presynaptic transmitter release and synaptic plasticity, and emerges early during adult neurogenesis. *Open Biol.* **13**, 230063 (2023).

11. G. Scheler, Logarithmic distributions prove that intrinsic learning is Hebbian. *F1000Research* [Preprint] (2017). Available at: https://f1000research.com/articles/6-1222 [Accessed 23 March 2022].

12. H. Uzan, S. Sardi, A. Goldental, R. Vardi, I. Kanter, Stationary log-normal distribution of weights stems from spontaneous ordering in adaptive node networks. *Sci. Rep.* **8**, 13091 (2018).

13. P. Zheng, C. Dimitrakakis, J. Triesch, Network Self-Organization Explains the Statistics and Dynamics of Synaptic Connection Strengths in Cortex. *PLOS Comput. Biol.* **9**, e1002848 (2013).

14. D. B. Chklovskii, B. W. Mel, K. Svoboda, Cortical rewiring and information storage. *Nature* **431**,

782–788 (2004).

15. G. Knott, A. Holtmaat, Dendritic spine plasticity—Current understanding from in vivo studies. *Brain Res. Rev.* **58**, 282–289 (2008).

16. M. Butz, F. Wörgötter, A. van Ooyen, Activity-dependent structural plasticity. *Brain Res. Rev.* **60**, 287–305 (2009).

17. J. C. Magee, C. Grienberger, Synaptic Plasticity Forms and Functions. *Annu. Rev. Neurosci.* **43**, 95–117 (2020).

18. P. Gong, C. van Leeuwen, Emergence of scale-free network with chaotic units. *Phys. Stat. Mech. Its Appl.* **321**, 679–688 (2003).

19. P. Gong, C. van Leeuwen, Evolution to a small-world network with chaotic units. *EPL Europhys. Lett.* **67**, 328 (2004).

20. A. Griffa, M. P. van den Heuvel, Rich-club neurocircuitry: function, evolution, and vulnerability. *Dialogues Clin. Neurosci.* **20**, 121–132 (2018).

21. X. Liao, A. V. Vasilakos, Y. He, Small-world human brain networks: Perspectives and challenges. *Neurosci. Biobehav. Rev.* **77**, 286–300 (2017).

22. O. Sporns, R. F. Betzel, Modular Brain Networks. *Annu. Rev. Psychol.* **67**, 613–640 (2016).

23. N. Jarman, E. Steur, C. Trengove, I. Y. Tyukin, C. van Leeuwen, Self-organisation of small-world networks by adaptive rewiring in response to graph diffusion. *Sci. Rep.* **7**, 1–9 (2017).

24. H. F. Kwok, P. Jurica, A. Raffone, C. van Leeuwen, Robust emergence of small-world structure in networks of spiking neurons. *Cogn. Neurodyn.* **1**, 39–51 (2007).

25. I. Rentzeperis, C. van Leeuwen, Adaptive rewiring evolves brain-like structure in weighted networks. *Sci. Rep.* **10**, 1–11 (2020).

26. I. Rentzeperis, C. van Leeuwen, Adaptive Rewiring in Weighted Networks Shows Specificity, Robustness, and Flexibility. *Front. Syst. Neurosci.* **15**, 13 (2021).

27. M. Rubinov, O. Sporns, C. van Leeuwen, M. Breakspear, Symbiotic relationship between brain structure and dynamics. *BMC Neurosci.* **10** (2009).

28. C. C. Tapia, V. A. Makarov, C. van Leeuwen, Basic principles drive self-organization of brain-like connectivity structure. *Commun. Nonlinear Sci. Numer. Simul.* **82**, 105065 (2020).

29. J. Li, I. Rentzeperis, C. van Leeuwen, Functional and spatial rewiring principles jointly regulate context-sensitive computation. *PLOS Comput. Biol.* **19**, e1011325 (2023).

30. R. Luna, J. Li, R. Bauer, C. van Leeuwen, Retinal waves in adaptive rewiring networks orchestrate

convergence and divergence in the visual system. *Netw. Neurosci.* 1–20 (2024). https://doi.org/10.1162/netn_a_00370.

31. I. Rentzeperis, S. Laquitaine, C. van Leeuwen, Adaptive rewiring of random neural networks generates convergent–divergent units. *Commun. Nonlinear Sci. Numer. Simul.* **107**, 106135 (2022).

32. A. Chapman, *Semi-Autonomous Networks* (Springer International Publishing, 2015).

33. W. Ren, R. W. Beard, E. M. Atkins, Information consensus in multivehicle cooperative control. *IEEE Control Syst.* **27**, 71–82 (2007).

34. C. Seguin, O. Sporns, A. Zalesky, Brain network communication: concepts, models and applications. *Nat. Rev. Neurosci.* **24**, 557–574 (2023).

35. F. Abdelnour, H. U. Voss, A. Raj, Network diffusion accurately models the relationship between structural and functional brain connectivity networks. *NeuroImage* **90**, 335–347 (2014).

36. L. Luo, Architectures of neuronal circuits. *Science* **373**, eabg7285 (2021).

37. A. J. Keller, *et al.*, A Disinhibitory Circuit for Contextual Modulation in Primary Visual Cortex. *Neuron* **108**, 1181-1193.e8 (2020).

38. C. A. Brittin, S. J. Cook, D. H. Hall, S. W. Emmons, N. Cohen, Volumetric reconstruction of main Caenorhabditis elegans neuropil at two different time points. [Preprint] (2018). Available at: https://www.biorxiv.org/content/10.1101/485771v1 [Accessed 1 September 2025].

39. M. Helmstaedter, *et al.*, Connectomic reconstruction of the inner plexiform layer in the mouse retina. *Nature* **500**, 168–174 (2013).

40. M. H. DeGroot, Reaching a Consensus. *J. Am. Stat. Assoc.* **69**, 118–121 (1974).

41. R. Olfati-Saber, R. M. Murray, Consensus problems in networks of agents with switching topology and time-delays. *IEEE Trans. Autom. Control* **49**, 1520–1533 (2004).

42. A. Litwin-Kumar, B. Doiron, Formation and maintenance of neuronal assemblies through synaptic plasticity. *Nat. Commun.* **5**, 5319 (2014).

43. J. Triesch, A. D. Vo, A.-S. Hafner, Competition for synaptic building blocks shapes synaptic plasticity. *eLife* **7**, e37836 (2018).

44. I. R. Fiete, W. Senn, C. Z. H. Wang, R. H. R. Hahnloser, Spike-Time-Dependent Plasticity and Heterosynaptic Competition Organize Networks to Produce Long Scale-Free Sequences of Neural Activity. *Neuron* **65**, 563–576 (2010).

45. J. N. Bourne, K. M. Harris, Coordination of size and number of excitatory and inhibitory

synapses results in a balanced structural plasticity along mature hippocampal CA1 dendrites during LTP. *Hippocampus* **21**, 354–373 (2011).

46. H. Markram, M. Tsodyks, Redistribution of synaptic efficacy between neocortical pyramidal neurons. *Nature* **382**, 807–810 (1996).

47. S. Royer, D. Paré, Conservation of total synaptic weight through balanced synaptic depression and potentiation. *Nature* **422**, 518–522 (2003).

48. F. Wilcoxon, Individual Comparisons by Ranking Methods. *Biom. Bull.* **1**, 80–83 (1945).

49. L. A. Waller, D. Smith, J. E. Childs, L. A. Real, Monte Carlo assessments of goodness-of-fit for ecological simulation models. *Ecol. Model.* **164**, 49–63 (2003).

50. A. Arenas, J. Duch, A. Fernández, S. Gómez, Size reduction of complex networks preserving modularity. *New J. Phys.* **9**, 176 (2007).

51. E. A. Leicht, M. E. J. Newman, Community Structure in Directed Networks. *Phys. Rev. Lett.* **100**, 118703 (2008).

52. V. Latora, M. Marchiori, Efficient Behavior of Small-World Networks. *Phys. Rev. Lett.* **87**, 198701 (2001).

53. C. A. Brittin, S. J. Cook, D. H. Hall, S. W. Emmons, N. Cohen, A multiscale brain map derived from whole-brain volumetric reconstructions. (2020). https://doi.org/10.5281/zenodo.4763083.

54. C. W. Lynn, Heavy_tailed_connectivity. (2025). Available at: https://github.com/ChrisWLynn/Heavy_tailed_connectivity [Accessed 20 August 2026].

55. B. Oksendal, *Stochastic differential equations: an introduction with applications* (Springer Science & Business Media, 2013).

56. J. Yang, S. A. Prescott, Homeostatic regulation of neuronal function: importance of degeneracy and pleiotropy. *Front. Cell. Neurosci.* **17** (2023).

57. A. Citri, R. C. Malenka, Synaptic Plasticity: Multiple Forms, Functions, and Mechanisms. *Neuropsychopharmacology* **33**, 18–41 (2008).

58. D. O. Hebb, *The organization of behavior: A neuropsychological theory* (Psychology press, 1949).

59. T. V. P. Bliss, T. Lømo, Long-lasting potentiation of synaptic transmission in the dentate area of the anaesthetized rabbit following stimulation of the perforant path. *J. Physiol.* **232**, 331–356 (1973).

60. G. Lynch, J. Larson, S. Kelso, G. Barrionuevo, F. Schottler, Intracellular injections of EGTA block

induction of hippocampal long-term potentiation. *Nature* **305**, 719–721 (1983).

61. T. V. Bliss, G. L. Collingridge, A synaptic model of memory: long-term potentiation in the hippocampus. *Nature* **361**, 31–39 (1993).

62. G. S. Lynch, T. Dunwiddie, V. Gribkoff, Heterosynaptic depression: a postsynaptic correlate of long-term potentiation. *Nature* **266**, 737–739 (1977).

63. R. M. Mulkey, R. C. Malenka, Mechanisms underlying induction of homosynaptic long-term depression in area CA1 of the hippocampus. *Neuron* **9**, 967–975 (1992).

64. W. B. Levy, O. Steward, Temporal contiguity requirements for long-term associative potentiation/depression in the hippocampus. *Neuroscience* **8**, 791–797 (1983).

65. W. Gerstner, R. Kempter, J. L. Van Hemmen, H. Wagner, A neuronal learning rule for sub-millisecond temporal coding. *Nature* **383**, 76–78 (1996).

66. H. Markram, J. Lübke, M. Frotscher, B. Sakmann, Regulation of synaptic efficacy by coincidence of postsynaptic APs and EPSPs. *Science* **275**, 213–215 (1997).

67. G. Bi, M. Poo, Synaptic modifications in cultured hippocampal neurons: dependence on spike timing, synaptic strength, and postsynaptic cell type. *J. Neurosci.* **18**, 10464–10472 (1998).

68. J. T. Trachtenberg, *et al.*, Long-term in vivo imaging of experience-dependent synaptic plasticity in adult cortex. *Nature* **420**, 788–794 (2002).

69. A. Holtmaat, K. Svoboda, Experience-dependent structural synaptic plasticity in the mammalian brain. *Nat. Rev. Neurosci.* **10**, 647–658 (2009).

70. C. Sala, M. Segal, Dendritic Spines: The Locus of Structural and Functional Plasticity. *Physiol. Rev.* **94**, 141–188 (2014).

71. F. Engert, T. Bonhoeffer, Dendritic spine changes associated with hippocampal long-term synaptic plasticity. *Nature* **399**, 66–70 (1999).

72. M. Maletic-Savatic, R. Malinow, K. Svoboda, Rapid dendritic morphogenesis in CA1 hippocampal dendrites induced by synaptic activity. *Science* **283**, 1923–1927 (1999).

73. M. Matsuzaki, N. Honkura, G. C. Ellis-Davies, H. Kasai, Structural basis of long-term potentiation in single dendritic spines. *Nature* **429**, 761–766 (2004).

74. Q. Zhou, K. J. Homma, M. Poo, Shrinkage of dendritic spines associated with long-term depression of hippocampal synapses. *Neuron* **44**, 749–757 (2004).

75. M. Cirunay, G. Ódor, I. Papp, G. Deco, Scale-free behavior of weight distributions of connectomes. *Phys. Rev. Res.* **7**, 013134 (2025).

76. B. Piazza, D. L. Barabási, A. Ferreira Castro, G. Menichetti, A.-L. Barabási, Physical Network Constraints Define the Lognormal Architecture of the Brain's Connectome. [Preprint] (2025). Available at: http://biorxiv.org/lookup/doi/10.1101/2025.02.27.640551 [Accessed 10 June 2025].

77. S. Takemura, *et al.*, A visual motion detection circuit suggested by Drosophila connectomics. *Nature* **500**, 175–181 (2013).

78. L. R. Varshney, B. L. Chen, E. Paniagua, D. H. Hall, D. B. Chklovskii, Structural Properties of the Caenorhabditis elegans Neuronal Network. *PLoS Comput. Biol.* **7**, e1001066 (2011).

79. J. G. White, E. Southgate, J. N. Thomson, S. Brenner, The Structure of the Nervous System of the Nematode Caenorhabditis elegans. *Philos. Trans. R. Soc. Lond. B. Biol. Sci.* **314**, 1–340 (1986).

80. G. Buzsáki, K. Mizuseki, The log-dynamic brain: how skewed distributions affect network operations. *Nat. Rev. Neurosci.* **15**, 264–278 (2014).

81. Ł. Kuśmierz, S. Ogawa, T. Toyoizumi, Edge of Chaos and Avalanches in Neural Networks with Heavy-Tailed Synaptic Weight Distribution. *Phys. Rev. Lett.* **125**, 028101 (2020).

82. R. Iyer, V. Menon, M. Buice, C. Koch, S. Mihalas, The Influence of Synaptic Weight Distribution on Neuronal Population Dynamics. *PLOS Comput. Biol.* **9**, e1003248 (2013).

83. G. M. Edelman, J. A. Gally, Degeneracy and complexity in biological systems. *Proc. Natl. Acad. Sci.* **98**, 13763–13768 (2001).

84. G. Tononi, O. Sporns, G. M. Edelman, Measures of degeneracy and redundancy in biological networks. *Proc. Natl. Acad. Sci.* **96**, 3257–3262 (1999).

85. E. Marder, J.-M. Goaillard, Variability, compensation and homeostasis in neuron and network function. *Nat. Rev. Neurosci.* **7**, 563–574 (2006).

86. E. Marder, Variability, compensation, and modulation in neurons and circuits. *Proc. Natl. Acad. Sci.* **108**, 15542–15548 (2011).

87. S. Marom, E. Marder, A biophysical perspective on the resilience of neuronal excitability across timescales. *Nat. Rev. Neurosci.* **24**, 640–652 (2023).

88. L. Albantakis, C. Bernard, N. Brenner, E. Marder, R. Narayanan, The Brain's Best Kept Secret Is Its Degenerate Structure. *J. Neurosci.* **44** (2024).

89. R. C. Froemke, M. Poo, Y. Dan, Spike-timing-dependent synaptic plasticity depends on dendritic location. *Nature* **434**, 221–225 (2005).

90. Y. Dan, M.-M. Poo, Spike Timing-Dependent Plasticity: From Synapse to Perception. *Physiol. Rev.* **86**, 1033–1048 (2006).

91. K. C. Bittner, *et al.*, Conjunctive input processing drives feature selectivity in hippocampal CA1 neurons. *Nat. Neurosci.* **18**, 1133–1142 (2015).

92. K. C. Bittner, A. D. Milstein, C. Grienberger, S. Romani, J. C. Magee, Behavioral time scale synaptic plasticity underlies CA1 place fields. *Science* **357**, 1033–1036 (2017).

93. A. D. Milstein, *et al.*, Bidirectional synaptic plasticity rapidly modifies hippocampal representations. *eLife* **10**, e73046 (2021).

94. C. Grienberger, J. C. Magee, Entorhinal cortex directs learning-related changes in CA1 representations. *Nature* **611**, 554–562 (2022).

95. J. B. Priestley, J. C. Bowler, S. V. Rolotti, S. Fusi, A. Losonczy, Signatures of rapid plasticity in hippocampal CA1 representations during novel experiences. *Neuron* **110**, 1978-1992.e6 (2022).

96. L. Z. Fan, *et al.*, All-optical physiology resolves a synaptic basis for behavioral timescale plasticity. *Cell* **186**, 543-559.e19 (2023).

97. J. C. Magee, Behavioral timescale synaptic plasticity: properties, elements and functions. *Nat. Neurosci.* **29**, 520–534 (2026).

98. T. Hige, Y. Aso, G. M. Rubin, G. C. Turner, Plasticity-driven individualization of olfactory coding in mushroom body output neurons. *Nature* **526**, 258–262 (2015).

99. A. M. Mittal, D. Gupta, A. Singh, A. C. Lin, N. Gupta, Multiple network properties overcome random connectivity to enable stereotypic sensory responses. *Nat. Commun.* **11**, 1023 (2020).

100. H. Asari, M. Meister, Divergence of visual channels in the inner retina. *Nat. Neurosci.* **15**, 1581–1589 (2012).

101. S. Hammer, A. Monavarfeshani, T. Lemon, J. Su, M. A. Fox, Multiple retinal axons converge onto relay cells in the adult mouse thalamus. *Cell Rep.* **12**, 1575–1583 (2015).

102. J. L. Morgan, D. R. Berger, A. W. Wetzel, J. W. Lichtman, The fuzzy logic of network connectivity in mouse visual thalamus. *Cell* **165**, 192–206 (2016).

103. J. S. Lund, A. Angelucci, P. C. Bressloff, Anatomical substrates for functional columns in macaque monkey primary visual cortex. *Cereb. Cortex* **13**, 15–24 (2003).

104. MICrONS Consortium, Functional connectomics spanning multiple areas of mouse visual cortex. *Nature* **640**, 435–447 (2025).

105. I. Rentzeperis, J. Li, R. Bauer, C. van Leeuwen, Opening the black box of neural variability: From noise to mechanisms. *Neurosci. Biobehav. Rev.* **186**, 106661 (2026).

106. D. Arendt, A. S. Denes, G. Jékely, K. Tessmar-Raible, The evolution of nervous system centralization. *Philos. Trans. R. Soc. B Biol. Sci.* **363**, 1523–1528 (2008).

107. K. J. Peterson, J. A. Cotton, J. G. Gehling, D. Pisani, The Ediacaran emergence of bilaterians: congruence between the genetic and the geological fossil records. *Philos. Trans. R. Soc. B Biol. Sci.* **363**, 1435–1443 (2008).

108. M. Ercsey-Ravasz, *et al.*, A Predictive Network Model of Cerebral Cortical Connectivity Based on a Distance Rule. *Neuron* **80**, 184–197 (2013).

109. H. F. Song, H. Kennedy, X.-J. Wang, Spatial embedding of structural similarity in the cerebral cortex. *Proc. Natl. Acad. Sci.* **111**, 16580–16585 (2014).

110. S. Horvát, *et al.*, Spatial Embedding and Wiring Cost Constrain the Functional Layout of the Cortical Network of Rodents and Primates. *PLOS Biol.* **14**, e1002512 (2016).

111. R. F. Betzel, *et al.*, Generative models of the human connectome. *NeuroImage* **124**, 1054–1064 (2016).

112. D. Akarca, P. E. Vértes, E. T. Bullmore, D. E. Astle, A generative network model of neurodevelopmental diversity in structural brain organization. *Nat. Commun.* **12**, 4216 (2021).

113. S. Oldham, *et al.*, Modeling spatial, developmental, physiological, and topological constraints on human brain connectivity. *Sci. Adv.* **8**, eabm6127 (2022).

114. A.-L. Barabási, R. Albert, Emergence of Scaling in Random Networks. *Science* **286**, 509–512 (1999).

115. G. Bi, M. Poo, Synaptic modification by correlated activity: Hebb's postulate revisited. *Annu. Rev. Neurosci.* **24**, 139–166 (2001).

116. A. Hayashi-Takagi, Increased heavy-tailed distribution of synaptic weights distorts neurocomputation in schizophrenia. *Curr. Opin. Neurobiol.* **94**, 103089 (2025).

117. G. G. Turrigiano, S. B. Nelson, Hebb and homeostasis in neuronal plasticity. *Curr. Opin. Neurobiol.* **10**, 358–364 (2000).

118. O. Sporns, R. Kötter, Motifs in Brain Networks. *PLoS Biol.* **2**, e369 (2004).

119. I. Magrans de Abril, J. Yoshimoto, K. Doya, Connectivity inference from neural recording data: Challenges, mathematical bases and research directions. *Neural Netw.* **102**, 120–137 (2018).

120. G. Gosti, *et al.*, A recurrent Hopfield network for estimating meso-scale effective connectivity in MEG. *Neural Netw.* **170**, 72–93 (2024).

121. J. M. Beggs, D. Plenz, Neuronal Avalanches in Neocortical Circuits. *J. Neurosci.* **23**, 11167–11177

(2003).

122. B. A. Olshausen, D. J. Field, Emergence of simple-cell receptive field properties by learning a sparse code for natural images. (1996).

123. I. Rentzeperis, D. Prandi, M. Bertalmío, A Neural Model for V1 That Incorporates Dendritic Nonlinearities and Backpropagating Action Potentials. *J. Neurosci.* **45** (2025).

124. X.-Y. Zhang, J. M. Moore, X. Ru, G. Yan, Geometric Scaling Law in Real Neuronal Networks. *Phys. Rev. Lett.* **133** (2024).

125. M. Kaiser, C. C. Hilgetag, A. van Ooyen, A Simple Rule for Axon Outgrowth and Synaptic Competition Generates Realistic Connection Lengths and Filling Fractions. *Cereb. Cortex* **19**, 3001–3010 (2009).

126. H. S. Seung, U. Sümbül, Neuronal Cell Types and Connectivity: Lessons from the Retina. *Neuron* **83**, 1262–1272 (2014).

127. C. Zhang, A. L. Kolodkin, R. O. Wong, R. E. James, Establishing Wiring Specificity in Visual System Circuits: From the Retina to the Brain. *Annu. Rev. Neurosci.* **40**, 395–424 (2017).

# Acknowledgments

The resources and services used in this work were provided by the VSC (Flemish Supercomputer Center), funded by the Research Foundation - Flanders (FWO) and the Flemish Government.

## Author contributions

J.L.: conceptualization, software, investigation, visualization, writing—original draft, writing—review & editing.

C.v.L.: conceptualization, writing—original draft, writing—review & editing, funding acquisition.

R.B.: software, writing—review & editing.

I.R.: conceptualization, software, visualization, writing—original draft, writing—review & editing.

# Funding

The research was supported by an Odysseus grant (G.0003.12) from the Flemish Organization for Science (F.W.O) to Cees van Leeuwen.

## Competing interests

The authors declare no competing interests.

Self-*regulated emergence of heavy-tailed weight distributions in evolving complex network architectures*

**SUPPLEMENTARY INFORMATION**

Jia Li, Cees van Leeuwen, Roman Bauer, Ilias Rentzeperis

# Supplementary Methods

## Notation and definitions

Connectivity is modeled as a directed, weighted graph $G = (V, E, W)$, where $V = \{1,2, \dots, n\}$ is the set of nodes, $E \subset V \times V$ the set of connections, and $W = \{w_{ij}: w_{ij} \geq 0, i, j \in V\}$ the set of connection weights. Each node represents a microcircuit consisting of a single excitatory neuron locally clustered with inhibitory neurons that provide inhibition to it. Therefore, inhibitory activity is not explicitly modeled. An ordered node pair $(i,j) \in E$ represents a directed connection from $j$ (tail) to $i$ (head), denoted as $j \to i$. For $(i,j) \in V \times V$, $w_{ij}$ is positive if $(i,j) \in E$ and is zero otherwise. We assume no self-loops in the network, i.e., $w_{ii} = 0$ for all $i \in V$. The number of nodes is $|V| = n$ and of connections $|E| = m$.

The nodes attached to the tails of the in-connections of node $i$ constitute the in-neighborhood of $i$, $N_{in}(i)$, with the remaining nodes, $V - i - N_{in}(i)$, being denoted as $N_{in}^c(i)$. Analogously, the nodes attached to the heads of the out-connections of $i$ constitute the out-neighborhood of $i$, $N_{out}(i)$, and the rest is denoted as $N_{out}^c(i)$. The in-degree of node $i$ is defined as the number of its in-connections, and its out-degree as the number of its out-connections. Its in-strength is the sum of its in-connection weights, $s_{in}(i) = \sum_j w_{ij}$ and its out-strength the sum of its out-connection weights, $s_{out}(i) = \sum_j w_{ji}$.

Unless otherwise stated, initial networks $G = (V, E, W)$ have $n = 100$ nodes and $m = 912$ connections. The connections initially are randomly assigned to pairs of nodes. Each connection has an initial weight sampled from a normal distribution, $N(1, 0.25^2)$. Negative weights (a highly unlikely occurrence as indicated by its probability: $3.17 * 10^{-5}$) are set to 0.05. The sampled weight values are subsequently normalized so that their sum equals the number of connections. The model is not density-invariant, but the main qualitative conclusions hold within a biologically plausible range of sparse connectivities (see Supplementary Note 2 for a sensitivity analysis).

## Derivation of solutions for consensus and advection dynamics

As defined in Equation (2), consensus dynamics at the node level is expressed as follows:

$$\dot{x}_i(t) = \sum_{\{\forall j | j \to i\}} w_{ij} \left( x_j(t) - x_i(t) \right)$$

In matrix form, for all nodes $n$ collectively, Equation (2) reads as follows:

$$\dot{\boldsymbol{x}}(t) = -L_{in} \boldsymbol{x}(t) \tag{S1}$$

where $L_{in}$ is the in-degree Laplacian, a variant of the classical Laplacian matrix used in undirected graphs. $L_{in}$ has the following entries:

$$l_{ij}^{in} = \begin{cases} \sum_{k=1}^{n} w_{ik}\,, if\ i = j \\ -w_{ij}, if\ i \neq j \end{cases}$$

The solution to Equation (S1) is Equation (3):

$$\boldsymbol{x}(t_{end}) = c(\tau)\boldsymbol{x}(t_{begin})$$

where $\tau = t_{end} - t_{begin}$, $c(\tau) = \exp(-L_{in}\tau)$, and $\boldsymbol{x}(t)$ the node concentrations at time $t$.

As defined in Equation (4), advection dynamics at the node level is expressed as follows:

$$\dot{x}_i(t) = \sum_{\{\forall j | j \to i\}} w_{ij} x_j(t) - \sum_{\{\forall k | i \to k\}} w_{ki} x_i(t)$$

This expression can be written in matrix form for all $n$ nodes collectively as:

$$\dot{\boldsymbol{x}}(t) = -L_{out}\boldsymbol{x}(t) \tag{S2}$$

where $L_{out}$ has entries:

$$l_{ij}^{out} = \begin{cases} \sum_{k=1}^{n} w_{ki}\,, if\ i = j \\ -w_{ij}, if\ i \neq j \end{cases}$$

$L_{out}$, the out-degree Laplacian, is another variant of the Laplacian matrix as now its $i_{th}$ diagonal entry corresponds to the out-strength of node $i$. The solution to Equation (S2) is Equation (5):

$$\boldsymbol{x}(t_{end}) = a(\tau)\boldsymbol{x}(t_{begin})$$

where $\tau = t_{end} - t_{begin}$, $a(\tau) = \exp(-L_{out}\tau)$, and $\boldsymbol{x}(t)$ the node concentrations at time $t$.

## Control algorithms

To evaluate the distinct contributions of adaptive rewiring and adaptive weight adjustment in our setting, we introduce two control conditions. First, to assess whether adaptive weight adjustment alone could generate heavy-tailed distributions and any characteristic complexity features in the absence of structural plasticity, we skip the adaptive rewiring steps. In this condition, the network's connectivity structure is initially random and held fixed throughout learning, allowing us to study the effect of weight dynamics on emergent network complexity features in isolation.

Second, to evaluate whether random rewiring could have similar effects to adaptive rewiring, we substitute the latter with the former. In the random rewiring step, when an in-connection of node $v$ is rewired, two nodes $k \in N_{in}(v)$ and $l \in N_{in}^{c}(v)$ are selected randomly. The in-connection $k \to v$ is cut and a new in-connection, $l \to v$, is added. An analogous process is applied for out-connections.

## Heavy-tailed distributions and power-law fit of the tail

To characterize the distribution of connection weights, we evaluate several candidate distributions, right-skewed ones (lognormal, Weibull, gamma, exponential, inverse Gaussian, and inverse gamma) as well as the normal distribution. By definition, a distribution is considered heavy-tailed if its tail decays more slowly than an exponential distribution. Among our candidate distributions, the lognormal, inverse gamma, and Weibull distributions (when the shape parameter is less than 1) qualify as heavy-tailed. Maximum likelihood estimation (MLE) is used to fit each distribution to connection weights. Model fit is assessed by the Kolmogorov-Smirnov (KS) statistic, where a smaller KS statistic indicates a closer fit between the model and the observed data. We apply the Wilcoxon signed-rank test (1) to compare the KS statistics of each candidate distribution against that of the one with the smallest mean KS statistic. The significance level is 0.05. We reject the null hypothesis when a $p$-value is below 0.05.

To qualify the upper tail of the weight distribution, we fit a power-law to the tail of connection weights exceeding the median. If a distribution follows a power-law in its upper tail, its $p(X = x)$ has the form

$$p(X = x) = C \cdot x^{-\alpha}$$

where $\alpha$ is the exponent that governs the rate of decay and $C$ the normalizing constant. A smaller $\alpha$ indicates a heavier tail. In log-log scale, this relationship appears as a linear trend. To construct the histogram of the probability distribution, we apply logarithmic binning, using bins that are uniformly spaced on a logarithmic scale. Unless otherwise stated, the interval between the median and the maximum of connection weights is divided into 20 bins. The power-law fit is obtained by linear regression on the log-transformed histogram using MLE.

## Monte Carlo test of goodness-of-fit

We use a Monte Carlo test of goodness-of-fit to assess the fit to the weight distributions of empirical neuronal connectomes, following the framework of Waller et al. (2). This framework was developed for stochastic simulation models based on local interactions. The null hypothesis indicates that “the data appear to be a typical realization of the model”.

The KS statistic is used to quantify the discrepancy between two distributions. The observed discrepancy, $KS_{obs}$, is computed by comparing the empirical weights with the reference distribution obtained by pooling weights across all simulated networks. The simulated discrepancies, $KS_i$, are computed by comparing weights of each simulated network with the reference distribution obtained by pooling weights across the remaining simulated networks. The Monte Carlo $p$-value is computed as

$$\frac{\#\{KS_i \geq KS_{obs}: i = 1, \dots, nsim\}}{nsim + 1}$$

where $nsim$ is the number of simulated networks. The significance level is set to 0.05. We reject the null hypothesis when a $p$-value is below 0.05.

## Connectivity metrics

We use four connectivity metrics to characterize the topology of weighted directed networks generated under control conditions: clustering coefficient, average efficiency, small-world index and modularity.

**Clustering coefficient.** The clustering coefficient of a node quantifies the extent to which its neighbors are connected to each other. The value ranges from 0 to 1, where higher values indicate more interconnected local neighborhoods. For weighted directed networks, the clustering coefficient of node $i$ is computed as

$$C(i) = \frac{\left[\left(A^{\left[\frac{1}{3}\right]} + (A^T)^{\left[\frac{1}{3}\right]}\right)^3\right]_{ii}}{2[d_i^{tot}(d_i^{tot} - 1) - 2d_i^{\leftrightarrow}]}$$

where $[\cdot]_{ij}$ denotes the $(i, j)$ element of a matrix, and $A^{\left[\frac{1}{3}\right]}$ denotes the element-wise cubic root of the adjacency matrix $A$ such that $\left[A^{\left[\frac{1}{3}\right]}\right]_{ij} = \left[w_{ij}^{\frac{1}{3}}\right]$, $d_i^{tot}$ is the total degree of node $i$, and $d_i^{\leftrightarrow}$ is the number of bidirectional connections between node $i$ and its neighbors (3). The clustering coefficient of a network is defined as

$$C = \frac{1}{n}\sum_{i=1}^{n} C(i)$$

**Efficiency.** The average efficiency quantifies the efficiency of information transmission over a network. The value is non-negative, where higher value indicate that nodes can transport information to the rest of the network more efficiently. We define the topological distance of a connection $v \to u$ as $d_{uv} = \frac{1}{w_{uv}}$, which can be interpreted as the difficulty of transmission (4). The length of a directed path is defined as the sum of topological distances of connections on this path. The shortest directed path length from node $j$ to node $i$, denoted $l_{ij}$, was defined as the minimum over all directed paths from $j$ to $i$. If there is no directed path from $j$ to $i$, $l_{ij} = \infty$. The average efficiency is then defined as the mean of the inversed shortest directed path lengths of all node pairs (5)

$$E = \frac{1}{n(n-1)} \sum_{i \neq j \in V} \frac{1}{l_{ij}}$$

**Small-world index.** The small-world index is originally defined as the ratio of the normalized clustering coefficient and the normalized average path length, each normalized against an equivalent random network (6). It captures the balance between local clustering and global integration, two features that are typically in tension. Higher values indicate that the network has relatively high local clustering and efficient communication across the network simultaneously. The average efficiency is inversely related to the shortest path length as shorter directed paths contribute larger values to the efficiency. Here we define the small-world index as the product of the clustering coefficient and the average efficiency. To assess the effect of adaptive weight adjustment, the two metrics were normalized against null networks generated by randomly permuting the weights across the existing edges.

$$S = \frac{C}{C_{null}} \times \frac{E}{E_{null}}$$

**Modularity.** The modularity index quantifies the extent to which a network can be divided into separate communities. Values range from -1 to 1. Negative values indicate fewer connections within communities than expected by random chance, whereas positive values indicate more connections within communities than expected. Higher values mean denser within-community connectivity.

We use the following definition for weighted directed networks

$$Q = \frac{1}{W}\sum_{i}\sum_{j}\left(w_{ij} - \frac{s_{in}(i)s_{out}(j)}{W}\right)\delta(C_i, Cj)$$

where $W = \sum_i \sum_j w_{ij}$, $s_{in}(i) = \sum_j w_{ij}$, $s_{out}(j) = \sum_i w_{ij}$, $\delta(C_i, C_j)$ is the Kronecker delta function, and $C_i$ is the community that node $i$ belongs to (7). The communities are detected by the algorithm proposed by Leicht and Newman (8).

# Supplementary Figures

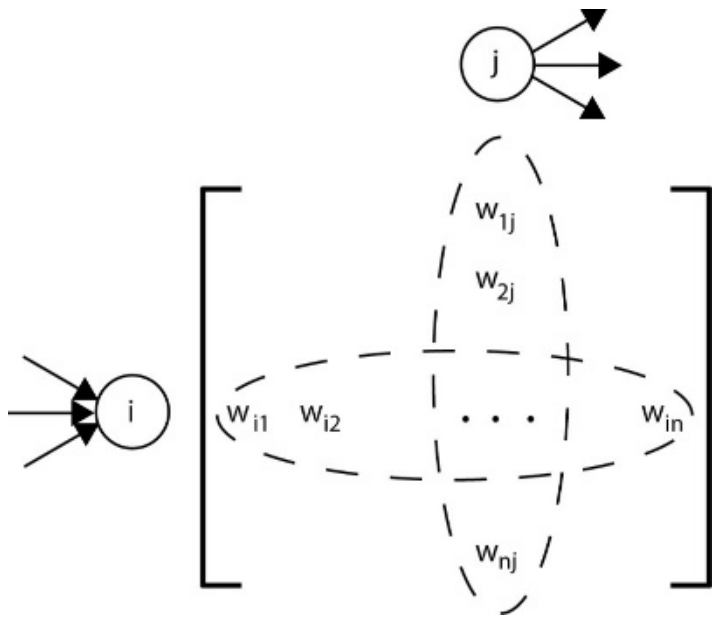

**Fig S1. The adjacency matrix of a weighted directed graph.** The $i$th row of $A$ provides the in-degree connections of node $x_i$, the $j$th column the out-degree connections of node $x_j$.

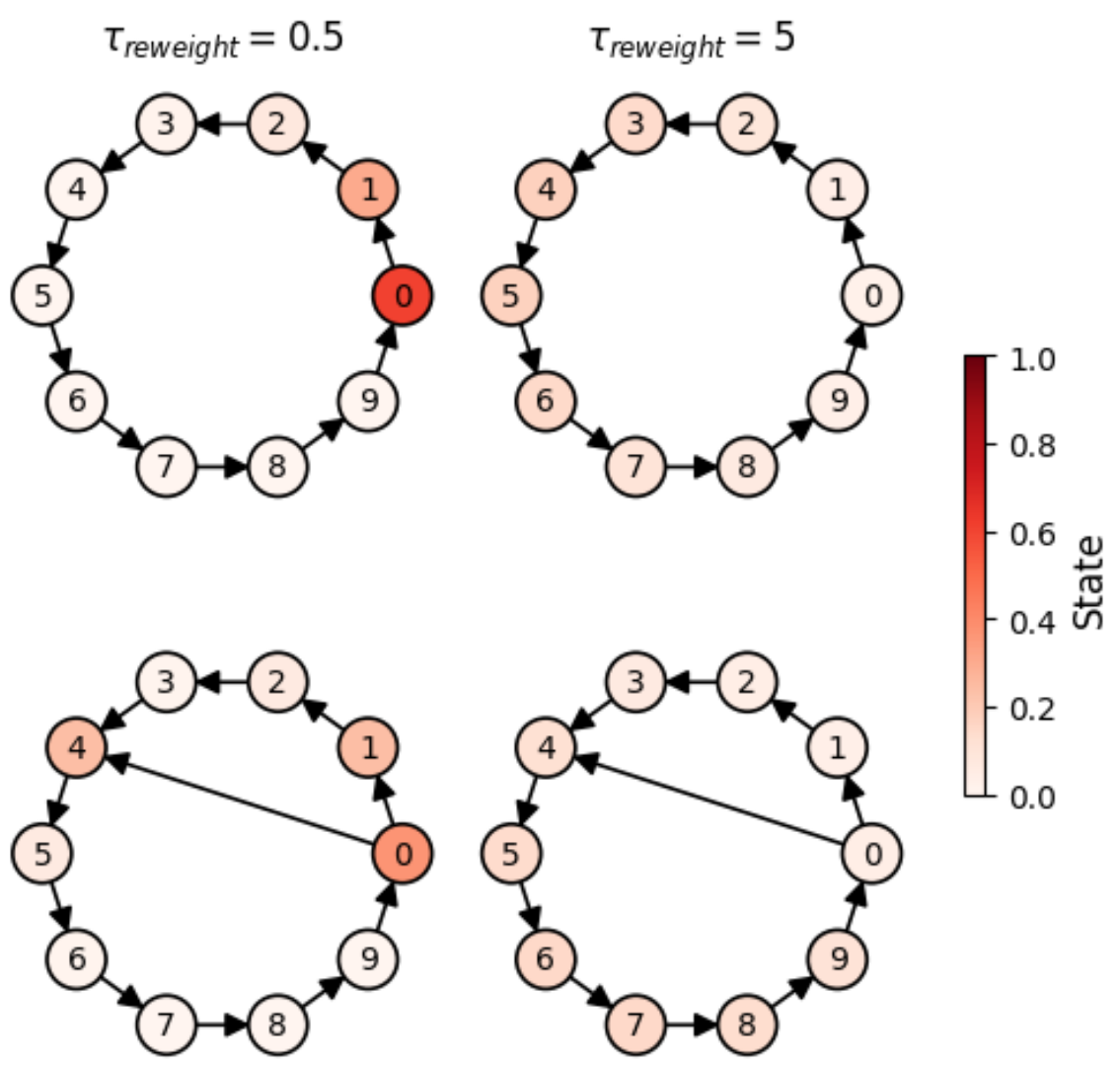

**Fig S2. Network with short path lengths exhibits faster diffusion.** Concentration diffuses on the network as time passes under advection dynamics. The source node was node 0 and was initialized with state 1. Other nodes were initialized with state 0.

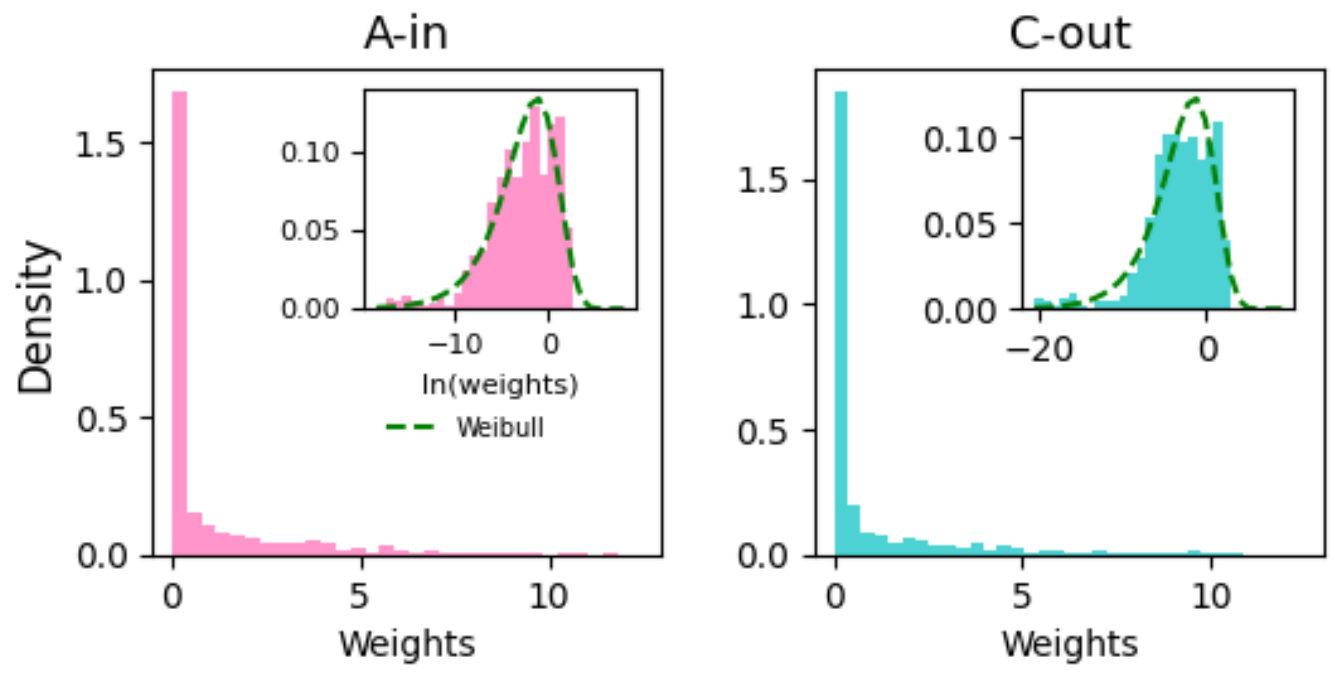

**Fig S3. Weight distributions are heavy-tailed at $\tau_{reweight} = 0.05$.** Each histograms shows the distribution of connection weights from a single representative simulated network. Weibull distributions provide the best fit. The estimated shape parameter is $0.34 \pm 0.06$ in the A-in condition and $0.37 \pm 0.08$ in the C-out condition.

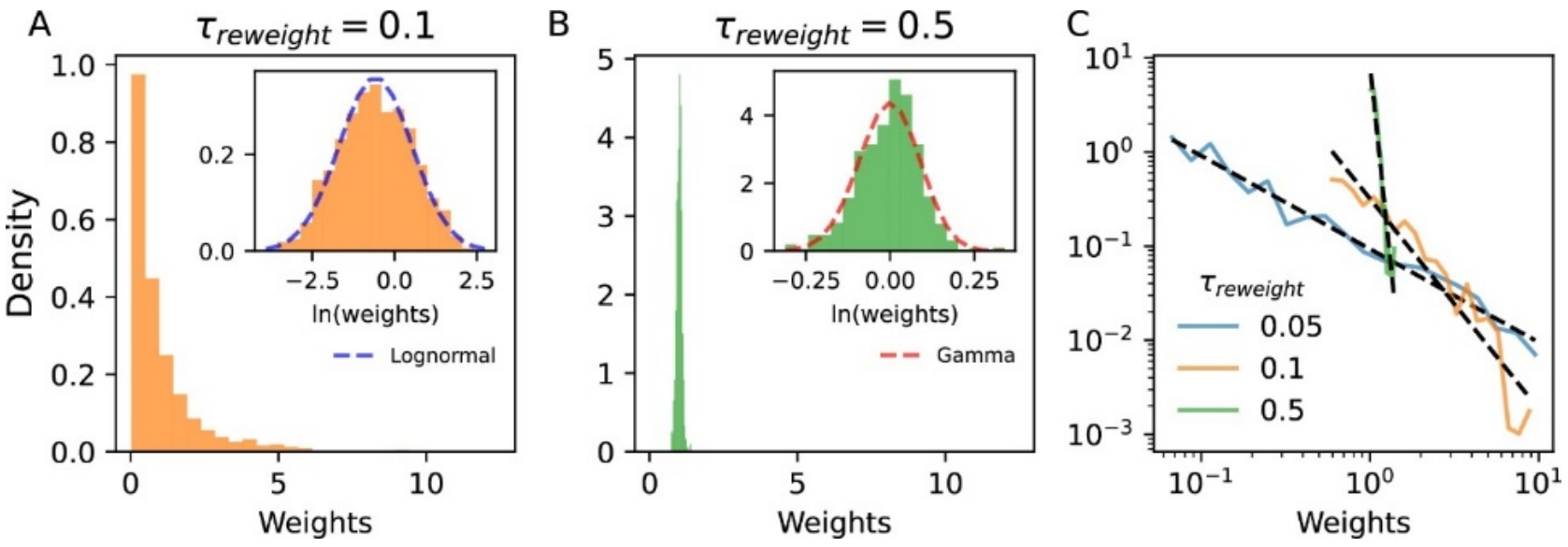

**Fig S4. Heavy-tailed weight distributions emerge only at small $\tau_{reweight}$ when using C-out adaptive weight adjustment alone.** (A, B) Distributions of connection weights of single representative simulated networks generated at different $\tau_{reweight}$: (A) $\tau_{reweight} = 0.1$, best fit by a lognormal distribution; (B) $\tau_{reweight} = 0.5$, best fit by a gamma distribution. (C) Power-law fits to the upper tails of the distributions of connection weights for $\tau_{reweight} = 0.05$, 0.1, 0.5. Colored lines represent upper tails of distributions of connection weights from single representative simulated networks, and black dashed lines are the best power-law fits.

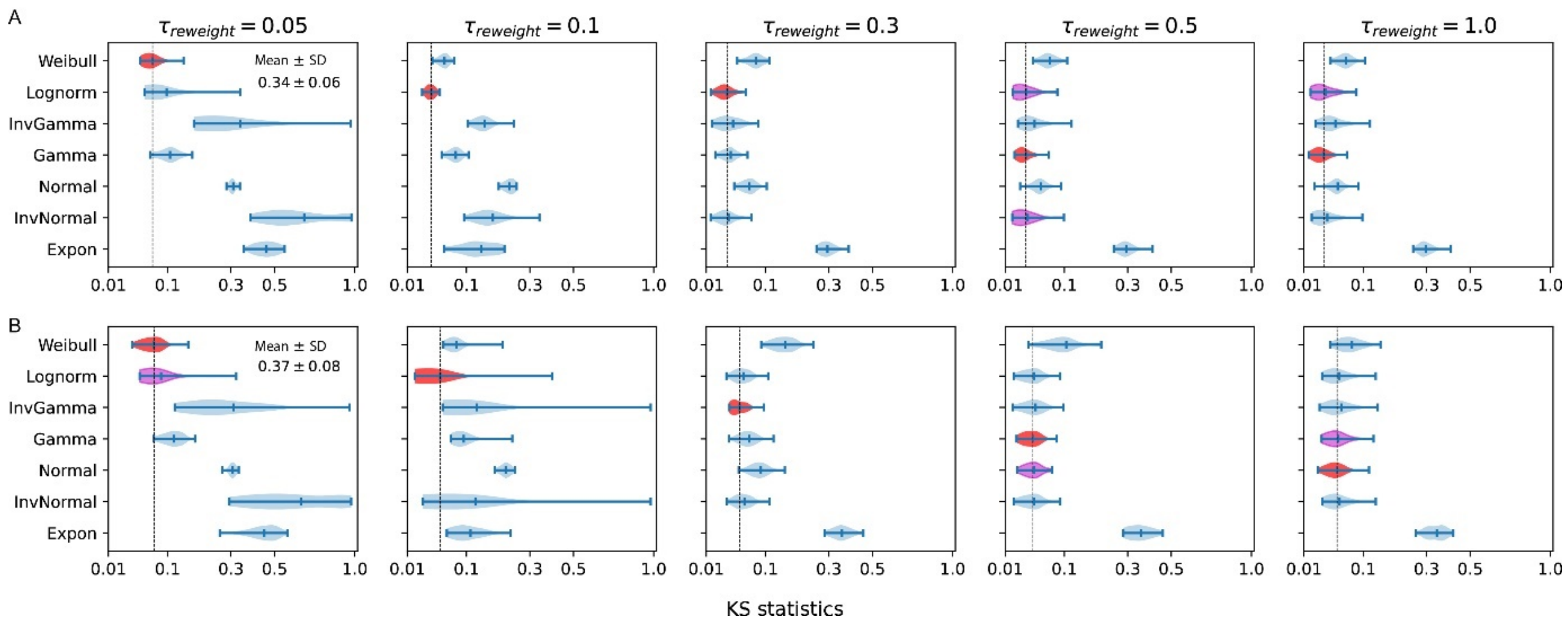

**Fig S5. KS statistics of candidate distributions fit across $\tau_{reweight}$ values.** (A) A-in condition. (B) C-out condition. Red indicates the distribution with the minimal mean (best fit) and the dashed line marks this minimum. Magenta highlights distributions with KS statistics statistically comparable to the best fit. "Mean ± SD" reports the average and standard deviation of shape parameters estimated from 30 Weibull fits. In both conditions, the best-fitting distribution shifts from heavy-tailed forms (lognormal, inverse gamma, and Weibull with shape parameter 1) to light-tailed distributions as $\tau_{reweighting}$ increases.

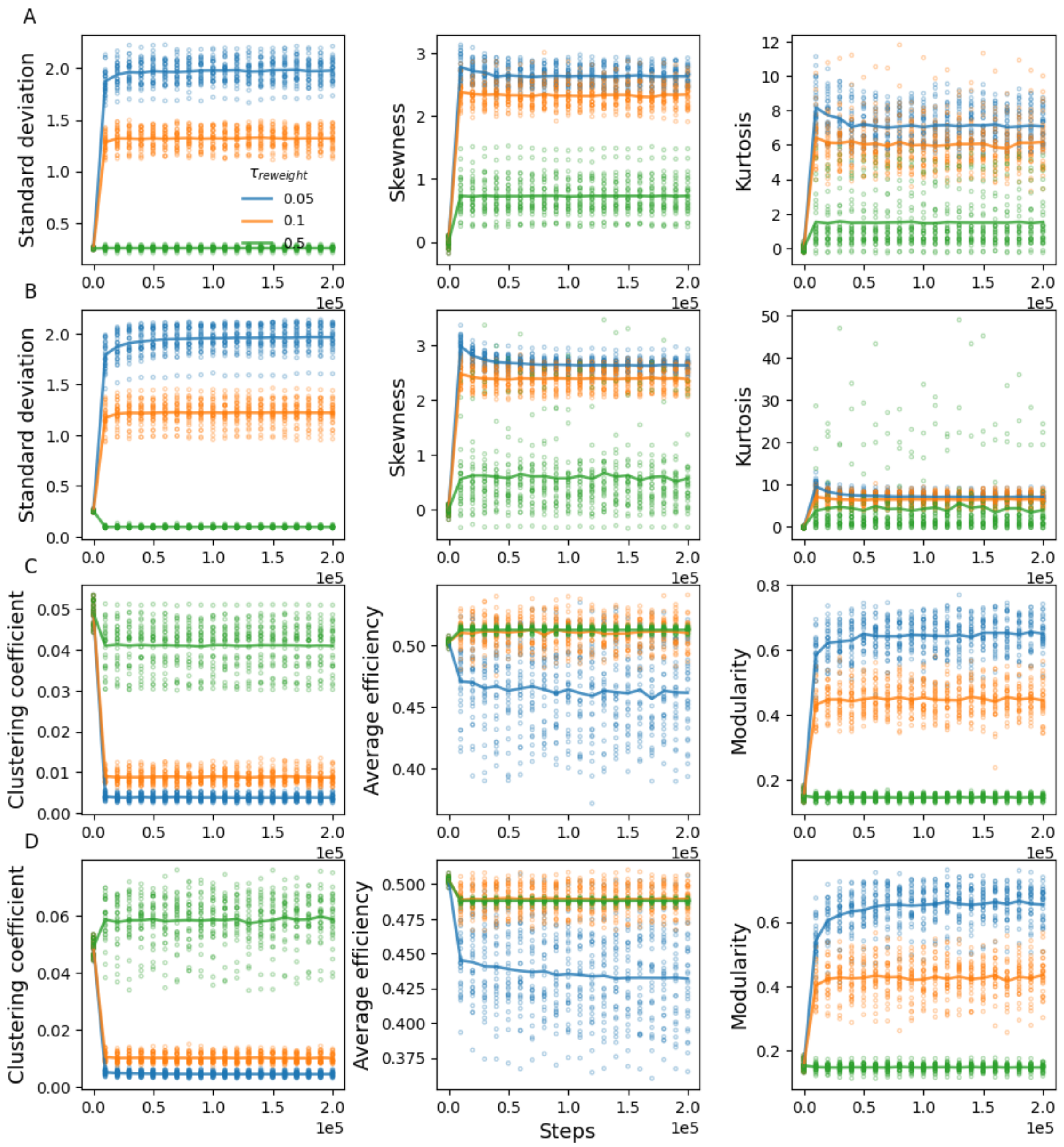

**Fig S6. Weight distribution and network topology reach stationarity within $2 \times 10^5$ steps of adaptive weight adjustment.** (A, B) Summary statistics of weight distributions as functions of simulation step across different $\tau_{reweight}$ in the A-in (A) and C-out (B) conditions. (C, D) Summary connectivity metrics as functions of simulation step for the same simulations in the A-in (C) and C-out (D) conditions. Points represent values from 30 individual simulated networks and lines show the means across networks at different steps.

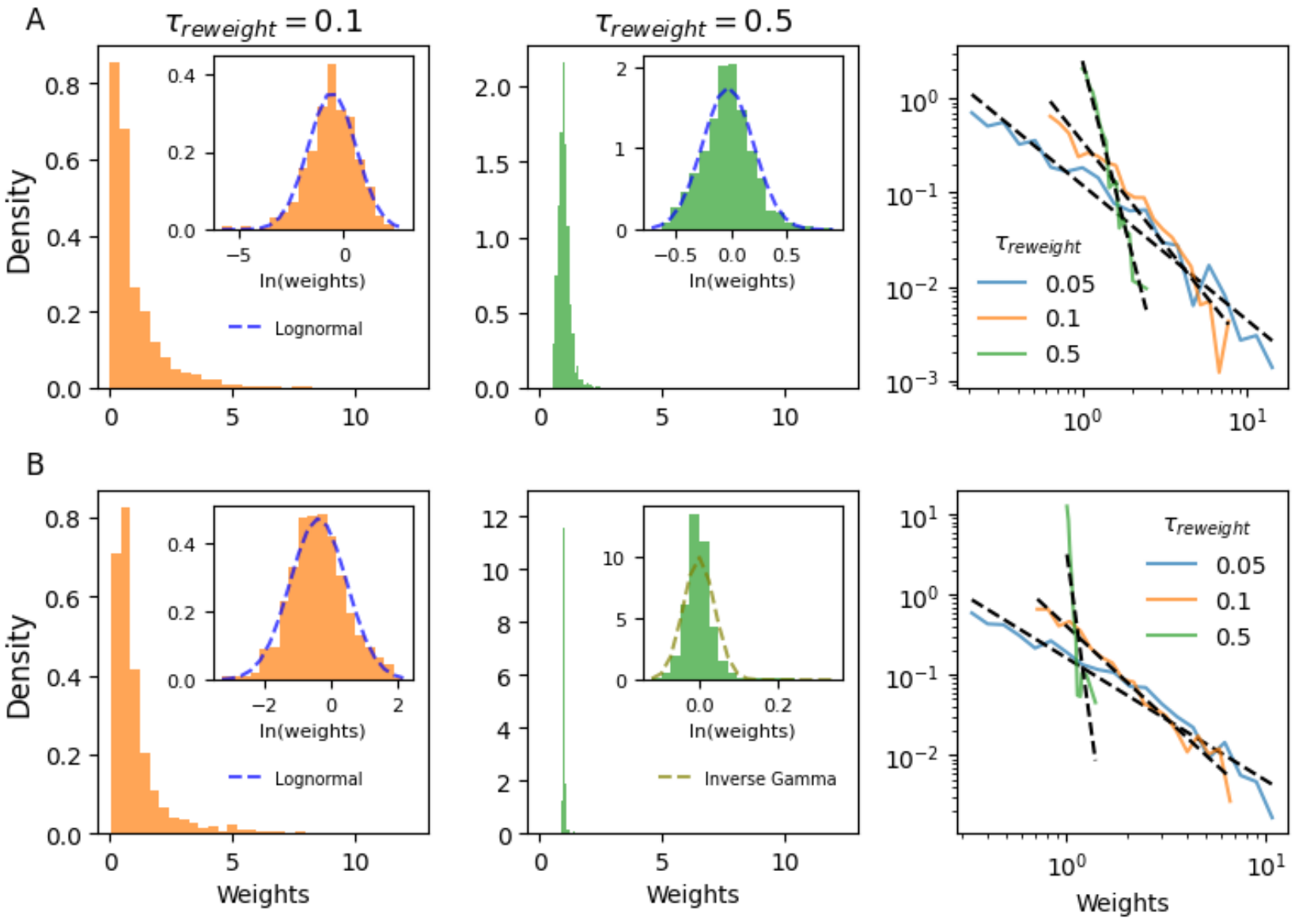

**Fig S7. Connection weights in networks generated by adaptive weight adjustment with random rewiring decay more rapidly as $\tau_{reweight}$ increases.** (A) A-in condition. (B) C-out condition. Results are from single representative simulated networks. Simulations were run for $N = 2 \times 10^5$ steps of adaptive weight adjustment. The corresponding stationarity analysis showed stabilization before the final simulation step used for analysis (Fig S9).

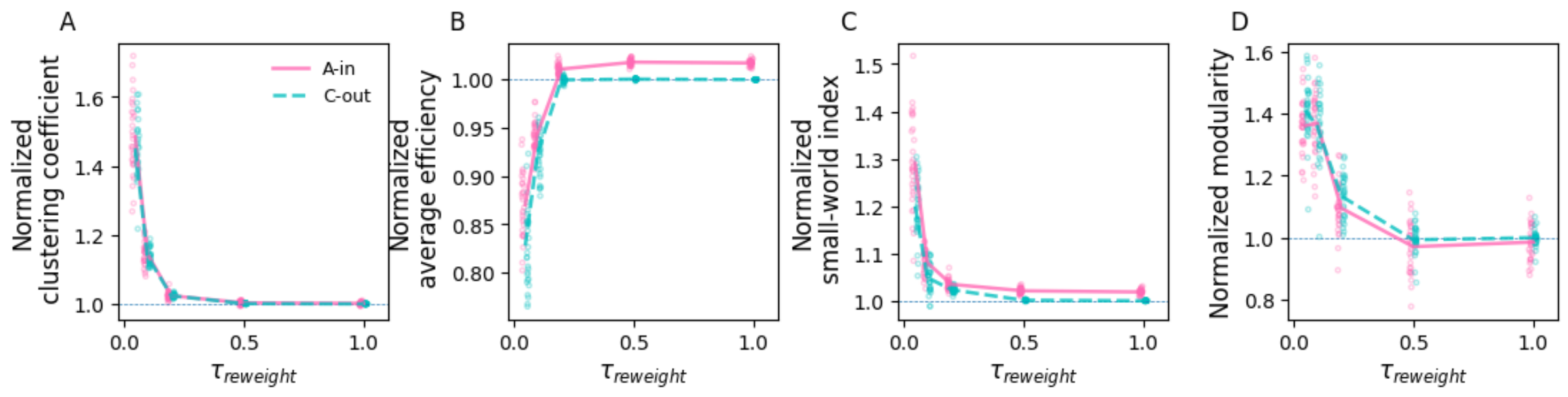

**Fig S8. Adaptive weight adjustment with random rewiring increases clustering coefficient, small-world, and modularity index and reduces efficiency when $\tau_{reweight}$ is small.** Points represent values from 30 individual simulated networks and lines connect the means across different $\tau_{reweight}$ values.

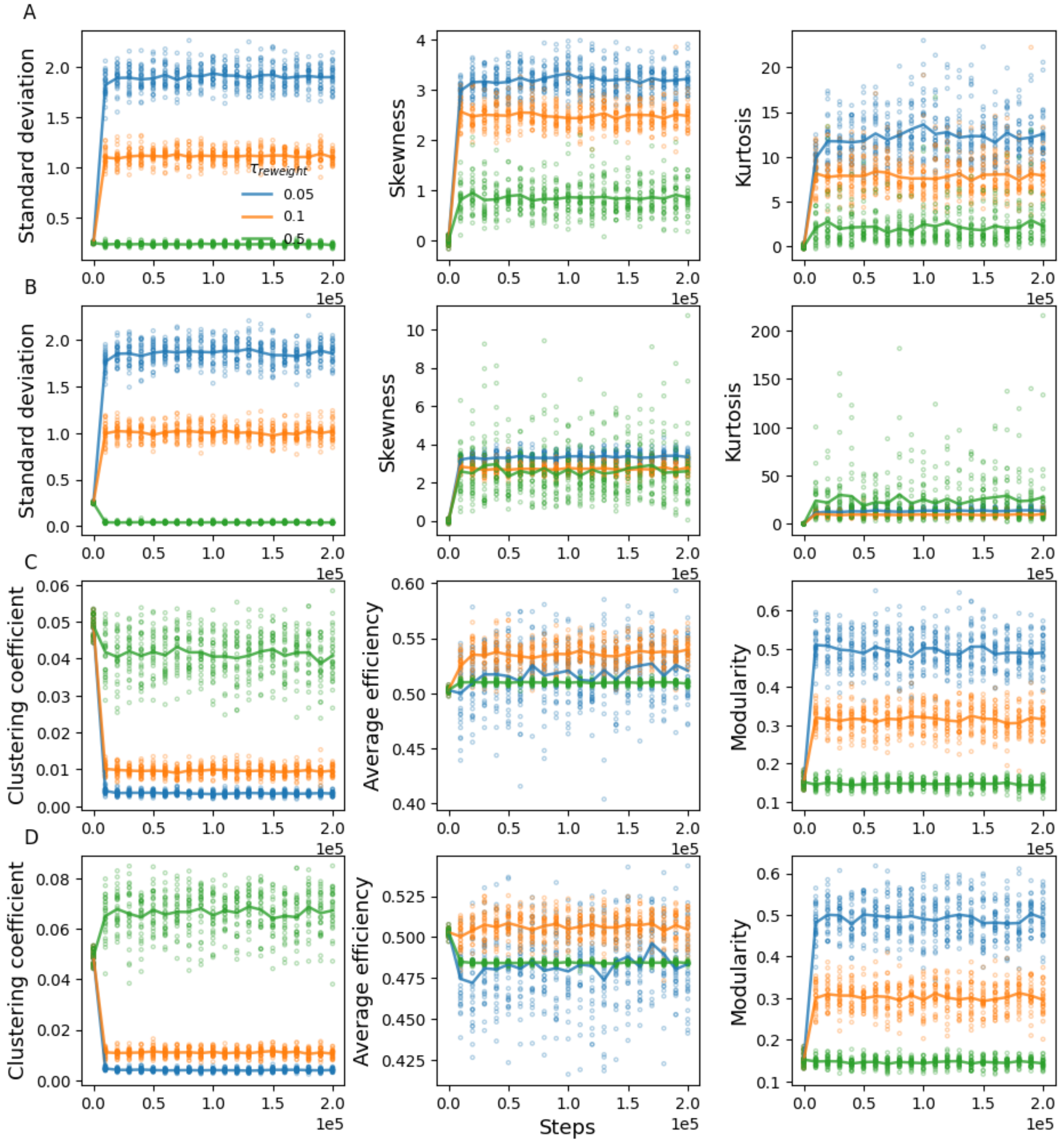

**Fig S9. Weight distributions and network topology reach stationarity during adaptive weight adjustment with intermittent random rewiring.** (A, B) Summary statistics of weight distributions as functions of the step of adaptive weight adjustment across different $\tau_{reweight}$ in the A-in (A) and C-out (B) conditions. (C, D) Summary connectivity metrics as functions of simulation step for the same simulations in the A-in (C) and C-out (D) conditions. Points represent values from 30 individual simulated networks and lines show the means across networks at different steps.

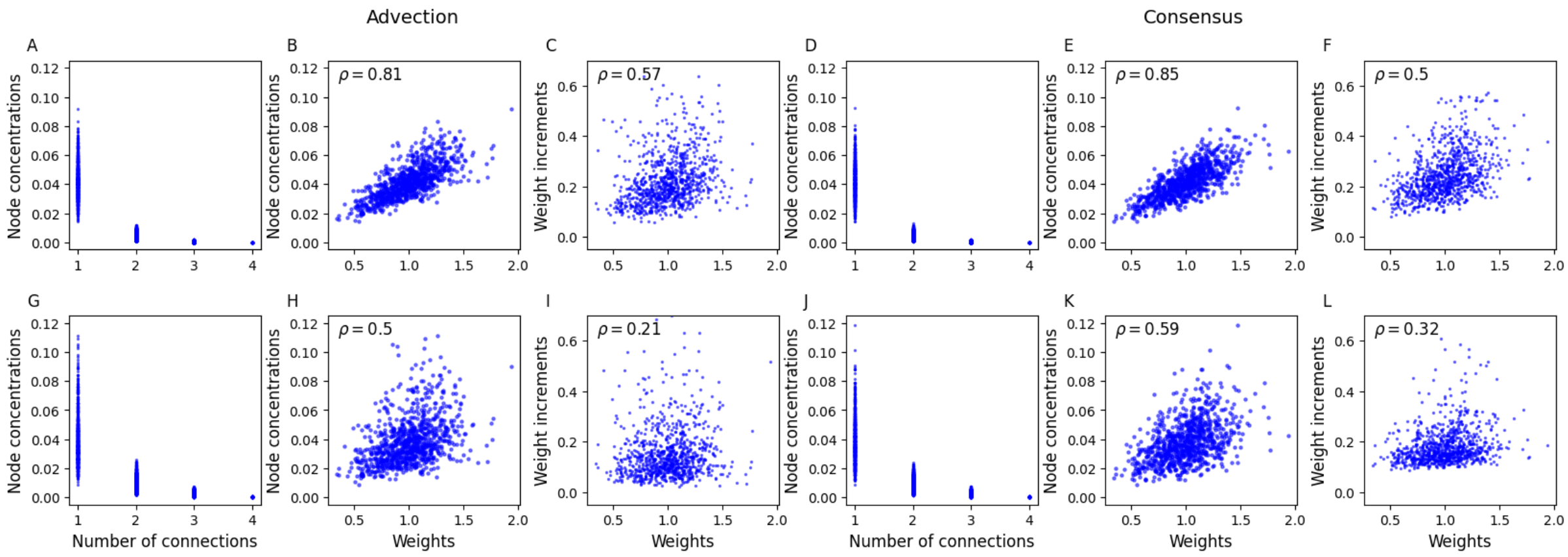

**Fig S10. Adaptive weight adjustment at small $\tau_{reweight}$ values resembles multiplicative dynamics.** (A, D) At $\tau_{reweight} = 0.1$, node concentrations decrease rapidly with increasing steps from the source of dynamics. (B, E) Concentrations of neighbors are strongly correlated with their connection weights to the candidate node. (C, F) Increments (Eq. 13) in connections are likewise correlated with the current weights. (G, J) A slight increase $\tau_{reweight}$ ($\tau_{reweight} = 0.2$) enables broader diffusion of concentrations across the network. (H, K) The correlation between neighbor concentrations and connection weights is reduced. (I, L) Weight increments become less dependent on the current weights. Correlation coefficients were computed for each node by grouping its outgoing connections (A-in condition) or incoming connections (C-out condition), then aggregated across nodes using Fisher's z-averaging (9). All panels are computed using an initial random network with normally distributed weights, as described in Methods. Node concentrations were obtained by using each node once as the source node and were then pooled across source nodes.

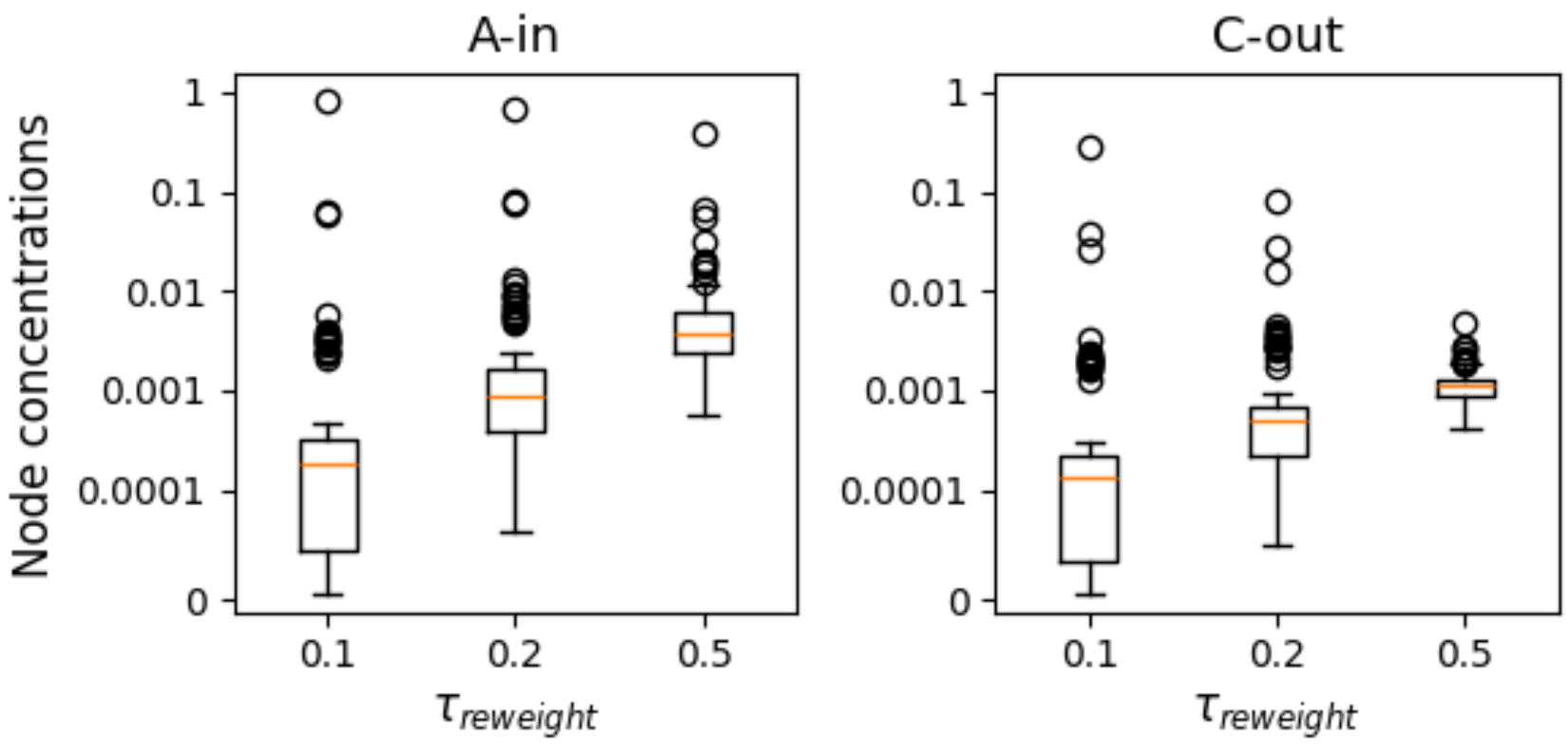

**Fig S11. Node concentrations homogenize as $\tau_{reweight}$ increases.** All panels were computed using the same initial random network to Fig S10. Node 0 was set to the source of dynamics.

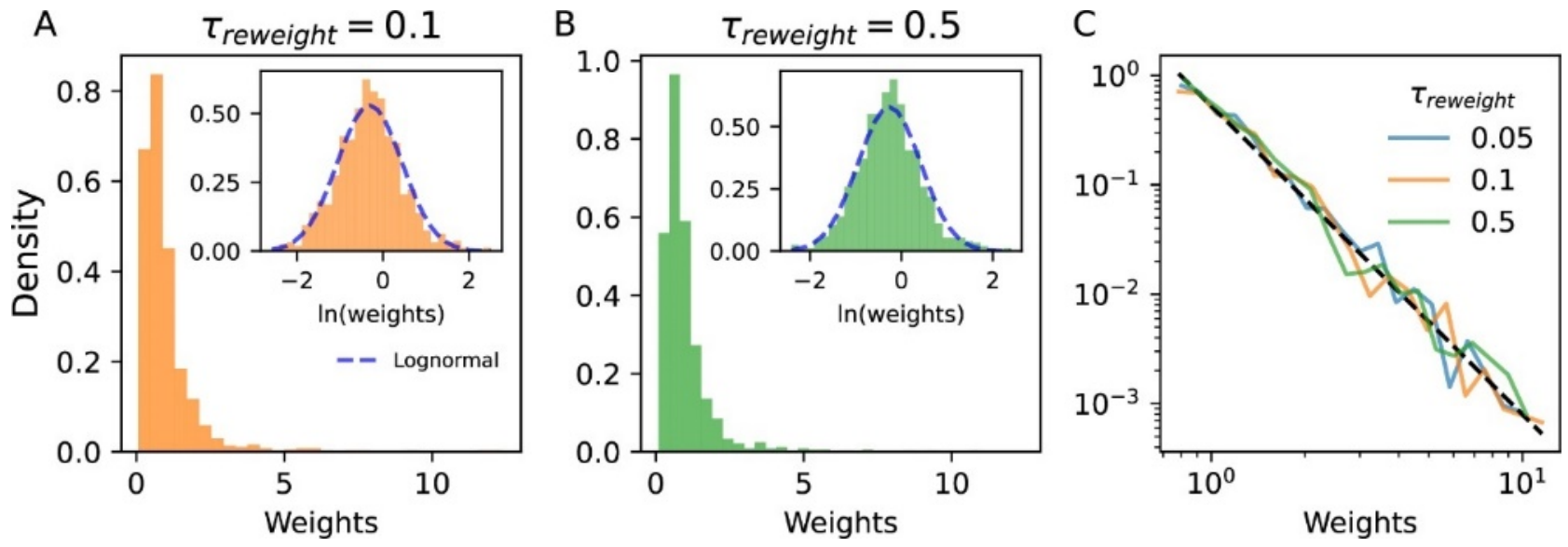

**Fig S12. Dual-adaptive algorithm produces heavy-tailed weights for a wider range of $\tau_{reweight}$ in the C-out condition.** (A, B) Distribution of connection weights of single representative simulated networks generated at $\tau_{reweight}$ is (A) 0.1 and (B) 0.5. (C) Power-law fit of upper tails of connection weights from networks generated at $\tau_{reweight} = 0.05, 0.1, 0.5$. Colored lines represent upper tails of distribution of connection weights from single representative simulated networks, and black dashed line is the best power-law fits for $\tau_{reweight} = 0.1$.

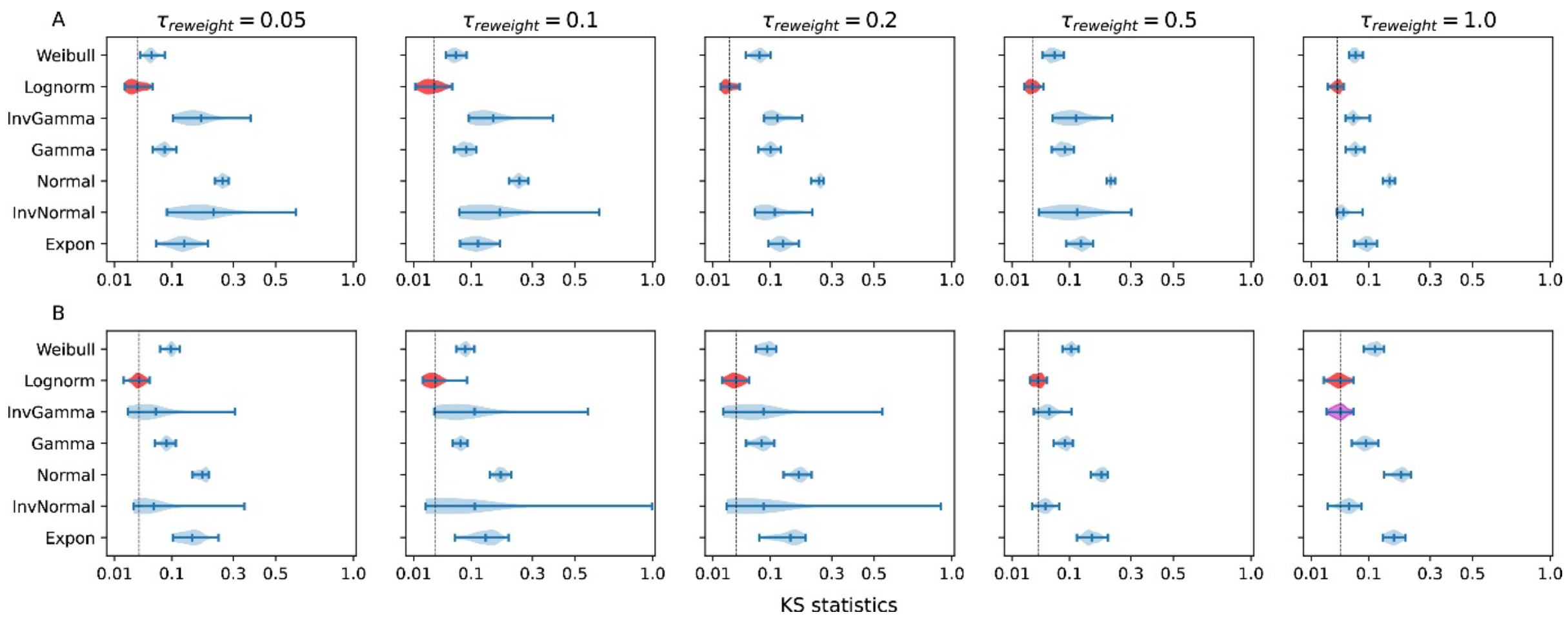

**Fig S13. KS statistics of candidate distributions fit across $\tau_{reweight}$.** (A) A-in condition. (B) C-out condition. Red indicates the distribution with the minimal mean (best fit) and the dashed line marks this minimum. Magenta highlights distributions with KS statistics statistically comparable to the best fit.

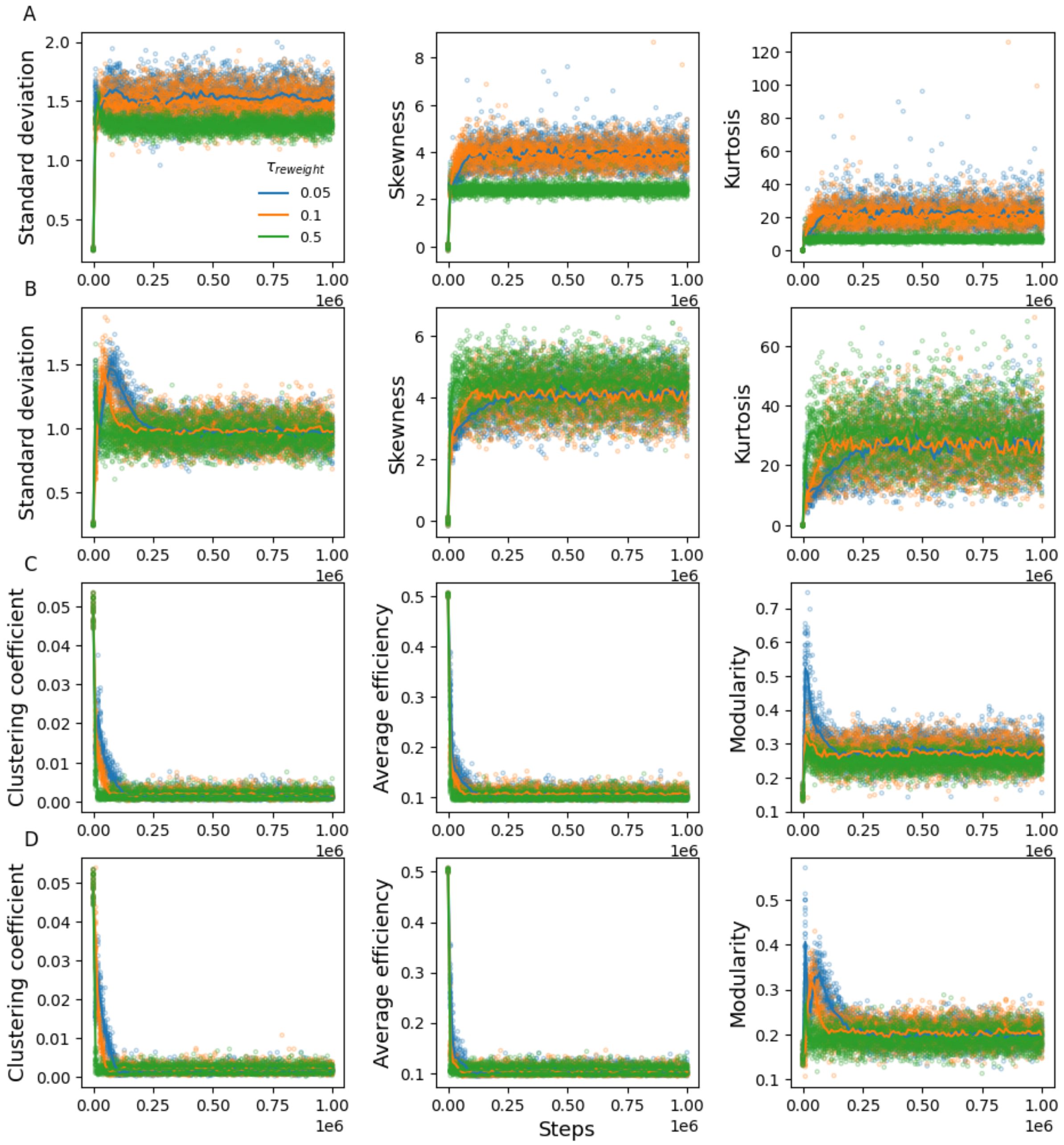

**Fig S14. Weight distributions and network topology reach stationarity during adaptive weight adjustment with intermittent adaptive rewiring.** (A, B) Summary statistics of weight distributions as functions of the step of adaptive weight adjustment across different $\tau_{reweight}$ in the A-in (A) and C-out (B) conditions. (C, D) Summary connectivity metrics as functions of simulation step for the same simulations in the A-in (C) and C-out (D) conditions. Points represent values from 30 individual simulated networks and lines show the means across networks at different steps.

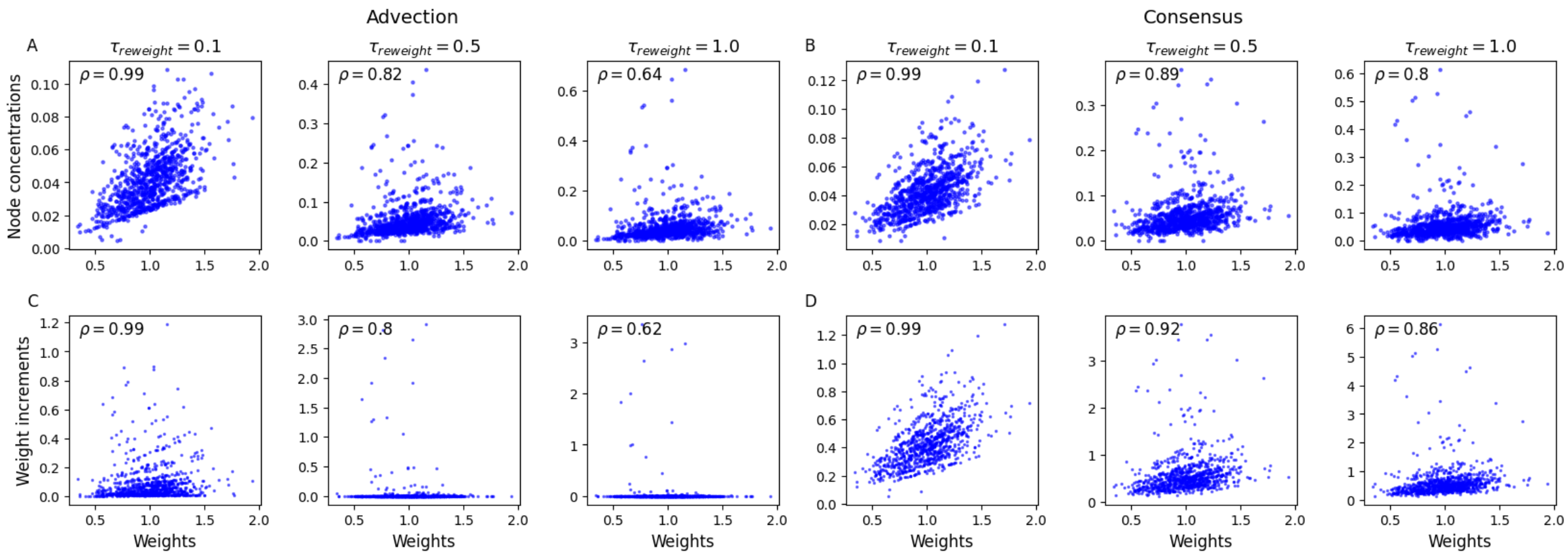

**Fig S15. Network topology induced by adaptive rewiring works against homogenizing tendency of the homeostatic nature of consensus and advection dynamics.** (A, B) The correlation between concentrations of neighbors and their connection weights to the source node persists as $\tau_{reweight}$ increases when concentrations follow (A) advection and (B) consensus dynamics. (C, D) Weight increments of connections are likewise correlated with the connection weights when concentrations follow (C) advection and (D) consensus dynamics. Correlation coefficients were computed for each node by grouping its outgoing connections (A-in condition) or incoming connections (C-out condition), then aggregated across nodes using Fisher's z-averaging (9). All panels are computed using a network generated by applying adaptive rewiring for $5 \times 10^5$ steps to the initial random network with normally distributed weights used in Fig. S10. Node concentrations were obtained by using each node once as the source node and were then pooled across source nodes.

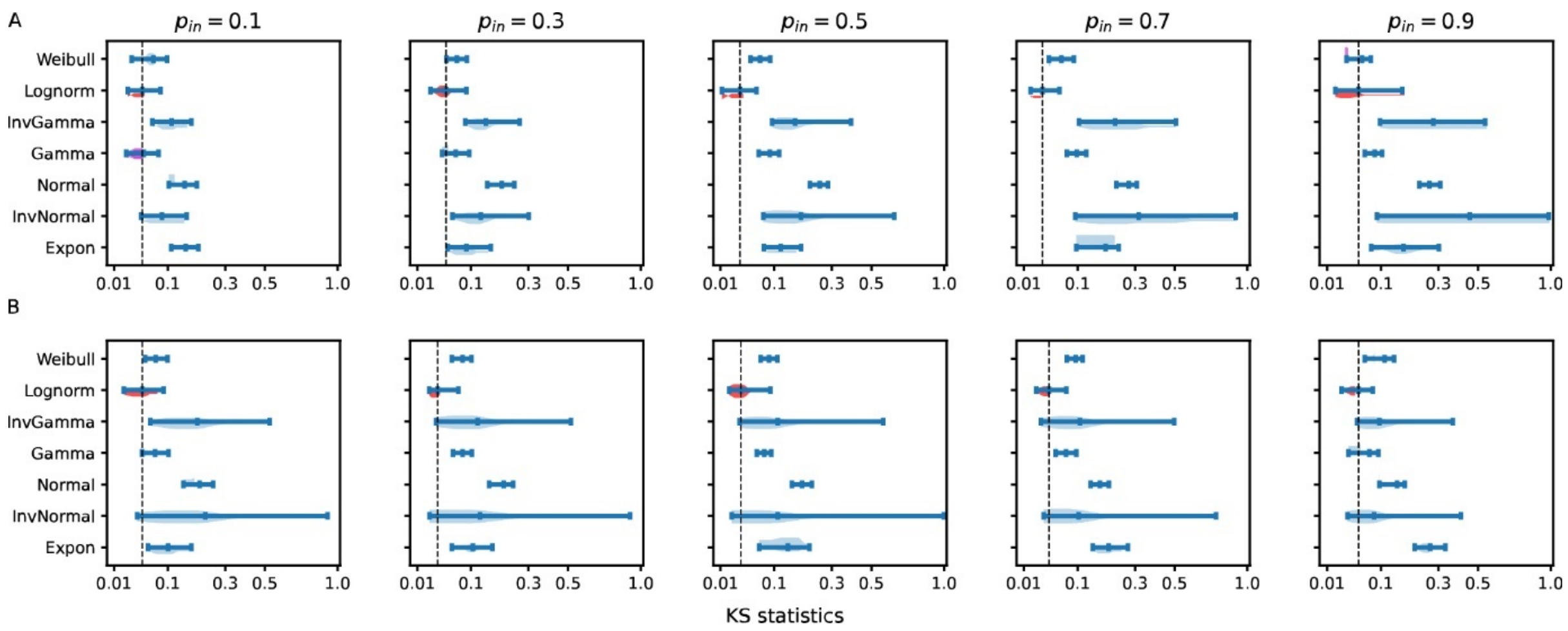

**Fig S16. KS statistics of candidate distributions fit across $p_{in}$.** (A) A-in condition. (B) C-out condition. $\tau_{reweight} = 0.1$. Red indicates the distribution with the minimal mean (best fit) and the dashed line marks this minimum. Magenta highlights distributions with KS statistics statistically comparable to the best fit.

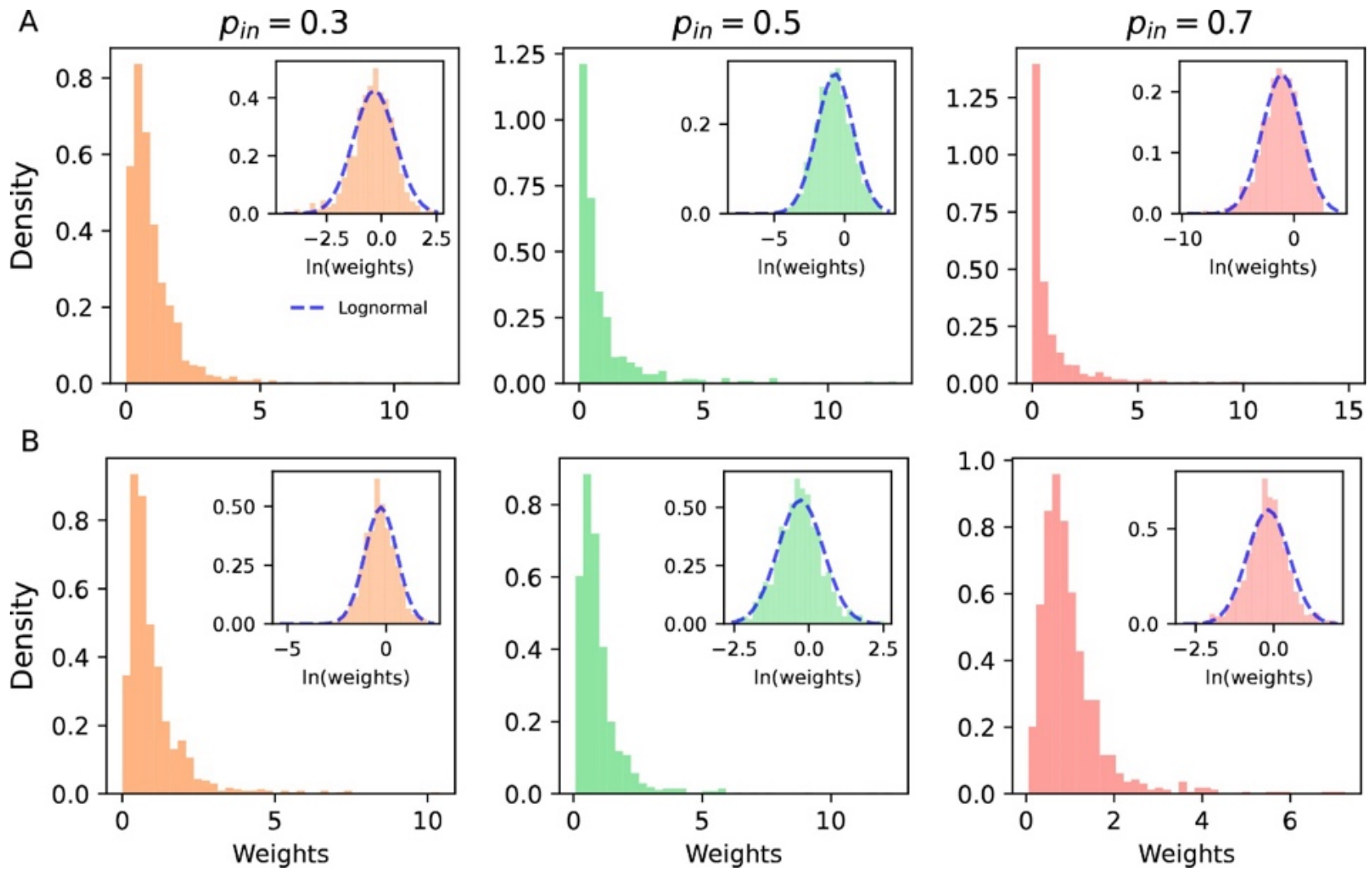

**Fig S17. Heavy-tailed weights from networks generated at $\tau_{reweight} = 0.1$ and various $p_{in}$.** (A) A-in condition. (B) C-out condition. Each histograms shows the distribution of connection weights from a single representative simulated network.

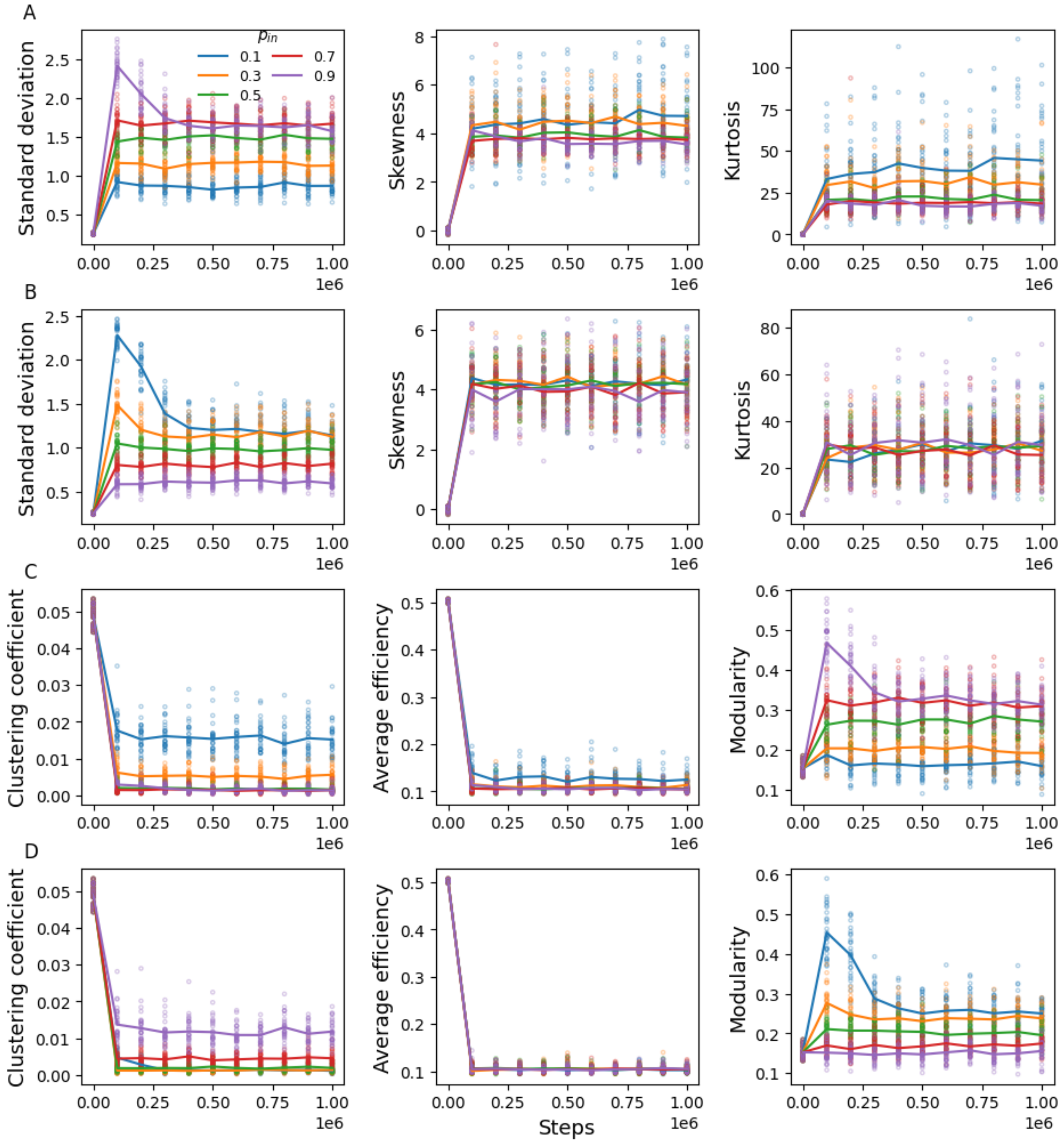

**Fig S18. Weight distributions and network topology reach stationarity during adaptive weight adjustment with intermittent adaptive rewiring under various $p_{in}$.** (A, B) Summary statistics of weight distributions as functions of the step of adaptive weight adjustment in the A-in (A) and C-out (B) conditions. (C, D) summary connectivity metrics as functions of simulation step for the same simulations in the A-in (C) and C-out (D) conditions. In these simulations, $\tau_{reweight} = 0.1$. Points represent values from 30 individual simulated networks and lines show the means across networks at different steps.

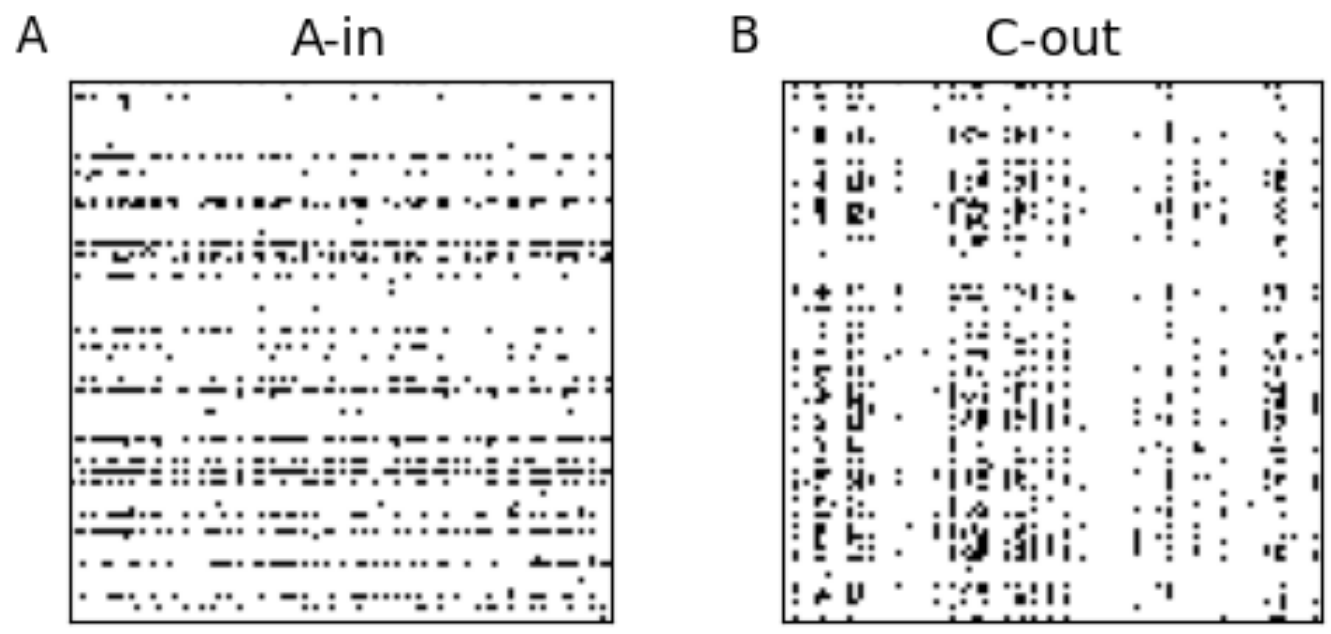

**Fig S19. Horizontal and vertical lines of the adjacency matrix representing converging and diverging hubs respectively in the emerged network.** (A, B) Adjacency matrices of single representative simulated networks generated by the settings in the caption of Fig. 8 when $\tau_{reweight} = 0.1$ and $r = 10$ in (A) A-in and (B) C-out conditions.

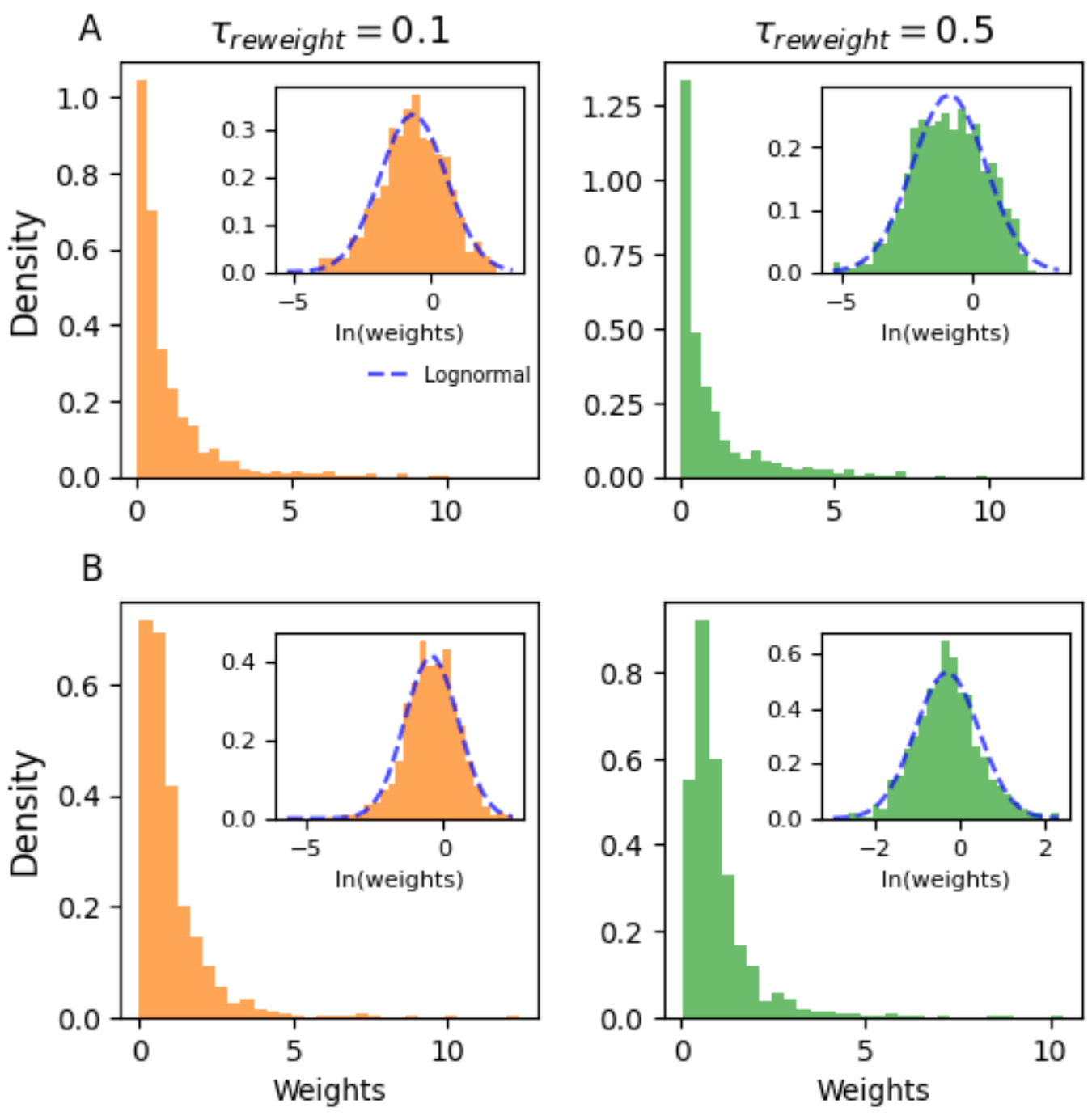

**Fig S20. Heavy-tailed weights from networks generated by the settings in the caption of Fig 8.** (A) A-in condition. (B) C-out condition. Each histograms shows the distribution of connection weights from a single representative simulated network.

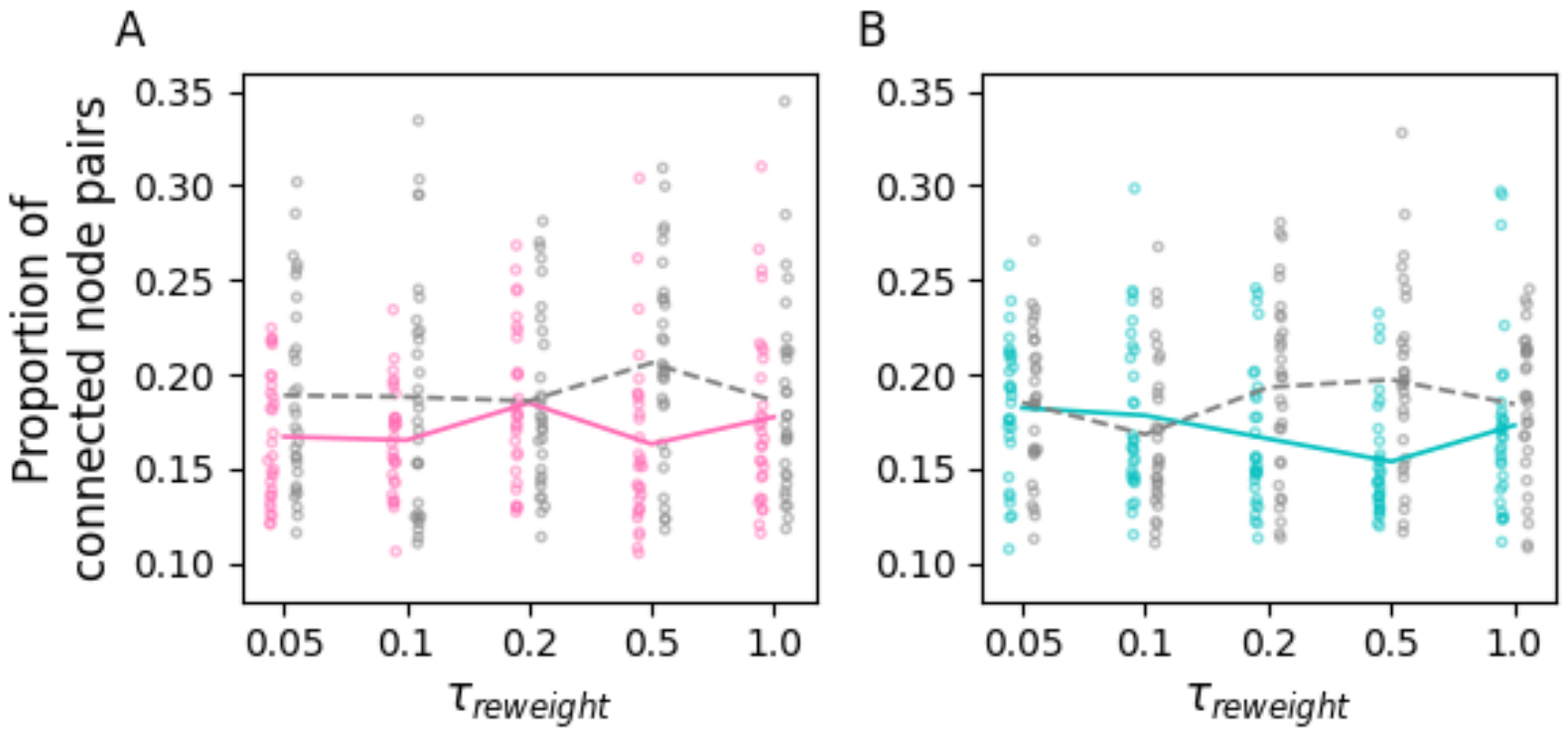

**Fig S21. Adaptive weight adjustment does not qualitatively change overall connectivity.** Proportion of connected node pairs in the network in the A-in (A) and C-out (B) conditions across different $\tau_{reweight}$. Colored points indicate 30 individual simulated networks, and colored lines connect the means across different $\tau_{reweight}$. Grey points and dashed grey lines show the corresponding values for fixed-weight networks.

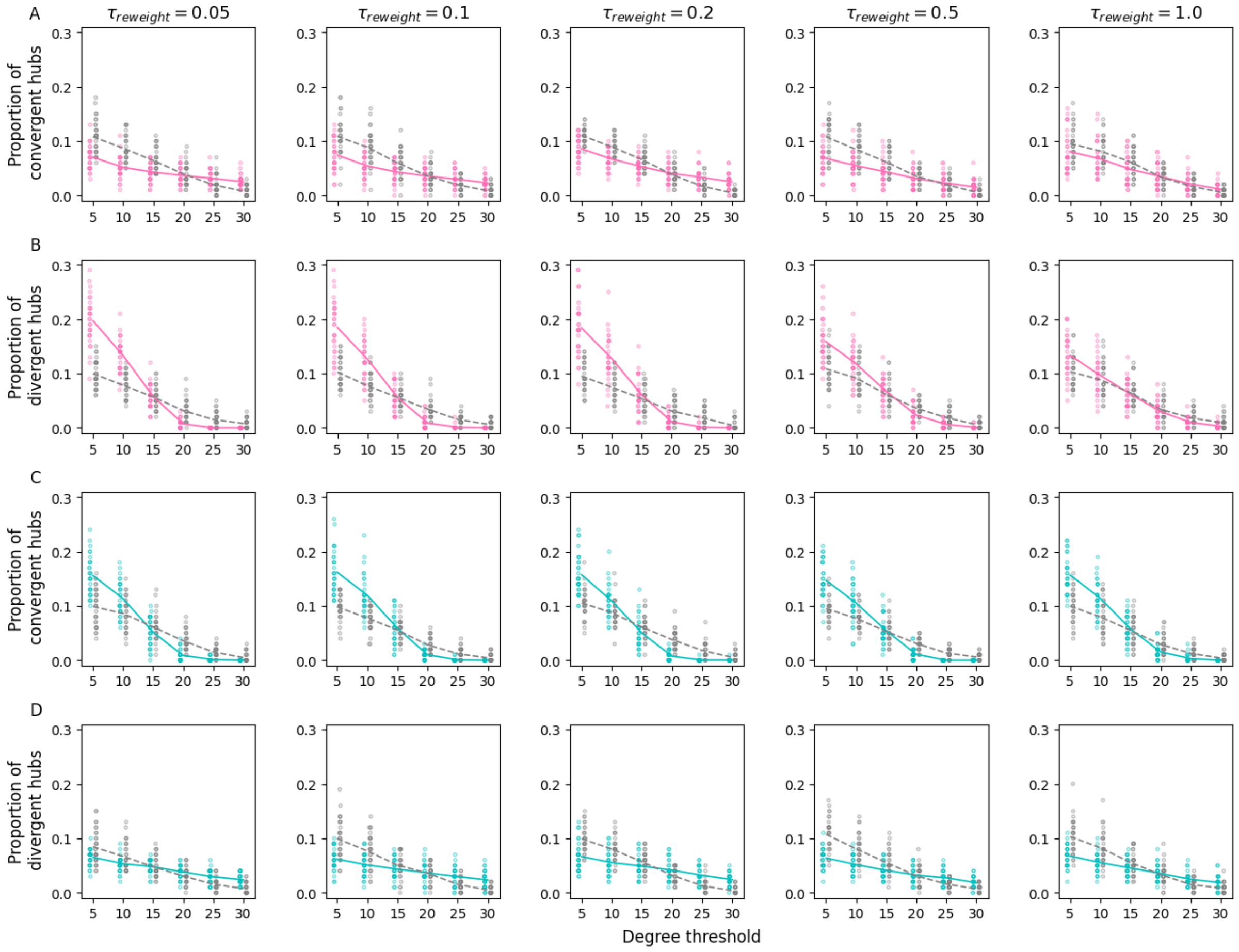

**Fig S22. The dual-adaptive algorithm alters the proportion of convergent and divergent hubs in a condition-dependent manner.** (A, B) Proportion of (A) convergent and (B) divergent hubs as a function of degree threshold for identifying hubs in the A-in condition across different $\tau_{reweight}$. (C, D) Corresponding results for the C-out condition. Colored points indicate 30 individual simulated networks, and colored lines connect the means across different $\tau_{reweight}$ values. Grey points and dashed grey lines show the corresponding values for fixed-weight networks.

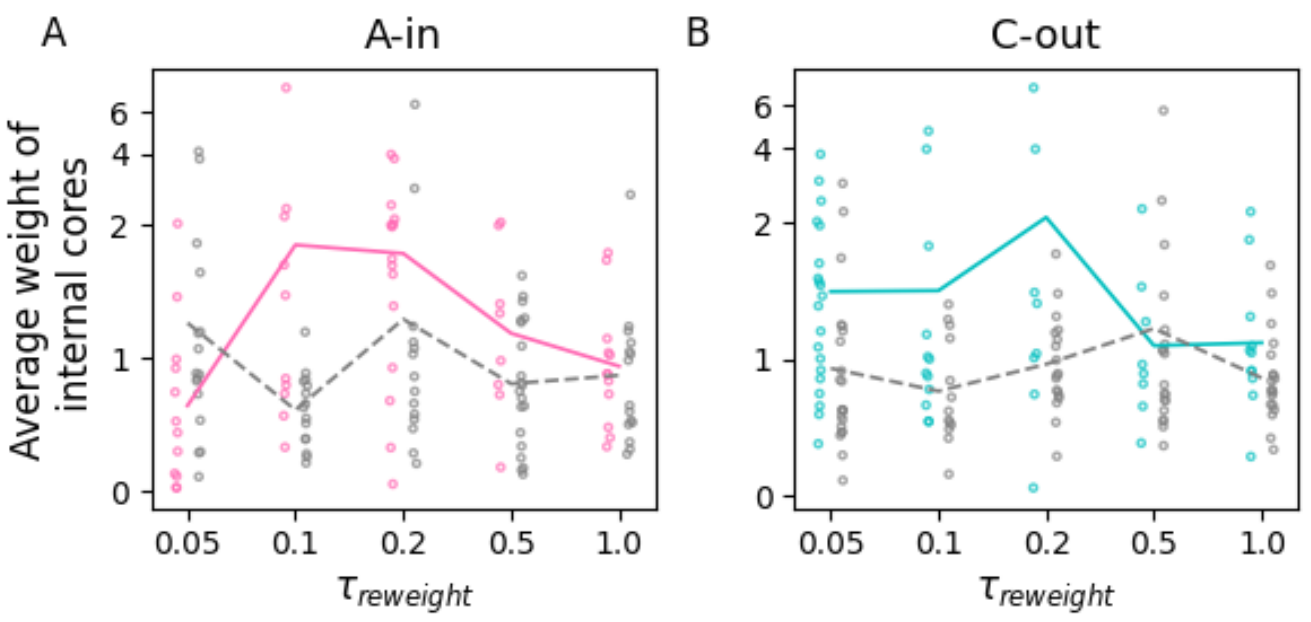

**Fig S23. Adaptive weight adjustment does not qualitatively increase average weight of the internal cores of convergent-divergent units.** Average weight of the internal cores in the A-in (A) and C-out (B) conditions across different $\tau_{reweight}$. Colored points indicate 30 individual simulated networks, and colored lines connect the means across different $\tau_{reweight}$. Grey points and dashed grey lines show the corresponding values for fixed-weight networks.

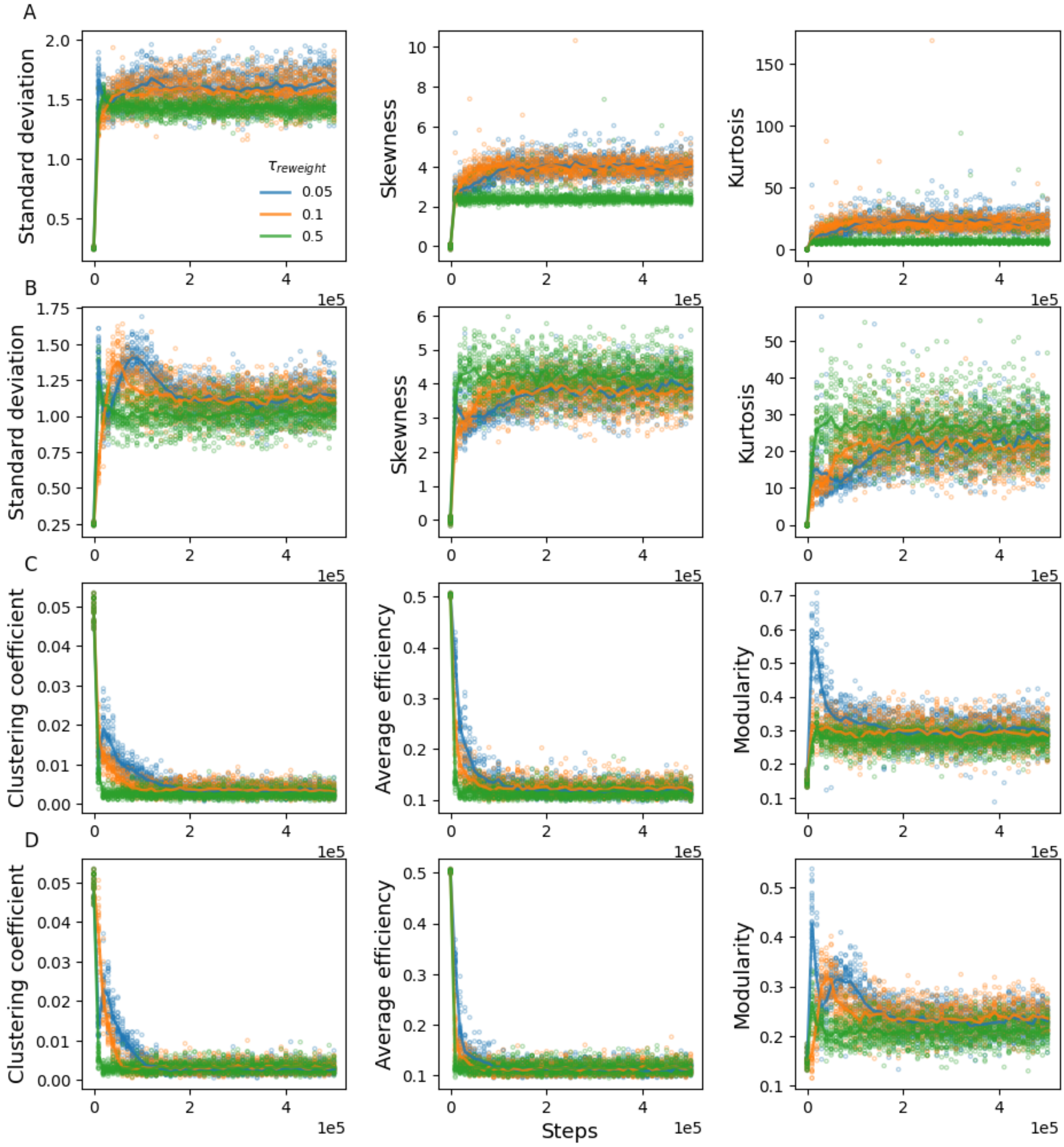

**Fig S24. Weight distributions and network topology of networks generated by the settings in the caption of Fig. 8 reach stationarity.** (A, B) Summary statistics of weight distributions as functions of the step of adaptive weight adjustment across different $\tau_{reweight}$ in the A-in (A) and C-out (B) conditions. (C, D) Summary connectivity metrics as functions of simulation step for the same simulations in the A-in (C) and C-out (D) conditions. Points represent values from 30 individual simulated networks and lines show the means across networks at different steps.

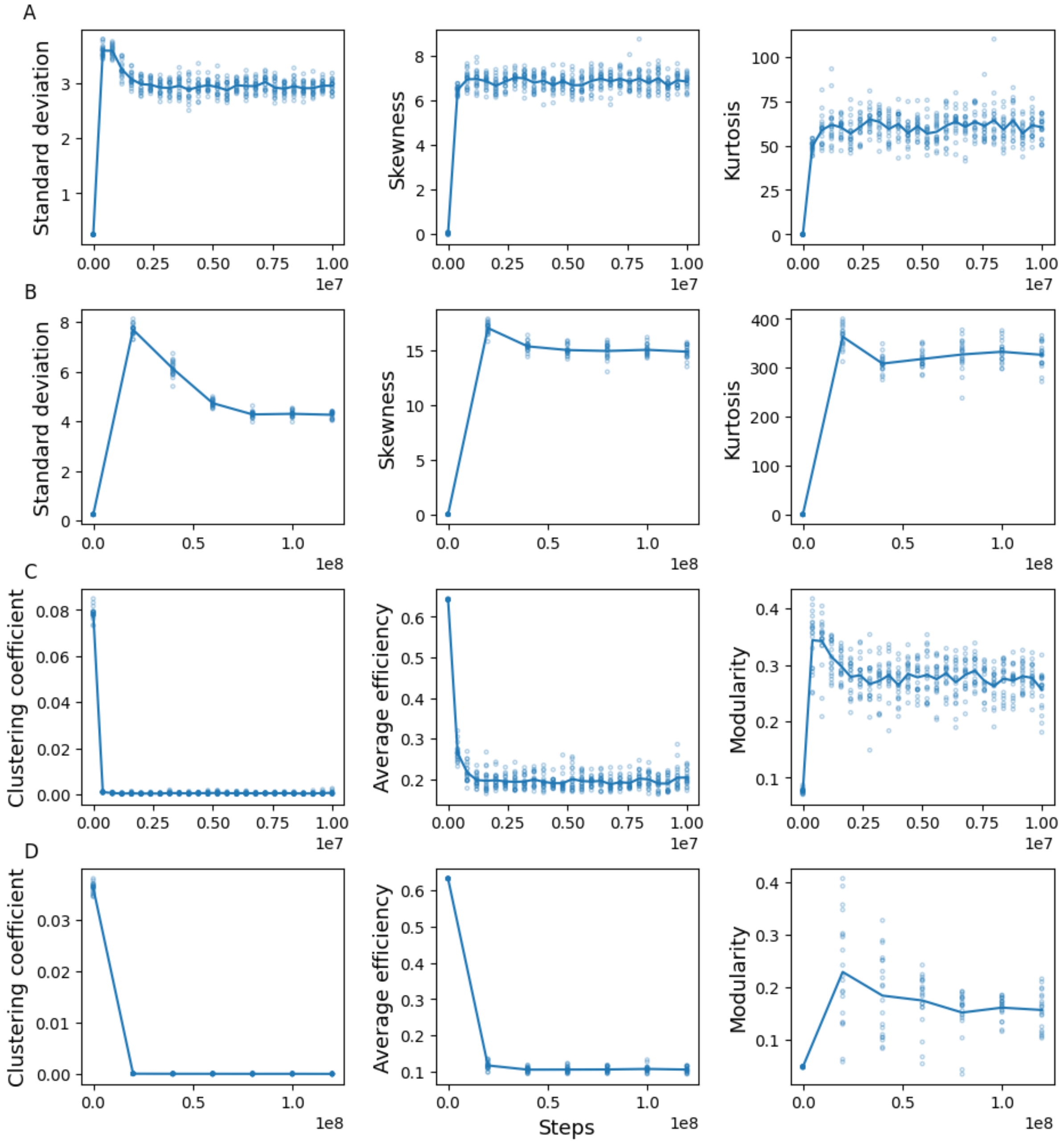

**Fig S25. Weight distributions and network topology reach stationarity in simulations fitted to empirical datasets.** (A, B) Summary statistics of the weight distributions as functions of simulation step for simulations fitted to the *C. elegans* connectome (A) and mouse retina connectome (B). (C, D) Summary connectivity metrics as functions of simulation step for the same simulations fitted to the C. elegans connectome (C) and mouse retina connectome (D). Points represent values from 20 individual simulated networks and lines connect the means across networks at different steps.

## Supplementary Note 1. Sensitivity analysis of the learning rate $\eta$

To assess whether the choice of $\eta$ affects the model outcomes, we ran simulations with $\eta = 1$ and 100, and compared them with the original setting $\eta = 10$ for $\tau_{reweight} = 0.05, 0.1$, and $0.5$. One instance was generated for each combination of $\eta$ and $\tau_{rewire}$. For the weight adjustment-alone condition and the weight adjustment-alone with intermittent random rewiring conditions, simulations were run for $N = \frac{2\times10^6}{\eta}$ steps of adaptive weight adjustment. For the dual-adaptive condition, simulations were run for $N = \frac{10^7}{\eta}$ steps of adaptive weight adjustment. The number of simulation steps was scaled inversely with $\eta$, so that the cumulative effect of adaptive weight adjustment was matched across different $\eta$.

For the weight adjustment-alone condition, the final weight distributions were essentially unchanged across $\eta$ values (Fig. S26). When weight adjustment was combined with either random or adaptive rewiring, smaller $\eta$ produced narrower weight distributions (Fig. S27, S28). However, the main qualitative conclusions remained unchanged. Connection weights in networks generated by adaptive weight adjustment with random rewiring decay rapidly as $\tau_{reweight}$ increases (Fig. S27), while this attenuation was not pronounced when adaptive rewiring was used (Fig. S28).

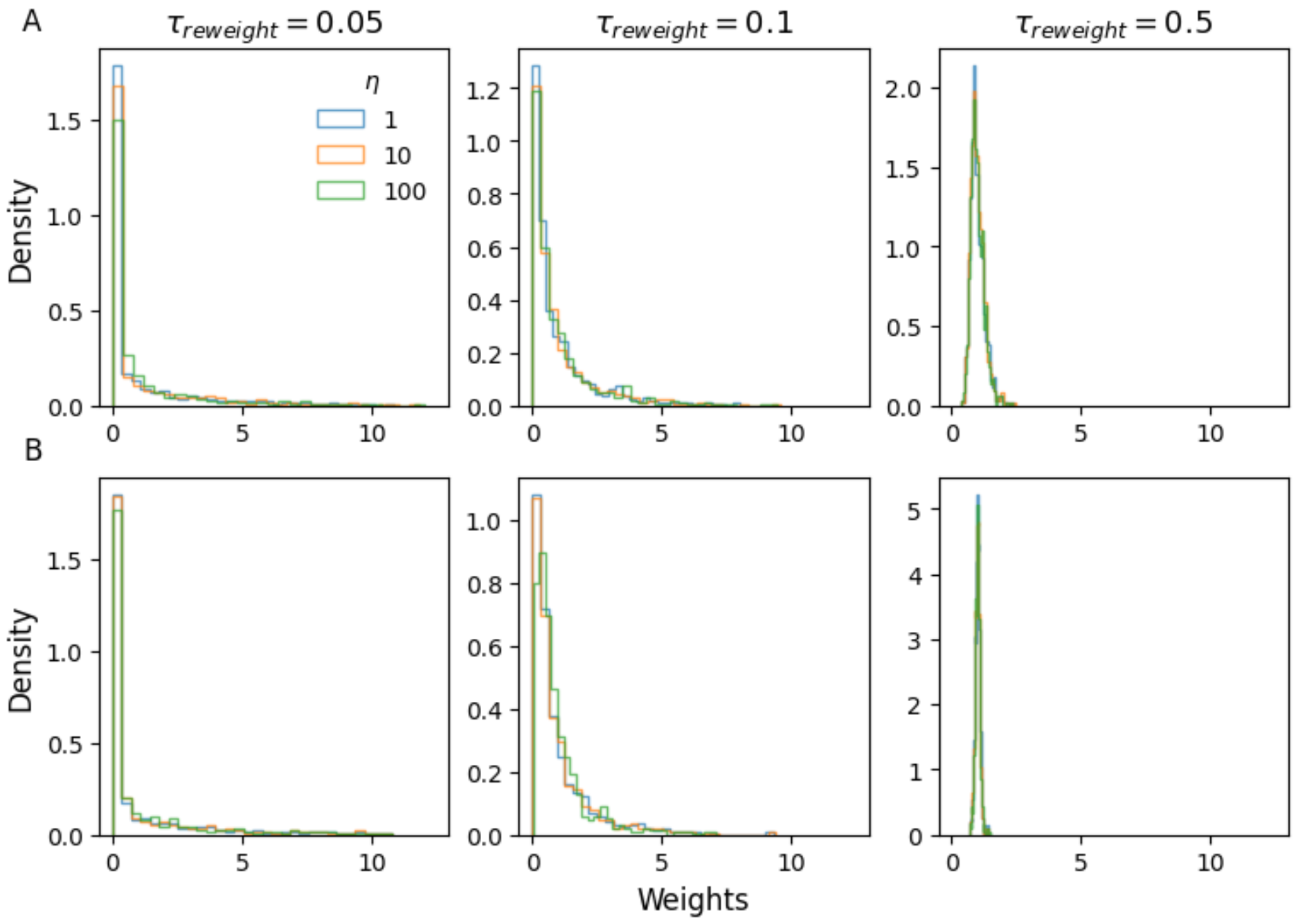

**Fig S26. Weight distribution is essentially unchanged across $\eta$ values when using adaptive weight adjustment alone.** (A, B) Weight distributions in the (A) A-in and (B) C-out conditions.

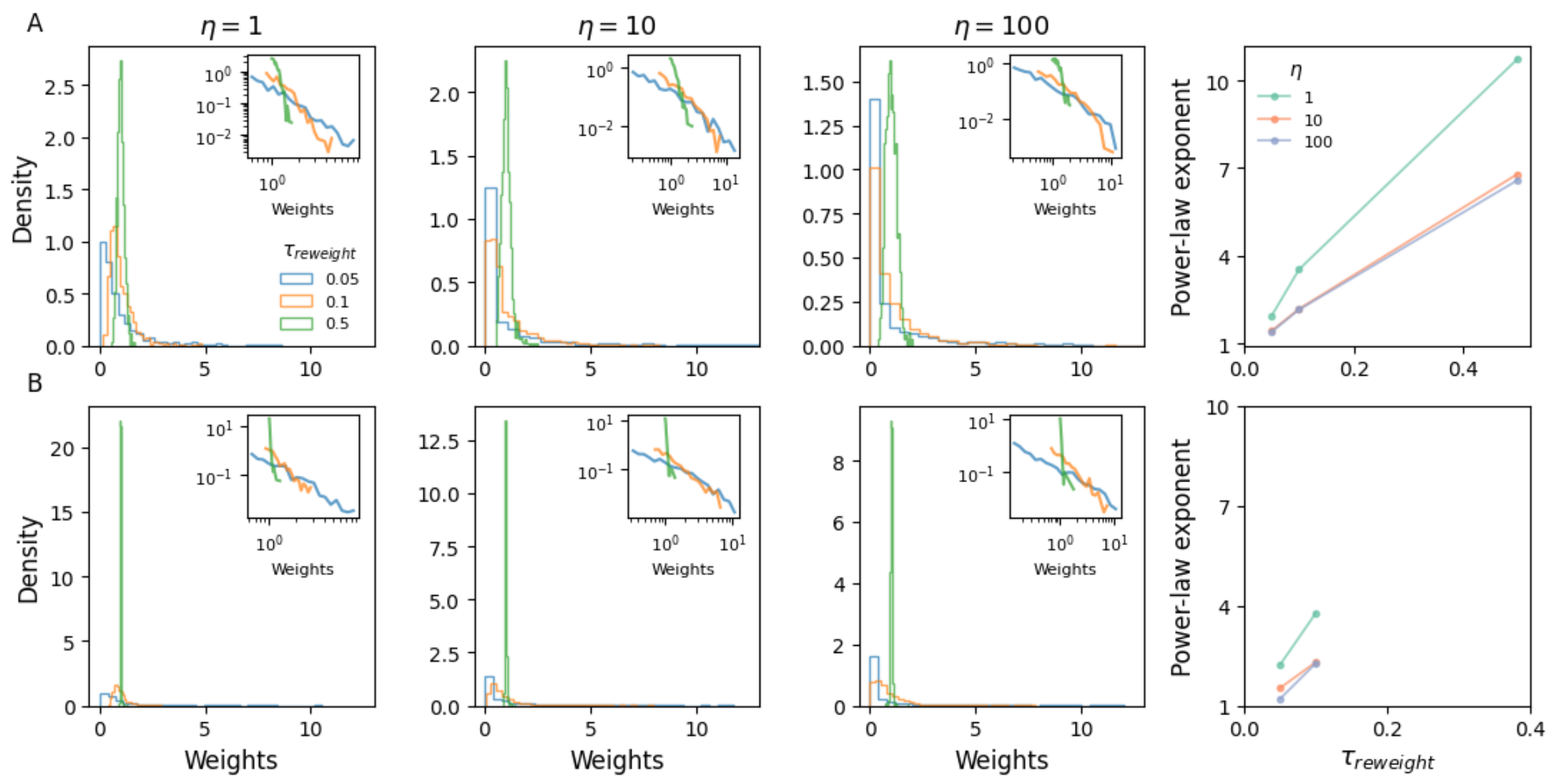

**Fig S27. Connection weights in networks generated by adaptive weight adjustment with random rewiring decay rapidly as $\tau_{reweight}$ increases.** (A, B) Weight distributions in the (A) A-in and (B) C-out conditions. Insets show the upper tails of connection weights from networks generated at $\tau_{reweight} = 0.05, 0.1, 0.5$. The rightmost panels show the estimated power-law exponents as a function of $\tau_{reweight}$. A point is drawn if the power-law fit achieved $R^2 > 0.85$.

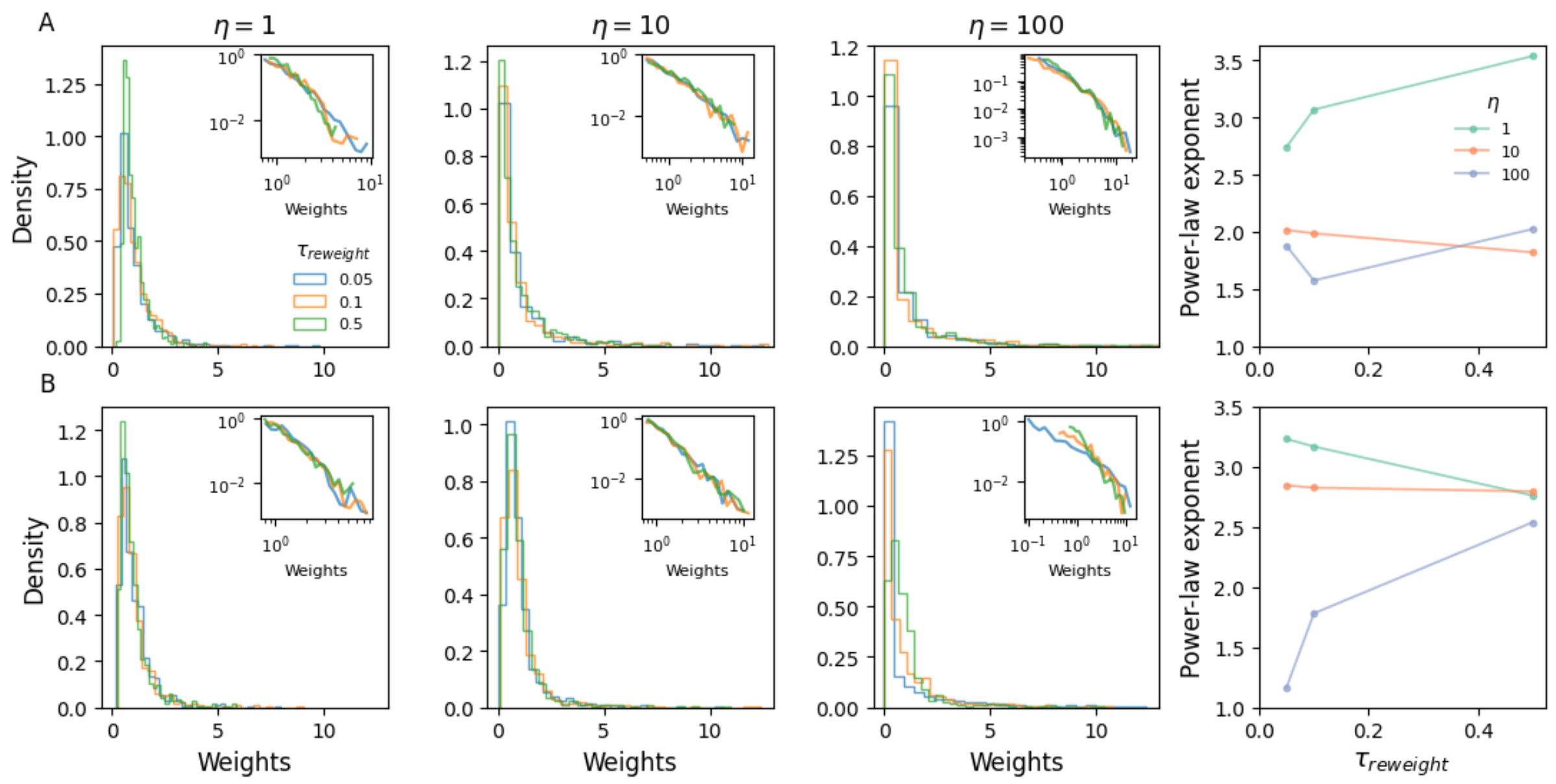

**Fig S28. Connection weights in networks generated by adaptive weight adjustment with adaptive rewiring show little attenuation as $\tau_{reweight}$ increases.** (A, B) Weight distributions in the (A) A-in and (B) C-out conditions. Insets show the upper tails of connection weights from networks generated at $\tau_{reweight} = 0.05, 0.1, 0.5$. The rightmost panels show the estimated power-law exponents as a function of $\tau_{reweight}$. A point is drawn if the power-law fit achieved $R^2 > 0.85$.

# Supplementary Note 2. Sensitivity analysis of network density

Biological recurrent excitatory circuits are generally sparse. In the experimental datasets and cortical microcircuit models we examined, recurrent excitatory connection probabilities were approximately in the range of 0.004–0.167 (10, 11). More extreme ultra-sparse cases have also been reported, such as near-zero local connectivity in the lateral hypothalamus (12).

To assess how network density affects the model outcomes, we ran simulations on networks ($n = 100$) with density 0.01, 0.05, 0.2 and 0.3 at $\tau_{reweight} = 0.05, 0.1,$ and $0.5$. Density 0.05 falls within the connection probability range reported for recurrent excitatory circuits, and density 0.2 was included as an upper end. Density 0.01 and 0.3 were used as sparse and dense boundary conditions, respectively. One instance was generated for each combination of network density and $\tau_{rewire}$. For the weight adjustment-alone condition and the weight adjustment-alone with intermittent random rewiring conditions, simulations were run for $N = 2 \times 10^5$ steps of adaptive weight adjustment. For the dual-adaptive condition, simulations were run for $N = 10^6$ steps of adaptive weight adjustment.

At densities 0.05 and 0.2, the main conclusions of the model remained qualitatively unchanged (Fig. S29, S30). Adaptive weight adjustment alone and in combination with random rewiring generated heavy-tailed weight distributions only for small values of $\tau_{reweight}$. By contrast, combining weight adjustment with adaptive rewiring produced heavy-tailed distributions across a wider range of $\tau_{reweight}$.

At two boundary cases, the model behavior depends on the underlying dynamics and rewiring processes. At the sparse boundary density 0.01, weight distributions showed longer right tails only when the network was rewired and the weights were adjusted using advection dynamics simultaneously (Fig. S31). At the dense boundary density 0.3, longer right tails were only observed when the network was rewired by adaptive rewiring and the weights were adjusted using advection dynamics simultaneously (Fig. S32).

Overall, these analyses indicate that the model is not density-invariant. Its main behavior is preserved within a biologically plausible density range, while very sparse and very dense networks represent boundary cases. However, the precise lower and upper boundaries should not be interpreted as universal thresholds. The number of connections that would render the random network weakly connected with probability one is $2\log(n)\,(n-1)$ (13). This corresponds to a density of approximately 0.09 when $n = 100$, but approximately 0.05 when $n = 200$. Thus, a fixed density value does not correspond to the same level of connectedness across different network sizes.

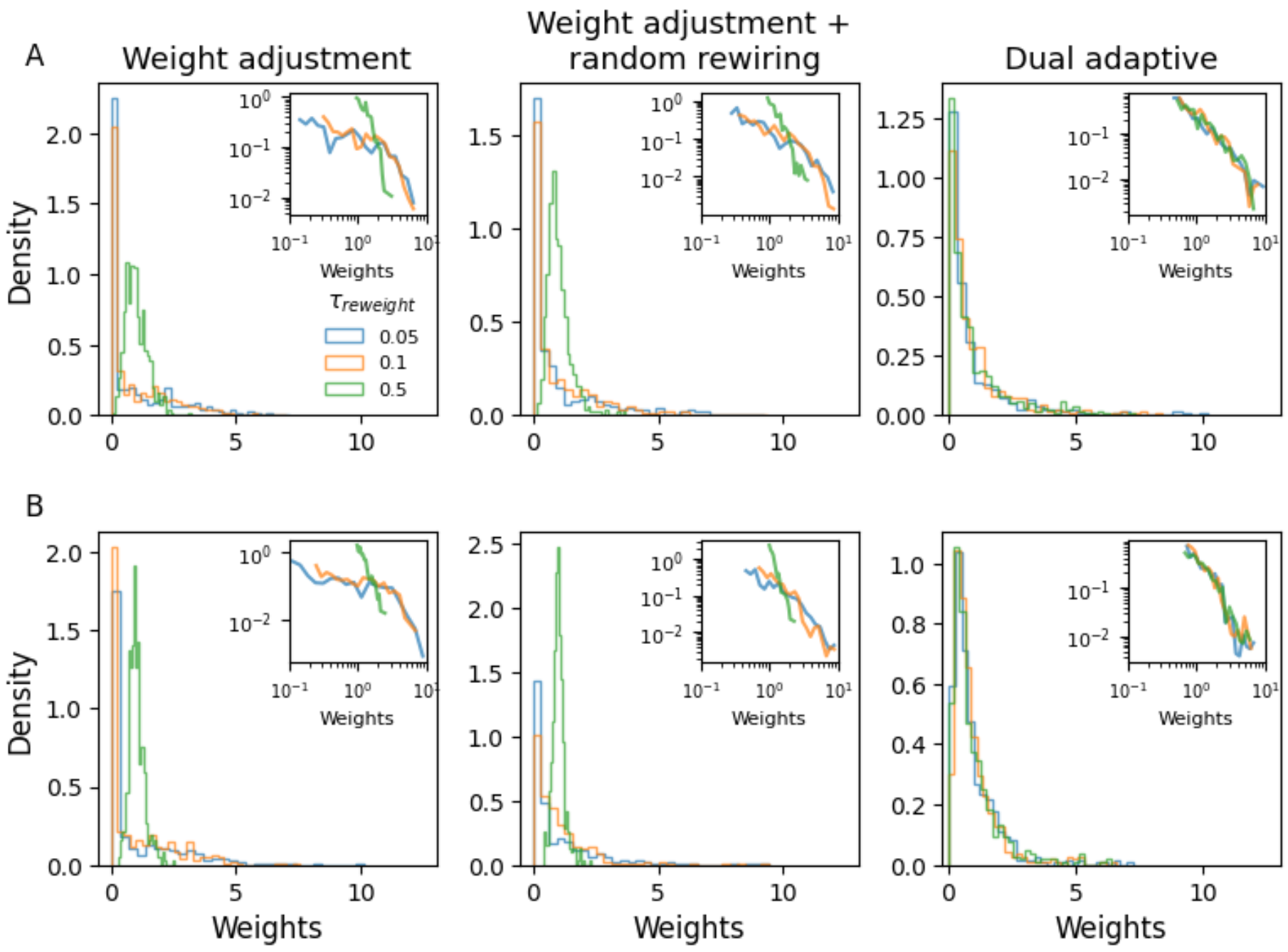

**Fig S29. Connection weights in networks with density = 0.05.** (A, B) Weight distributions in the A-in (A) and C-out (B) conditions. Insets show the upper tails of connection weights from networks generated at $\tau_{reweight} = 0.05, 0.1, 0.5$.

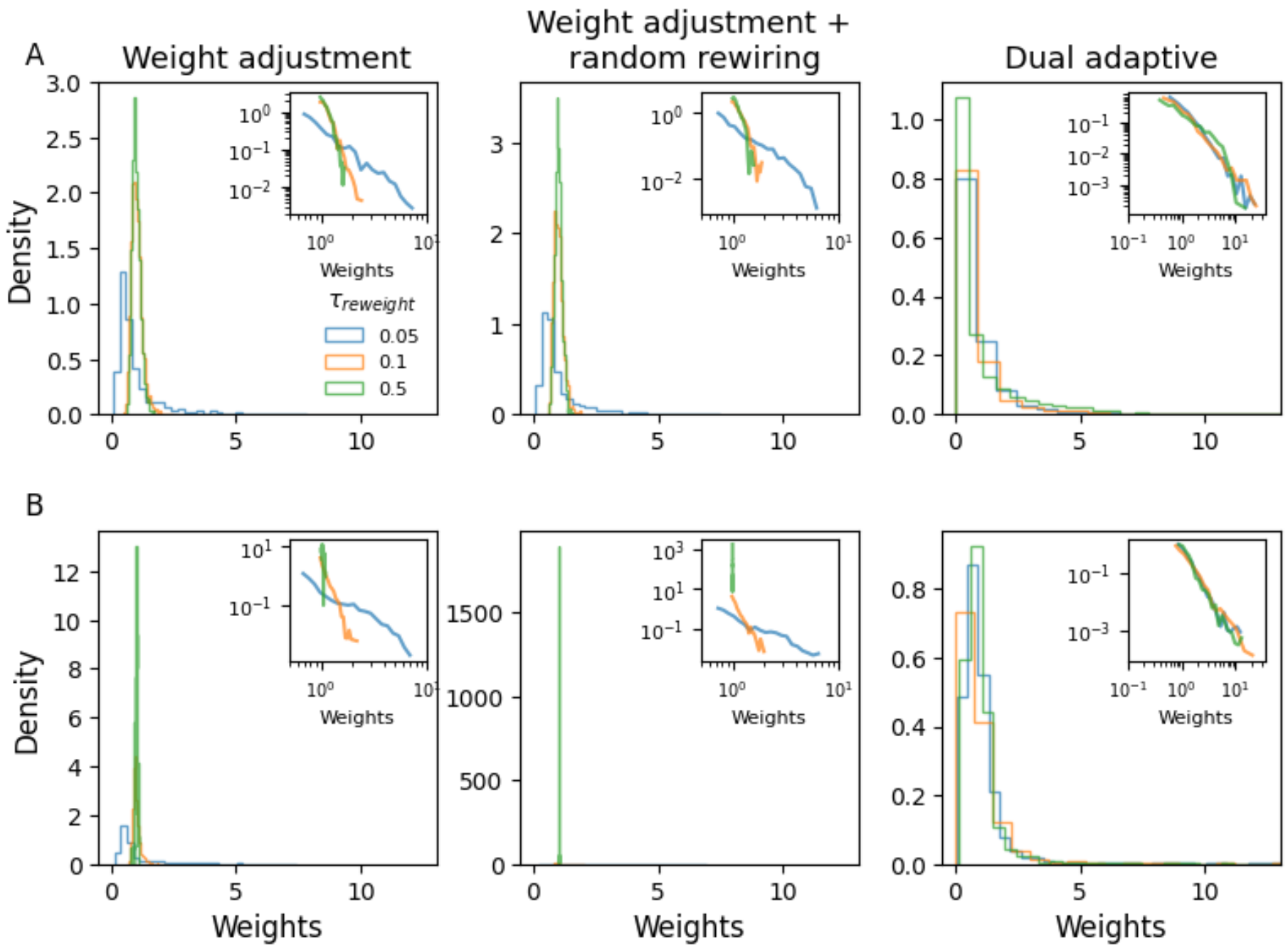

**Fig S30. Connection weights in networks with density = 0.2.** (A, B) Weight distributions in the A-in (A) and C-out (B) conditions. Insets show the upper tails of connection weights from networks generated at $\tau_{reweight} = 0.05, 0.1, 0.5$.

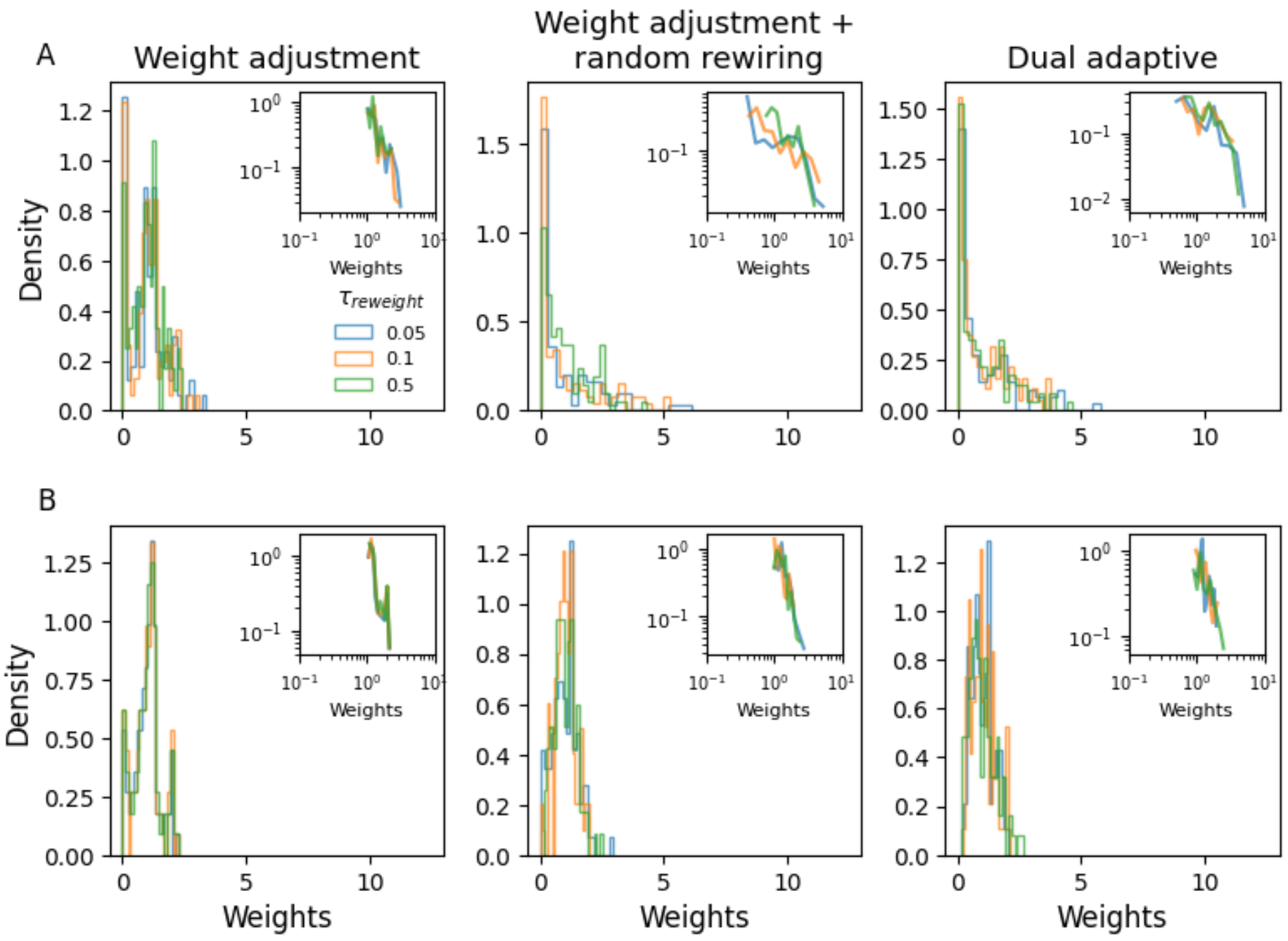

**Fig S31. Connection weights in networks with density = 0.01.** (A, B) Weight distributions in the A-in (A) and C-out (B) conditions. Insets show the upper tails of connection weights from networks generated at $\tau_{reweight} = 0.05, 0.1, 0.5$. For the power-law fit to weights exceeding the median, the interval between the median and the maximum connection weight was divided into 10 logarithmically spaced bins because of the small number of connections.

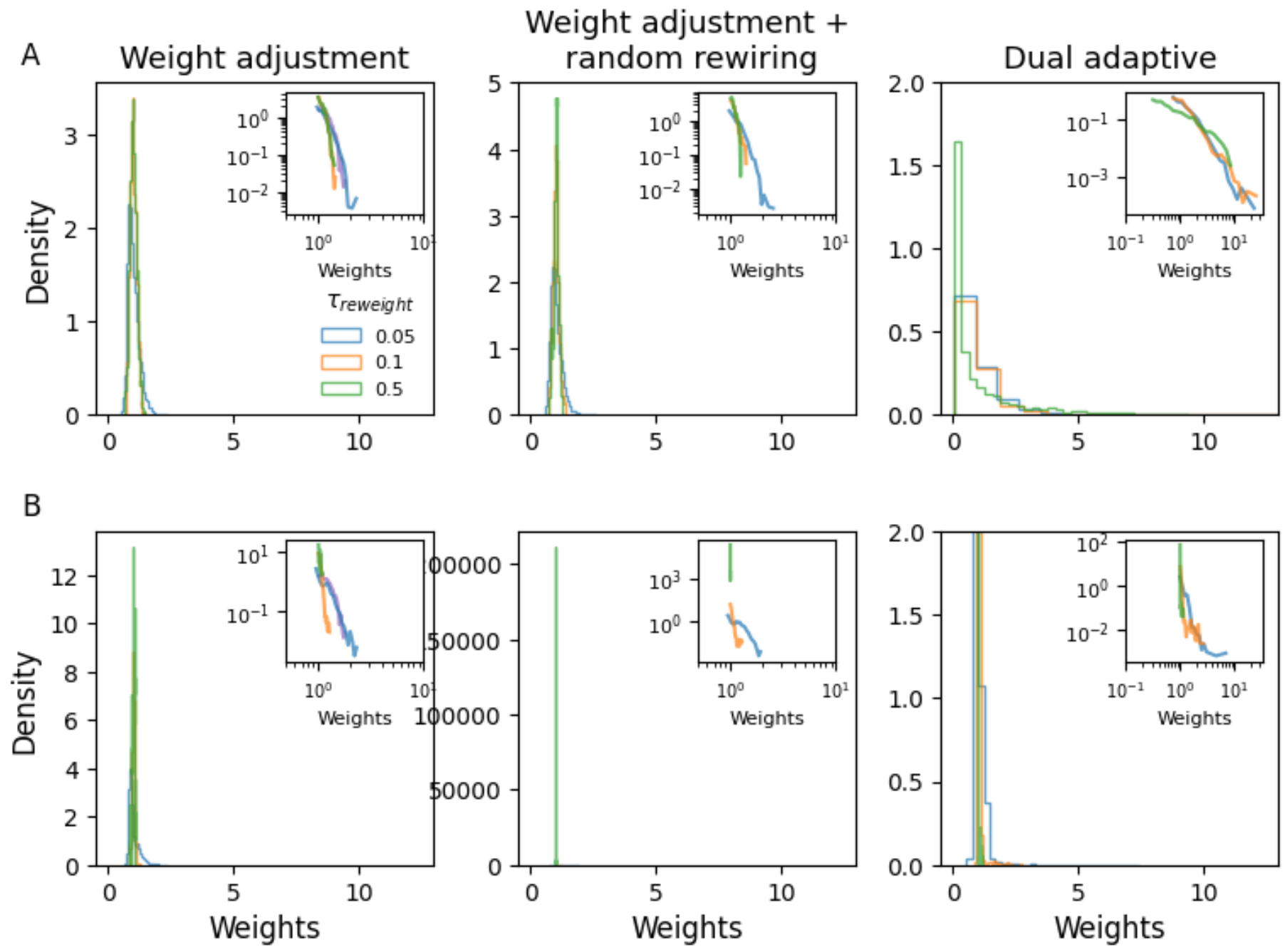

**Fig S32. Connection weights in networks with density = 0.3.** (A, B) Weight distributions in the A-in (A) and C-out (B) conditions. Insets show the upper tails of connection weights from networks generated at $\tau_{reweight} = 0.05, 0.1, 0.5$.

## Supplementary Table

| | Number of nodes | Number of connections |
|---|---|---|
| *C. elegans* | 185 | 5245 |
| Mouse retina | 1076 | 90009 |

**Table S1. Properties of neuronal networks.** The table summarizes the numbers of nodes and connections in the *C. elegans* and mouse retina connectomes used in this study.

## Supplementary References

1. F. Wilcoxon, Individual Comparisons by Ranking Methods. *Biom. Bull.* **1**, 80–83 (1945).

2. L. A. Waller, D. Smith, J. E. Childs, L. A. Real, Monte Carlo assessments of goodness-of-fit for ecological simulation models. *Ecol. Model.* **164**, 49–63 (2003).

3. G. Fagiolo, Clustering in complex directed networks. *Phys. Rev. E* **76**, 026107 (2007).

4. T. Opsahl, F. Agneessens, J. Skvoretz, Node centrality in weighted networks: Generalizing degree and shortest paths. *Soc. Netw.* **32**, 245–251 (2010).

5. V. Latora, M. Marchiori, Efficient Behavior of Small-World Networks. *Phys. Rev. Lett.* **87**, 198701 (2001).

6. M. D. Humphries, K. Gurney, Network 'Small-World-Ness': A Quantitative Method for Determining Canonical Network Equivalence. *PLOS ONE* **3**, e0002051 (2008).

7. A. Arenas, J. Duch, A. Fernández, S. Gómez, Size reduction of complex networks preserving modularity. *New J. Phys.* **9**, 176 (2007).

8. E. A. Leicht, M. E. J. Newman, Community Structure in Directed Networks. *Phys. Rev. Lett.* **100**, 118703 (2008).

9. R. A. Fisher, On the 'probable error' of a coefficient of correlation deduced from a small sample. *Metron* **1**, 3–32 (1921).

10. T. C. Potjans, M. Diesmann, The Cell-Type Specific Cortical Microcircuit: Relating Structure and Activity in a Full-Scale Spiking Network Model. *Cereb. Cortex* **24**, 785–806 (2014).

11. S. C. Seeman, *et al.*, Sparse recurrent excitatory connectivity in the microcircuit of the adult mouse and human cortex. *eLife* **7**, e37349 (2018).

12. D. Burdakov, M. M. Karnani, Ultra-sparse Connectivity within the Lateral Hypothalamus. *Curr. Biol.* **30**, 4063-4070.e2 (2020).

13. Erdös, P., A. Rényi, On random graphs I. *Publ. Math. Debr.* (1959).